\documentclass[a4paper,11pt]{article}

\usepackage{amssymb}
\usepackage{amsmath}
\usepackage{amscd}
\usepackage{latexsym}
\usepackage{epsfig}
\usepackage{color}
\usepackage{fancyhdr}
\usepackage[bf,footnotesize]{caption}
\usepackage{graphicx}
\usepackage{booktabs}
\usepackage{a4wide}
\usepackage{bbm} 
\usepackage{mathrsfs}
\usepackage{dsfont}
\usepackage{wasysym}
\usepackage{array}
\usepackage[all]{xy}
\usepackage{subfigure}

\usepackage{slashed}
\usepackage[numbers,sort&compress]{natbib}
\usepackage[colorlinks=true,a4paper=true,backref=false, linktocpage=true,
citecolor=blue,urlcolor=blue,linkcolor=blue,pdfpagemode=UseOutlines]{hyperref}

\hypersetup{%
  bookmarksnumbered=true,
  pdftitle = {},
  pdfsubject = {},
  pdfauthor = {U.~Wenger},
  pdfkeywords = {}
}

\graphicspath{{Figures_ExactResults/}}

\def\ln{\mathop{\rm ln}}                
\def\Dslash{\hbox{D}\kern-0.6em\raise0.15ex\hbox{/}} 

\def\bi{\begin{itemize}}
\def\ei{\end{itemize}}
\def\be{\begin{equation}}
\def\ee{\end{equation}}

\newcommand{\psibar}{\overline{\psi}}
\newcommand{\e}{\mathrm{e}}

\newcommand{\Tr}{\mathrm{Tr}}

\renewcommand{\O}{\mathcal{O}}

\newcommand{\Z}{\mathcal{Z}}

\newcommand{\C}{\mathcal{C}}

\date{\empty}

\begin{document}
\title{Supersymmetric quantum mechanics on the lattice:\\
II. Exact results}

\author{David Baumgartner and Urs Wenger\vspace{0.5cm}
\\
Albert Einstein Center for Fundamental Physics,\\
Institute for Theoretical Physics, University of Bern, \\
Sidlerstrasse 5, CH--3012 Bern, Switzerland\vspace{0.5cm}
\\
}

\maketitle

\begin{abstract}
  Simulations of supersymmetric field theories with spontaneously
  broken supersymmetry require in addition to the ultraviolet
  regularisation also an infrared one, due to the emergence of the
  massless Goldstino.  The intricate interplay between ultraviolet and
  infrared effects towards the continuum and infinite volume limit
  demands careful investigations to avoid potential problems. In this
  paper -- the second in a series of three -- we present such an
  investigation for ${\cal N}=2$ supersymmetric quantum mechanics
  formulated on the lattice in terms of bosonic and fermionic
  bonds. In one dimension, the bond formulation allows to solve the
  system exactly, even at finite lattice spacing, through the
  construction and analysis of transfer matrices. In the present paper
  we elaborate on this approach and discuss a range of exact results
  for observables such as the Witten index, the mass spectra and Ward
  identities.
\end{abstract}

\section{Introduction}

Regularising supersymmetric quantum field theories on a lattice in
order to investigate their nonperturbative properties remains to be a
challenging and demanding task. Besides the fact that the discreteness
of the space-time lattice explicitly breaks the Poincar\'e symmetry,
and hence supersymmetry itself, it can also be broken by specific
choices of the boundary conditions, in particular also by the finite
temperature. As a consequence, the effects from the ultraviolet and
infrared lattice regularisation are sometimes difficult to separate
from each other.  In addition, the restoration of supersymmetry in the
continuum and infinite volume limit is in general a delicate process
which requires careful fine-tuning or highly involved discretisation
schemes, both of which are sometimes difficult to control.  For these
reasons a complete and thorough understanding of the intricate
interplay between infrared and ultraviolet effects, when removing the
corresponding lattice regulators, is a crucial prerequisite for any
investigation of supersymmetric field theories on the lattice.

Supersymmetric quantum mechanics is a simple system which nevertheless
contains many of the important ingredients characterising
supersymmetric field theories. Moreover, in the path integral
formalism the system differs little from field theories in higher
dimensions and it is sufficiently involved to show similar complexity
and complications.  Hence, supersymmetric quantum mechanics provides
an adequate playground to address all the delicate questions and
issues mentioned above. In this paper -- the second in a series of
three -- we present exact results for ${\cal N} = 2$ supersymmetric
quantum mechanics discretised on the lattice using the bond
formulation. This formulation is based on the hopping expansion of the
original bosonic and fermionic degrees of freedom and is described in
detail in the first paper of our series
\cite{Baumgartner:2014nka}. For the fermions the bond formulation is
more appropriately termed fermion loop formulation since the fermionic
bond configurations turn out to be closed fermionic loops. In the case
of ${\cal N}=2$ supersymmetric quantum mechanics the fermion loop
formulation is particularly simple, since there are only two different
fermion loop configurations, namely one containing exactly one fermion
loop winding around the lattice in temporal direction, and one without
any fermion loop. The latter corresponds to the bosonic sector with
fermion number $F=0$ and the former to the fermionic sector with
$F=1$. This separation into the canonical sectors with fixed fermion
number forms the basis for the solution of the fermion sign problem
emerging in numerical Monte Carlo simulations of the quantum
mechanical system with broken supersymmetry. For a detailed discussion
of this issue we refer to the first paper in our series
\cite{Baumgartner:2014nka}.

In the present paper we make use of the fact that in the bond
formulation the weights of the bond configurations are completely
localised and the local bond configuration states can be enumerated locally
due to the discreteness of the new degrees of freedom. It is hence
straightforward to construct a transfer matrix which in turn can be
used to express the sum over all bond configurations, i.e., the
partition function, as the trace over an appropriate product of the
transfer matrix. As a consequence of the natural separation into
bosonic and fermionic contributions the transfer matrix block
diagonalises naturally into blocks with fixed fermion number, and this
simplifies the calculations considerably.  The transfer matrices
do not depend on the imaginary time coordinate and hence contain all the
physics of the system. It is therefore sufficient to understand the
spectral properties of the transfer matrices and calculate physical
observables such as the mass gaps directly from the eigenvalues of the
transfer matrices. More complicated observables such as correlation
functions and Ward identities can be calculated exactly using modified
transfer matrices which include appropriate source terms.

As discussed above, the exact results at finite lattice spacing are
most useful to gain a better understanding of the interplay between
the various limits required in any lattice calculation, not restricted
to supersymmetric quantum mechanics, in order to remove the infrared
and ultraviolet regulators. In particular, we can study in detail how
and under which circumstances supersymmetry is restored in the
continuum and thermodynamic limit, and how, in the case of broken
supersymmetry, the Goldstino mode emerges.

The outline of the paper is as follows. In Section \ref{sec:transfer
  matrices}, we derive explicitly the construction of the transfer
matrices for supersymmetric quantum mechanics starting from the bond
formulation of the lattice system. We then work out the calculation of
various observables such as correlation functions in
\ref{subsec:correlation functions}, the mass gaps from the eigenvalues
of the transfer matrices in \ref{subsec:mass_gaps}, and discuss some
Ward identities which may be used to investigate the restoration of
supersymmetry in the continuum in \ref{subsec:WI}.  After these
technical considerations, we present our exact results in Section
\ref{sec:results_count} for various observables of interest, such as
the Witten index in \ref{subsec:Zp_Za_results} and correlation
functions in \ref{subsec:correlation functions results}.  In addition,
we demonstrate in detail how the supersymmetry is recovered in the
continuum by means of energy spectra in \ref{subsec:mass_gaps_results}
and Ward identities in \ref{sec:Ward identitities}, and finally
present an exact calculation of the ground state energy in
\ref{subsec:ground state energy results}. For each quantity we discuss
in turn the results using the standard discretisation including the
counterterm \cite{Giedt:2004vb,Bergner:2007pu} and the results for the
$Q$-exact action \cite{Catterall:2000rv}. We do so as far as possible
for systems with unbroken and broken supersymmetry.  Finally, in
Section \ref{sec:conclusion} we summarise our results and close with
some conclusions, while in appendix \ref{app:technical aspects}
we provide some technical details concerning the numerical calculation
of the transfer matrices.
\section{The transfer matrix approach}
\label{sec:transfer matrices}
In order to introduce the notation we briefly recall the Euclidean
action for supersymmetric quantum mechanics involving the bosonic
field $\phi$ and the fermionic fields $\psibar$ and $\psi$,
\begin{equation}
  S(\phi,\psibar,\psi) = \int_0^\beta dt \left\{ \frac{1}{2} \left(\frac{d\phi(t)}{dt} \right)^2 + \frac{1}{2}P^\prime(\phi(t))^2 + \psibar(t)\left( \partial_t + P^{\prime \prime}(\phi(t)) \right)\psi(t) \right\}.
\label{S_continuum}
\end{equation}
The action depends on the superpotential $P(\phi)$ and is invariant
under the two supersymmetry transformations
\begin{equation}
\begin{array}{rclcrcl}
  \delta_1 \phi &=& \overline{\epsilon} \psi, &\quad& \delta_2 \phi &=& \psibar \epsilon, \\ 
  \delta_1 \psi &=& 0, &\quad&  \delta_2 \psi &=&  \left( \dot{\phi} - P^\prime \right) \epsilon,\\
  \delta_1 \psibar &=& -\overline{\epsilon} \left( \dot{\phi} + P^\prime \right), &\quad&  \delta_2 \psibar &=& 0.
\end{array}
\label{susy_transf_cont}
\end{equation}
Throughout our series of papers we use the superpotential 
\be
P_u(\phi) = \frac{1}{2}\mu \phi^2 + \frac{1}{4}g \phi^4
\label{eq:superpotential P_u}
\ee
as an example for unbroken supersymmetry with an additional parity
symmetry $\phi \rightarrow -\phi$ and 
\be
P_b(\phi) = -\frac{\mu^2}{4 \lambda}\phi + \frac{1}{3} \lambda
\phi^3
\label{eq:superpotential P_b}
\ee
for broken supersymmetry with an additional combined parity and charge
conjugation symmetry $\phi \rightarrow -\phi, \psi \rightarrow
\psibar, \psibar \rightarrow \psi$.

After choosing a suitable discretisation of the derivatives,
supersymmetric quantum mechanics can be formulated on the lattice in
terms of bosonic and fermionic bond occupation numbers $n_i^b(x) \in
\mathbbm{N}_0$ and $n^f(x)=0,1$, respectively, connecting sites $x$
and $x+1$. We refer to our first paper \cite{Baumgartner:2014nka} for
further details and explanations.  In particular, the partition
function can be written as a sum over all allowed, possibly
constrained, bond configurations $\C = \{n^b_i(x),n^f(x)\}$ in the
configuration space $\Z$,
\begin{equation}
Z = \sum_{\C \subset \Z} W_F(\C) \, 
\label{eq:Z from bond configurations}
\end{equation}
where the weight $W_F(\C)$ of a configuration is given by
\begin{equation}
W_F(\C) = \prod_x \left( \prod_i \frac{w_i^{n^b_i(x)}}{n^b_i(x)!} \right)
\prod_x  Q_F(N(x)) \, .
\label{eq:configuration weight}
\end{equation}
Here, $w_i$ is the weight of a bosonic bond $b_i$ with $i \in \{j
\rightarrow k \,| \, j,k \in \mathbbm{N}\}$, while $F=0,1$ is the
fermion number determined by the fermionic bond configuration
$\{n^f(x)\}$.  The site weight $Q_F$ is given by
\begin{equation}
\label{eq:site weight}
 Q_F(N(x)) = \int_{-\infty}^\infty d \phi \ \phi^N(x) \e^{-V(\phi)} M(\phi)^{1 - F}
\end{equation}
where 
\begin{equation}
N(x) = \sum_{j,k} \left(j\cdot n^b_{j \rightarrow k}(x) + k\cdot
  n^b_{j \rightarrow k}(x-1) \right)
\label{eq:site occupation number}
\end{equation}
is the site occupation number, i.e., the total number of bosonic bonds
connected to site $x$. Finally, the potential $V(\phi)$ and the
monomer term $M(\phi)$ in eq.(\ref{eq:site weight}) depend on the
superpotential $P(\phi)$ and the specifics of the chosen
discretisation.

\subsection{Transfer matrices and partition functions}
\label{subsec:transfer matrices and partition functions}
We now express the bond formulation of supersymmetric quantum
mechanics on the lattice in terms of transfer matrices.  For the
construction we start by considering a bond configuration $\C$ in the
configuration space $\Z$ contributing to the partition function
$Z$. The degrees of freedom are now expressed by means of bond
occupation numbers $\{n^b_i(x),n^f(x)\}$ for the bosonic and fermionic
bonds. These bonds connect nearest neighbouring lattice sites and it
is hence natural to define bond states associated with the bonds of
the lattice. The states are characterised by the fermionic and bosonic
bond occupation numbers and are hence written as
$|n^f(x),\{n^b_{i}(x)\} \rangle$, where the coordinate $x$ refers to
the bond connecting the sites $x$ and $x+1$. The transfer matrix
$T(x)$ then describes the transition of the bond state at $x-1$ to the
bond state at $x$.  Since the fermionic occupation number $n^f$, and
hence the fermion number $F$, is conserved at each site, the transfer
matrix decomposes into block diagonal form, each block representing
separately the bosonic and fermionic sector.  So the separation of
bond configurations into the bosonic and fermionic sectors $\Z_0$ and
$\Z_1$, respectively, reflects itself in the block structure of the
transfer matrix, and from now on it is sufficient to discuss
separately the submatrices $T^F(x)$ with fixed fermionic bond
occupation number $n^f=F$.

\begin{figure}[h]
\centering
\includegraphics[width=0.7\textwidth]{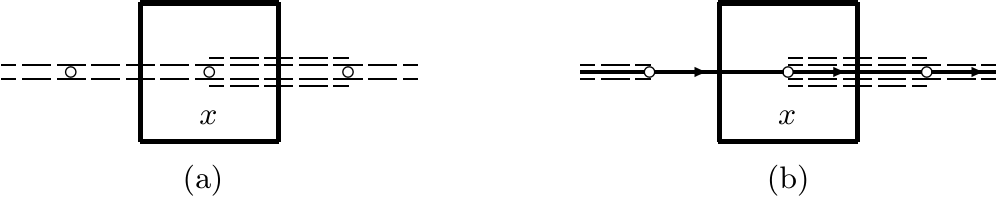}
\caption{Graphical representation of the transfer matrix for a system
  with only one type of bosonic bond $b_{1\rightarrow 1}$ with
  occupation numbers $n^b_{1\rightarrow 1}$ (dashed lines). Plot (a)
  represents the entry $T^0_{2,4}$ of the transfer matrix in the
  bosonic sector and plot (b) shows the entry $T^1_{0,4}$ of the
  transfer matrix in the fermionic sector. The occupation of the
  fermionic bond is represented by the directed full line.}
\label{fig:ill_transfer_T0}
\end{figure}

In figure \ref{fig:ill_transfer_T0} we give two examples for the
characterisation of the transfer matrix for a system with only one
type of bosonic bond $b_{1\rightarrow 1}$ with corresponding
occupation numbers $n^b_{1\rightarrow 1}$ (dashed lines) in each of
the sectors $F = 0,1$.
The occupation of the fermionic bond is represented by the directed
full line, cf.~the first paper of our series
\cite{Baumgartner:2014nka} for further explanations on the graphical
notation. In fact, since the characterisation of the set of states is
independent of the coordinate $x$, it is sufficient to characterise
the states just by $|n^f,\{n^b_{i}\} \rangle$ and hence the transfer
matrix does not depend on $x$. As a consequence, the complete system
for fixed fermion number $F$ is characterised by just one transfer
matrix and all the physical information on the system can be extracted
from it. This is a rather remarkable property of the bond formulation
and stems from the fact that the reformulation of the continuous
degrees of freedom into discrete ones allows a complete and explicit
enumeration of all states. However, since there are no upper limits on
the bosonic bond occupation numbers, the two matrices $T^F, F=0,1$ are
infinitely large.
  
For a lattice consisting of $L_t$ lattice points the partition
function for both the fermionic and the bosonic sector can be
calculated independently in terms of $T^F$ as
\begin{equation}
Z_F = \Tr \left[ \left( T^F \right)^{L_t} \right]
\label{eq:ZF from T}
\end{equation}
where the transfer matrix multiplications now sum over all possible
bond configurations and the matching of the bond configurations at the
boundary is ensured by taking the trace.  After diagonalisation of the
transfer matrices one can calculate the partition functions
equivalently via the eigenvalues $\lambda_k^F$ of $T^F$,
\begin{equation}
 Z_F = \sum_{k} \left(\lambda_k^F \right)^{L_t}.
\end{equation}
Eventually, the partition functions in the two sectors can then be
combined as usual into partition functions with periodic and
antiperiodic boundary conditions for the fermion as
\begin{equation}
 Z_{p} = Z_0 - Z_1, \qquad Z_{a} = Z_0 + Z_1.
\end{equation}

Let us now write down the transfer matrix elements connecting the
incoming state $|F,\{m_i^b\} \rangle$ with the outgoing state
$|F,\{n_i^b\} \rangle$. This is straightforwardly done by comparing
eq.(\ref{eq:ZF from T}) with eq.(\ref{eq:Z from bond configurations})
and (\ref{eq:configuration weight}). Explicitly, we have
\begin{equation}
 T_{\{m_i^b\},\{n_i^b\}}^F = \sqrt{\prod_i\frac{w_i^{m_i^b}}{m^b_i!}\frac{w_i^{n_i^b}}{n^b_i!}} \ Q_F\left( N\right)
\label{eq:transfer matrix generic}
\end{equation}
where the site occupation number is given by $N = \sum_{j,k} \left(j
  \cdot n_{j\rightarrow k}^b + k \cdot m_{j\rightarrow k}^b\right)$.
Here we choose to distribute the contributions $w^n/n!$ from the
incoming and outgoing bonds symmetrically, but in principle one could
choose any distribution, e.g.~taking into account only contributions
from the forward bonds.

To be more concrete, we now specify the general expression for the
transfer matrices explicitly for the two discretisations discussed in
detail in the first paper of our series \cite{Baumgartner:2014nka} and
for which we present exact results in this paper. The standard
discretisation including the counterterm involves only one type of
bosonic bond $b_{1\rightarrow 1}$ carrying weight $w_{1\rightarrow 1}
= 1$ and the bond states can simply be labelled by the occupation
number $n \equiv n_{1\rightarrow 1}^b$. Explicitly, denoting the
incoming state by $m\equiv m_{1\rightarrow 1}^b$ and the outgoing by
$n\equiv n_{1\rightarrow 1}^b$ the transfer matrix can be written as
\begin{equation}
 T_{m,n}^F = \sqrt{\frac{1}{m! \cdot n!}} \
 Q_F\left(m+n\right) \, .
\label{eq:transfer matrix standard}
\end{equation}
For the $Q$-exact discretisation in addition to the bond
$b_{1\rightarrow 1}$ with weight $w_{1\rightarrow 1}$ we have the new
type of bond $b_{1\rightarrow \nu}$ with weight $w_{1\rightarrow \nu}$
where $\nu=3$ for the superpotential $P_u$ and $\nu=2$ for the
superpotential $P_b$. The explicit expressions for the weights are
given in our first paper \cite{Baumgartner:2014nka}. Labelling the
incoming state by $m\equiv\{m^b_{1\rightarrow 1}, m^b_{1\rightarrow
  \nu}\}$ and the outgoing by $n\equiv\{n^b_{1\rightarrow
  1},n^b_{1\rightarrow \nu}\}$ we have
\begin{equation}
 T^F_{m,n} = \sqrt{\frac{(w_{1 \rightarrow 1})^{m^b_{1 \rightarrow 1}+n^b_{1 \rightarrow 1}}}{(m^b_{1 \rightarrow 1}!)(n^b_{1 \rightarrow 1}!)}} \sqrt{\frac{(w_{1 \rightarrow \nu})^{m^b_{1 \rightarrow \nu}+n^b_{1 \rightarrow \nu}}}{(m^b_{1 \rightarrow \nu}!)(n^b_{1 \rightarrow \nu}!)}} \ Q_F(N) 
\label{eq:transfer matrix Q-exact}
\end{equation}
where $N = n_{1 \rightarrow 1}^b + n_{1 \rightarrow \nu}^b + m_{1
  \rightarrow 1}^b + \nu \cdot m_{1 \rightarrow \nu}^b$.

Note that depending on the additional symmetries present in the
system, the bond configuration space may factorise further into
sectors of fixed quantum numbers associated to the symmetries. This
leads to an additional block structure in the transfer matrices $T^F$
and it is then sufficient to discuss each of the submatrices
separately. As an example we mention the $\mathbb{Z}_2$ parity
symmetry present in the system with superpotential $P_u$ in
eq.(\ref{eq:superpotential P_u}). In that case the configuration space
decomposes into configurations containing exclusively either even or
odd bond occupation numbers. Consequently, this yields a decomposition
of the transfer matrices into submatrices associated with the parity
quantum numbers $\pm 1$.

Before discussing how various observables can be expressed in terms of
the transfer matrices or their eigenvalues, we need to emphasise that
one faces several numerical challenges when constructing and
evaluating the transfer matrices. Firstly, as already mentioned, the
matrices have infinite extent due to the fact that the bosonic bond
occupation numbers are not limited and in practice one therefore needs
to truncate the state space. Since the bond occupation numbers
introduce a natural ordering of the states, it is straightforward to
choose a cutoff such that the results are not affected. We discuss the
technical aspects of this procedure in detail in appendix
\ref{app:technical aspects cutoff}. Secondly, the evaluation of the
site weights tends to become numerically unstable for large values of
the site occupation number. We will deal with this numerical problem
in detail in the third paper of our series
\cite{Baumgartner:2015zna}. Thirdly, the numerical calculation of the
transfer matrix elements can be rather delicate if the bond occupation
numbers involved become large. We discuss strategies for a numerically
stable determination of the transfer matrix elements in appendix
\ref{app:technical aspects construction}.

\subsection{Correlation functions}
\label{subsec:correlation functions}
\begin{figure}
\centering
\includegraphics[width=0.45\textwidth]{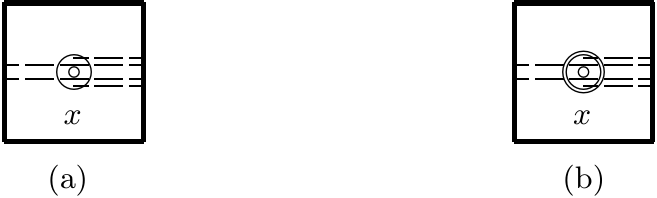}
\caption{Graphical representation of the bosonic transfer matrix at a
  site with additional bosonic sources. Plot (a) represents the
  matrix element $T^0_{2,4}(1)$ and plot (b) the matrix element
  $T^0_{2,4}(2)$ of a site with one and two additional sources,
  respectively.}
\label{ill_transfer_T0(1)}
\end{figure}

Next, we extend the concept of transfer matrices to the calculation of
correlation functions. Recalling from our first paper how the
two-point functions are calculated in the bond language, we realise
that the transfer matrix approach provides a perfect tool for the
exact calculation of the bosonic as well as the fermionic two-point
function.  We first consider the bosonic case. To get a contribution
to the expectation value of $\langle \phi_{x_1}^{j} \phi_{x_2}^{k}
\rangle$, we have to add additional bosonic field variables at the
sites $x_1$ and $x_2$. The transfer matrices at these sites experience
a corresponding modification and the graphical representation of the
modified transfer matrix with additional bosonic sources is shown in
figure \ref{ill_transfer_T0(1)} where we use the symbol $\ocircle$ for
each additional source.
The additional variables affect the weight of the configuration via
the occupation number \mbox{$N(x) \rightarrow N(x)+
  j\cdot\delta_{x,x_1} + k\cdot\delta_{x,x_2}$}. Thus, we introduce
modified transfer matrices which allow for additional bosonic sources
by defining
\begin{equation}
 T_{\{m_i^b\},\{n_i^b\}}^F(p)
\equiv T_{\{m_i^b\},\{n_i^b\}}^F(\phi^p) =
\sqrt{\prod_i\frac{w_i^{m_i^b}}{m^b_i!}\frac{w_i^{n_i^b}}{n^b_i!}} \
Q_F\left( N + p \right)\, ,
\end{equation}
such that we can calculate the non-normalised expectation value of an arbitrary
$n$-point correlation function by using the transfer matrices $T^F(p)$, i.e.,
\begin{equation}
\langle \! \langle \phi_{x_1}^{p_1} \ldots \phi_{x_n}^{p_n} \rangle \!
\rangle_F  = \Tr \left[ \prod_x T^F\left(\sum_{i=1}^n p_i \cdot
\delta_{x,x_i}\right) \right] \, .
\end{equation}
The originally defined transfer matrices in eq.(\ref{eq:transfer
  matrix generic}) correspond to transfer matrices with no additional
sources, $T^F(0) \equiv T^F$.

As a concrete example we now specify the non-normalised bosonic
two-point correlation function $g_F^b(x_2-x_1) = \langle \! \langle
\phi_{x_1} \phi_{x_2} \rangle \!  \rangle_F$. Defining $t = (x_2 -
x_1) \ \mathrm{mod} \ L_t$ and using translational invariance it reads
\begin{eqnarray}\label{corr_bos_t}
 \!\!\!\!\!\!\!\!\!g_F^b(t) & = &\left\{ \begin{array}{ll}
\Tr \left[ T^F(1)\left( T^F(0)\right)^{t - 1}T^F(1) \left(T^F(0) \right)^{L_t - t - 1}  \right] & \mathrm{if} \ t \neq 0, \\[8pt]
\Tr \left[ T^F(2)\left(T^F(0)\right)^{L_t - 1}  \right] & \mathrm{if} \ t = 0.
\end{array} \right.
\end{eqnarray}
For the connected part of the bosonic correlation function we also
need the expectation value of $\phi$. From the previous considerations
it is easy to see that the non-normalised expectation value for any
moment of $\phi$ can be calculated as
\begin{equation}
 \langle \! \langle \phi^p \rangle \! \rangle_F = \Tr \left[ T^F(p) \left( T^F(0) \right)^{L_t - 1} \right].
\end{equation}
Eventually, the connected part of the bosonic correlation function for
each sector is given by
\begin{equation}\label{corr_sectors_bos}
 C^b_{0,1}(t) = \frac{g^b_{0,1}(t)}{Z_{0,1}} - \left(\frac{\langle \!
     \langle \phi \rangle \! \rangle_{0,1}}{Z_{0,1}}\right)^2\, ,
\end{equation}
while for periodic and antiperiodic boundary conditions it is
calculated according to the discussion in our first paper
\cite{Baumgartner:2014nka}, i.e.,
\begin{eqnarray}\label{corr_ferm_t_boundary}
 C^b_{p} (t) & = & \frac{g_0^b(t) - g_1^b(t)}{Z_0 - Z_1} - \left(\frac{ \langle \! \langle \phi \rangle \! \rangle_0 - \langle \! \langle \phi \rangle \! \rangle_1}{Z_0 - Z_1} \right)^2,\\
 C^b_{a} (t) & = & \frac{g_0^b(t) + g_1^b(t)}{Z_0 + Z_1} - \left(\frac{ \langle \! \langle \phi \rangle \! \rangle_0 + \langle \! \langle \phi \rangle \! \rangle_1}{Z_0 + Z_1} \right)^2
\end{eqnarray}
for periodic and antiperiodic boundary conditions, respectively.

\begin{figure}%
\centering
\includegraphics[width=0.8\textwidth]{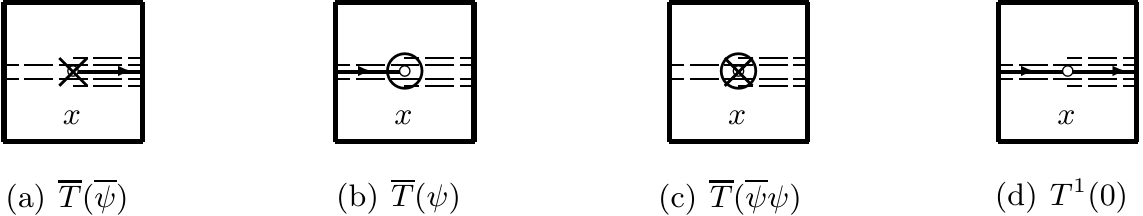}
\caption{Graphical representation of the transfer matrices with
  additional fermionic sources.  We use the symbol $\times$ for the
  fermionic source $\psibar$ and $\ocircle$ for the fermionic sink
  $\psi$. All examples are for the matrix element
  $\overline{T}_{2,4}$. The weights for the matrix elements (a)--(c) are
  the same and equal to the weight of (d).  }
\label{fig_T_bar_matrices}%
\end{figure}

To construct the fermionic correlation function in the transfer matrix
approach, we need to recall the structure of a bond configuration
contributing to the fermionic two-point function from our first
paper. In analogy to the bosonic case, we introduce new transfer
matrices which take into account the additional fields $\psibar$ and
$\psi$. In particular, we define a transfer matrix representing a site
with a fermionic source and sink $\overline{T}(\psibar \psi)$, one
representing a site with a fermionic source, $\overline{T}(\psibar)$,
and one with a fermionic sink, $\overline{T}(\psi)$. As usual, single
additional fermionic variables have to be paired with a fermionic
bond, an outgoing one to the right for a site with a source variable
$\psibar$ and an incoming one from the left for a site with a sink
variable $\psi$. The graphical representation for these transfer
matrices is shown in figure \ref{fig_T_bar_matrices} where we denote
the fermionic source $\psibar$ by a bold $\times$ and the fermionic
sink $\psi$ by a bold $\ocircle$.
Using again $t = (x_1 - x_2) \ \mathrm{mod} \ L_t$ the non-normalised
fermionic correlation function $g^f(x_1 - x_2) = \langle \! \langle
\psi_{x_1} \psibar_{x_2} \rangle \! \rangle$ can be composed of these
matrices by
\begin{eqnarray}
 \!\!\!\!\!\!\!\!\!g^f(t) & = &\left\{ \begin{array}{ll}
\Tr \left[ \overline{T}(\psibar)\left( T^1(0)\right)^{t - 1} \overline{T}(\psi) \left(T^0(0) \right)^{L_t - t - 1}  \right] & \mathrm{if} \ t \neq 0, \\[8pt]
\Tr \left[ \overline{T}(\psibar\psi)\left(T^0(0)\right)^{L_t - 1}  \right] & \mathrm{if} \ t = 0.
\end{array} \right.
\end{eqnarray}
Of course this expression can easily be generalised to take into
account more complicated fermionic operators such as $\psi \phi^p$ and
$\psibar \phi^p$. The additional presence of the bosonic variable
$\phi^p$ simply increases the site occupation number according to the
discussion on the bosonic correlation functions.

Since the weight of a site saturated with an additional fermionic
source or sink paired with a fermionic bond is the same as the weight
of a site saturated with two fermionic bonds or a source and a sink
variable, the newly introduced transfer matrices all have the same
entries as the transfer matrix $T^1(0)$, i.e.,
\begin{equation}
\overline{T}(\psibar) = \overline{T}(\psi) = \overline{T}(\psibar\psi)
= T^1(0)\, .
\end{equation}
Therefore, the definition of new matrices for sites with additional
fermionic variables is in fact obsolete in practice and the presence
of a fermionic source or sink expresses itself by a change from $T^0$
to $T^1$ and vice versa. The non-normalised fermionic two-point
function can hence be written in terms of the matrices $T^1(0)$ and
$T^0(0)$ as
\begin{equation}
 g^f(t) = \Tr \left[ \left(T^1(0)\right)^{t + 1}\left( T^0(0) \right)^{L_t - t - 1}\right].
\end{equation}
Yet, the formation of the fermionic correlation function is a little
more subtle than the one of the bosonic correlation function. The
translation invariance of the two-point function together with the
cyclic invariance of the trace amounts to the fact that $g^f(t)$ is a
superposition of \emph{all possible} configurations with an open
fermionic string where the fermionic source and the sink are separated
by the distance $t$. For a given bosonic bond configuration, there are
thus $L_t$ different configurations with an open fermionic string. For
$t$ of them, the fermionic string crosses the boundary and for
antiperiodic boundary conditions, we have to account for those as they
pick up a negative sign. Keeping track of all the signs correctly, the
fermionic correlation functions for periodic and antiperiodic boundary
conditions, respectively, read
\begin{equation}
 C^f_p(t) = \frac{g^f(t)}{Z_0 - Z_1}, \qquad C^f_a(t) = \frac{L_t -
   2t}{L_t}\frac{g^f(t)}{Z_0 + Z_1} \, .
\label{corr_ferm_t}
\end{equation}
From our discussion of the fermionic two-point function in our first
paper \cite{Baumgartner:2014nka} we remember that it is really only
defined in the bosonic sector $F = 0$ and we have
\begin{equation}
C^f_0 = \frac{g^f(t)}{Z_0}.
\label{corr_sectors_ferm}
\end{equation}
On the other hand, we can interpret the open fermion string of length
$t$ as an open antifermion string of complementary length $L_t-t$ on
the background of bond configurations in sector $F=1$. This
interpretation becomes evident when one calculates the energy or mass
gaps from the correlation functions in terms of the eigenvalues of the
transfer matrices which we are going to do in the next section.

\subsection{Mass gaps}\label{subsec:mass_gaps}
Observables closely related to the correlation functions are of course
the energy or mass gaps. It is well known that in the transfer matrix
formalism these mass gaps can be calculated directly from the ratios
of eigenvalues of the transfer matrices,
cf.~\cite{PhD_Baumgartner:2012} for the explicit calculation in our
supersymmetric quantum mechanics setup.  Ordering the eigenvalues of
the transfer matrix $T^F$ according to
\begin{equation}                                                       \lambda_0^F > \lambda_1^F > \dots ,
\end{equation}
the calculation of the $i$-th fermionic mass gap in the bosonic sector
yields
\begin{equation}
 m^f_i = -\ln(\lambda_i^1/\lambda_0^0).
\end{equation}
By interpreting the expectation value $\langle \psibar_t \psi_0
\rangle = C^{\overline{f}}(t)$ as the correlator of the antifermion
$\overline{f}$ in the fermionic sector $F=1$, we can similarly
calculate its mass via
\begin{equation}
 m^{\overline{f}}_i = -\ln(\lambda_i^0/\lambda_0^1),
\end{equation}
and we see that the masses of the fermion and antifermion are the same
-- at least in the continuum -- up to a minus sign. Of course, this is
in accordance with the standard quantum mechanical interpretation of
an antiparticle as a particle with negative energy propagating
backward in time, and so this confirms our interpretation of the open
fermion string as a propagating fermion in sector $F=0$ or as a
complementary antifermion in sector $F=1$.  The bosonic mass gaps are
defined in each sector $F = 0,1$ individually and are calculated as
\begin{equation}
 m^b_{i,F} = -\ln(\lambda_i^F/\lambda_0^F).
\end{equation}

\begin{figure}[t]
\begin{center}
\includegraphics[width=0.7\textwidth]{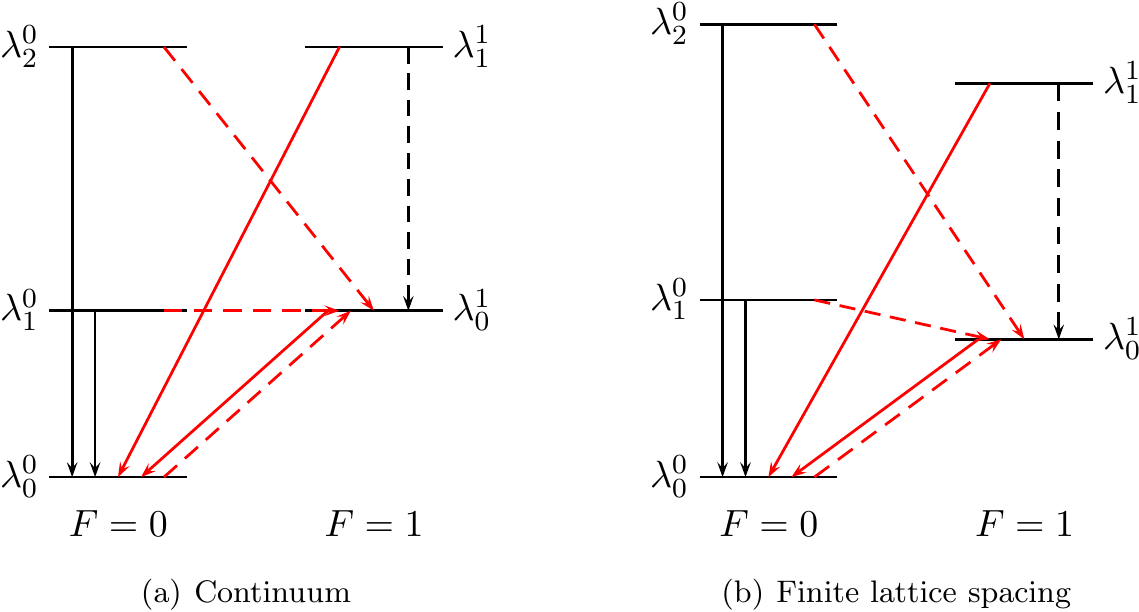}
\end{center}
\caption{Unbroken supersymmetric quantum mechanics. The energy levels
  and the respective mass gaps in the continuum (a), and for finite
  lattice spacing (b) where the energy levels are shifted w.r.t.~to
  the ones in the continuum due
  to discretisation artefacts. 
}
\label{fig:energy_levels_u_c}
\end{figure}

\begin{figure}[t]
\centering
  \includegraphics[width = 0.7\textwidth]{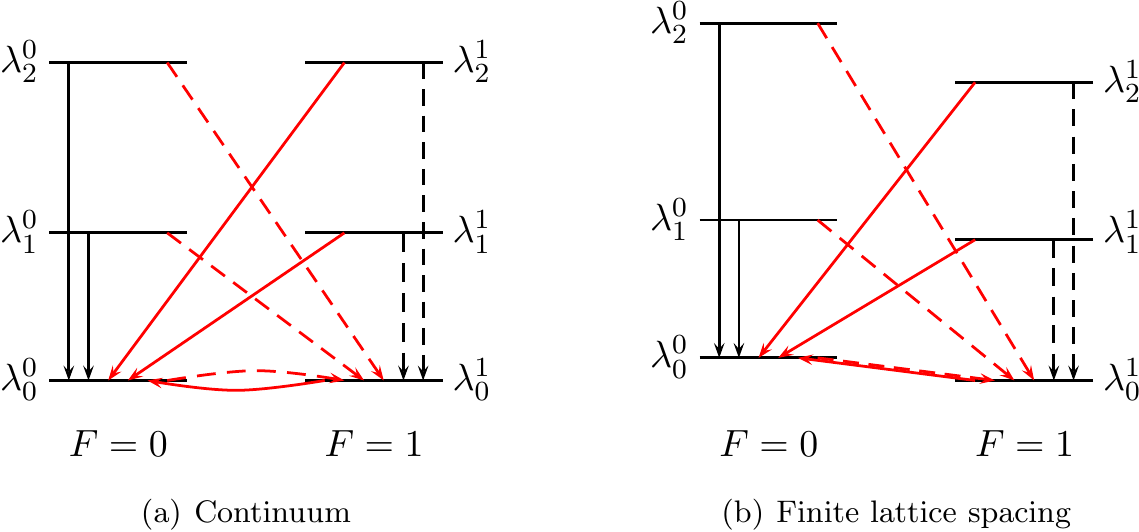}
\caption{Broken supersymmetric quantum mechanics. The energy levels
  and the respective mass gaps in the continuum (a), and for finite
  lattice spacing (b)  where the  energy levels are shifted w.r.t.~to
  the ones in the continuum due
  to discretisation artefacts. Note that the plot (b) illustrates a
  situation in which the fermionic vacuum is favoured at finite
  lattice spacing as compared to the bosonic vacuum.
}
\label{fig:mass_extr_b}
\end{figure}

It is useful to illustrate schematically which mass gap is measured
with respect to which vacuum via the ratios of the eigenvalues. In
figures \ref{fig:energy_levels_u_c} and \ref{fig:mass_extr_b} the mass
gaps in the bosonic sector, i.e., with respect to the bosonic vacuum,
are depicted by full lines while the mass gaps in the fermionic
sector, i.e., with respect to the fermionic vacuum, are drawn as
dashed lines. Bosonic mass gaps $m^b_{i,F}$ with bosonic quantum
numbers are further differentiated from the fermionic mass gaps
$m^{f,\overline{f}}_{i}$ with fermionic quantum numbers as black
versus red lines. Figure \ref{fig:energy_levels_u_c} illustrates a
system with unbroken supersymmetry and a unique bosonic ground state
in the continuum (a) and at finite lattice spacing (b), while figure
\ref{fig:mass_extr_b} illustrates a system with broken supersymmetry
and hence two degenerate bosonic and fermionic ground states, again in
the continuum (a) and at finite lattice spacing (b). The shifts in the
energy levels w.r.t.~to the ones in the continuum are due to
discretisation artefacts of $\O(a)$ and are expected to disappear in
the continuum.  Illustration (b) in figure \ref{fig:mass_extr_b}
represents a situation in which the fermionic vacuum is favoured at
finite lattice spacing as compared to the bosonic vacuum.  Moreover,
it is interesting to note that unless the vacua are degenerate, there
is always one negative fermionic mass gap, namely the one measured
from the energetically lower to the higher vacuum.

\subsection{Ward identities}\label{subsec:WI}
One of the main goals of our efforts in supersymmetric quantum
mechanics is to gain a precise understanding of whether and how
supersymmetry is restored in the continuum limit. For such
investigations Ward identities are most useful and many of our exact
results discussed in this paper refer to various Ward identities which
can be derived for the different discretisations we consider.  A Ward
identity can be derived by rewriting the expectation value of an
observable $\O(\phi)$ in the path integral formulation for the
transformed variable, $\phi \rightarrow \phi^\prime = \phi + \delta
\phi$, assuming that the measure of the path integral is invariant
under this variation, $\mathcal{D} \phi^\prime = \mathcal{D}
\phi$. Since the physics cannot depend on the shift of the integration
variable, we find to leading order in $\delta$
\begin{eqnarray*}
 \langle \O \rangle & = & \frac{1}{Z} \int \mathcal{D}\phi^\prime \
 \O(\phi^\prime) \ \e^{-S(\phi^\prime)} \\ 
  & = & \frac{1}{Z} \int \mathcal{D}\phi \ (\O(\phi) + \delta \O(\phi) ) \ \e^{-S(\phi)}(1 - \delta S(\phi)) \\ 
   & = & \langle \O \rangle + \langle \delta \O \rangle - \langle \O
   \delta S \rangle \, ,  
\end{eqnarray*}
and therefore the relation
\begin{equation}
 \langle \delta \O \rangle = \langle \O \delta S \rangle
\end{equation}
must hold for any observable. Now, if the action is invariant under
the transformation $\delta$, the r.h.s.~of the equation vanishes,
yielding
\begin{equation}
 \langle \delta \O \rangle = 0
\label{WI_0}
\end{equation}
as a condition to test whether the symmetry is indeed restored in the
continuum.

As a first example, we consider the observable $\O = \psibar$. Its
variation under the lattice version of the supersymmetry
transformation $\delta_1$ in eq.(\ref{susy_transf_cont}) results in
the simple Ward identity
\begin{equation}
W_0 \equiv \langle  (\nabla^- \phi + P^\prime) \rangle = 0 \, .%
\end{equation}
Thus, the vanishing of the expectation value of the first derivative
of the superpotential $\langle P^\prime \rangle$ in the continuum
indicates restoration of supersymmetry.  Note that the variation of
the operator $\O=\psibar$ under the supersymmetry transformation
$\delta_2$ yields $\delta_2 \psibar = 0$ by definition.

As a second example we consider the observable $\O = \psibar_x
\phi_y$. Its variation under the supersymmetry transformation
$\delta_{1}$ yields Ward identities which connect bosonic and
fermionic correlation functions. In particular, we obtain
\begin{equation}
 W_1(y - x) \equiv \langle \psibar_x \psi_y \rangle +
 \langle (\nabla^-\phi + P^\prime)_x \phi_y \rangle = 0
 \, ,
\end{equation}
while the variation of the operator under the supersymmetry
transformation $\delta_2$ vanishes trivially. Analogously, one can use
the observable $\O = \psi_x \phi_y$ which under the supersymmetry
transformation $\delta_2$ yields a similar set of Ward identities,
\begin{equation}
W_2(x - y) \equiv \langle \psi_x \psibar_y \rangle +
\langle (\nabla^-\phi - P^\prime)_x \phi_y \rangle = 0
\, ,
\label{eq_WI_II}
\end{equation}
while the variation of the operator under the other supersymmetry
transformation $\delta_1$ vanishes trivially.

Let us now be more specific and calculate the Ward identities $W_0,
W_1$ and $W_2$ explicitly for the two superpotentials $P_u$ and $P_b$
employed in our investigation. Using the translational invariance of
the lattice, for the superpotential $P_u$ we find the Ward identities
\begin{align}
   W_0 &= \langle P_u^\prime\rangle = \mu \langle \phi \rangle + g \langle \phi^3 \rangle \, , \label{eq:WI unbroken}
\\
   W_1(t) &= -\langle \psi_t \psibar_0 \rangle + (1 + \mu) \langle
   \phi_t \phi_0 \rangle - \langle \phi_{t + 1} \phi_0 \rangle + g
   \langle \phi_t \phi^3_0 \rangle \, ,\\
   W_2(t) &= \langle \psi_t \psibar_0 \rangle + (1 - \mu) \langle
   \phi_t \phi_0 \rangle - \langle \phi_{t + 1} \phi_0 \rangle - g
   \langle \phi_t^3 \phi_0 \rangle \, ,
\end{align}
while for the superpotential $P_b$, we obtain analogously
\begin{align}
  W_0 &= \langle P_b^\prime \rangle = -\frac{\mu^2}{4 \lambda} + \lambda
  \langle \phi^2  \rangle \, ,\label{eq:WI broken}
\\
  W_1(t) &= - \langle \psi_t \psibar_0 \rangle + \langle \phi_t \phi_0
   \rangle 
 - \langle \phi_{t + 1} \phi_0 \rangle - \frac{\mu^2}{4 \lambda} 
\langle \phi \rangle + \lambda \langle \phi_t \phi^2_0 \rangle \, ,\\
  W_2(t) &= \langle \psi_t \psibar_0 \rangle + \langle \phi_t \phi_0
  \rangle - \langle \phi_{t + 1} \phi_0 \rangle + \frac{\mu^2}{4
    \lambda} \langle \phi \rangle - \lambda \langle \phi_t^2 \phi_0
  \rangle \, .
\end{align}
With this we conclude the discussion of the observables which we
investigate in the following, and we now proceed to the discussion of
the results.

%
%
\section{Exact results}\label{sec:results_count}

In this section, we present our exact lattice results for the action
with counterterm as well as for the $Q$-exact action by employing the
transfer matrix technique. For the two superpotentials $P_u$ and
$P_b$, the actions are given explicitly in the first paper of our
series \cite{Baumgartner:2014nka}. For our further discussion it is
useful to recall that the continuum limit is taken by fixing the
dimensionful parameters $\mu, g, \lambda$ and $L$ while taking the
lattice spacing $a \rightarrow 0$. In practice, the dimensionless
ratios $f_u = g/\mu^2, f_b = \lambda/\mu^{3/2}$ fix the couplings and
$\mu L$ the extent of the system in units of $\mu$, while $a \mu$ and
$a/L$ are subsequently sent to zero.  We perform our calculations for
couplings $f_u$ and $f_g$ which lie well outside of the perturbative
regime in order to assess the systematics of the nonperturbative
lattice calculations.
Finally, we also recall that for antiperiodic fermionic boundary
conditions the finite extent $\mu L$ corresponds to finite inverse
temperature in units of $\mu$ and the limit $\mu L \rightarrow \infty$
is therefore required to recover the system at zero temperature.

\subsection{The ratio $Z_{p}/Z_{a}$ and the Witten index $W$}
\label{subsec:Zp_Za_results}
We start by calculating the ratio $Z_{p}/Z_{a}$. At zero temperature
this ratio is equal to the Witten index and represents therefore an
important indicator for whether supersymmetry is broken or not. In
quantum mechanics, whether or not supersymmetry is broken is not a
dynamical question, but depends solely on the asymptotic form of the
superpotential.  For unbroken supersymmetry, the bosonic vacuum lies
well below the fermionic one (or vice versa). Thus, in the zero
temperature limit $\mu L \rightarrow \infty$ only the bosonic sector
contributes to the partition function while the fermionic contribution
$Z_1$ vanishes, such that
\begin{equation}
 W = \lim_{\mu L \rightarrow \infty} \frac{Z_{p}}{Z_{a}} = \lim_{\mu L \rightarrow \infty} \frac{Z_0 - Z_1}{Z_0 + Z_1}
 \longrightarrow 1.
\end{equation}
For finite extent $\mu L$ (nonzero temperature), there are
nonvanishing contributions from the fermionic vacuum, i.e., the
partition function $Z_1$ is no longer zero due to quantum (thermal)
fluctuations, resulting in a ratio $Z_{p}/Z_{a} < 1$.  To leading
order in the inverse temperature, the asymptotic dependence is
governed by the energy gap $m_0^f$ between the fermionic and bosonic
vacuum,
\begin{equation}
\frac{Z_p}{Z_a} \sim \frac{1}{1+2 e^{-m_0^f L}} \,.
\label{eq:asymptotic Zp/Za}
\end{equation}

For broken supersymmetry on the other hand, both vacua are equally
preferable in the continuum and all bosonic and fermionic energy
levels are degenerate. Therefore we have $Z_0 = Z_1$ and the Witten
index goes to zero,
\begin{equation}
W = \frac{Z_{p}}{Z_{a}} = \frac{Z_0 - Z_1}{Z_0 + Z_1}
 \longrightarrow 0
\label{eq:Zp/Za broken continuum}
\end{equation}
independent of the extent or temperature of the system.  Using our
exact lattice calculation we can now investigate how these continuum
expectations are modified at finite lattice spacing and how the
continuum limit is eventually realised.

\begin{figure}
 \centering
 \includegraphics[width = 0.8\textwidth]{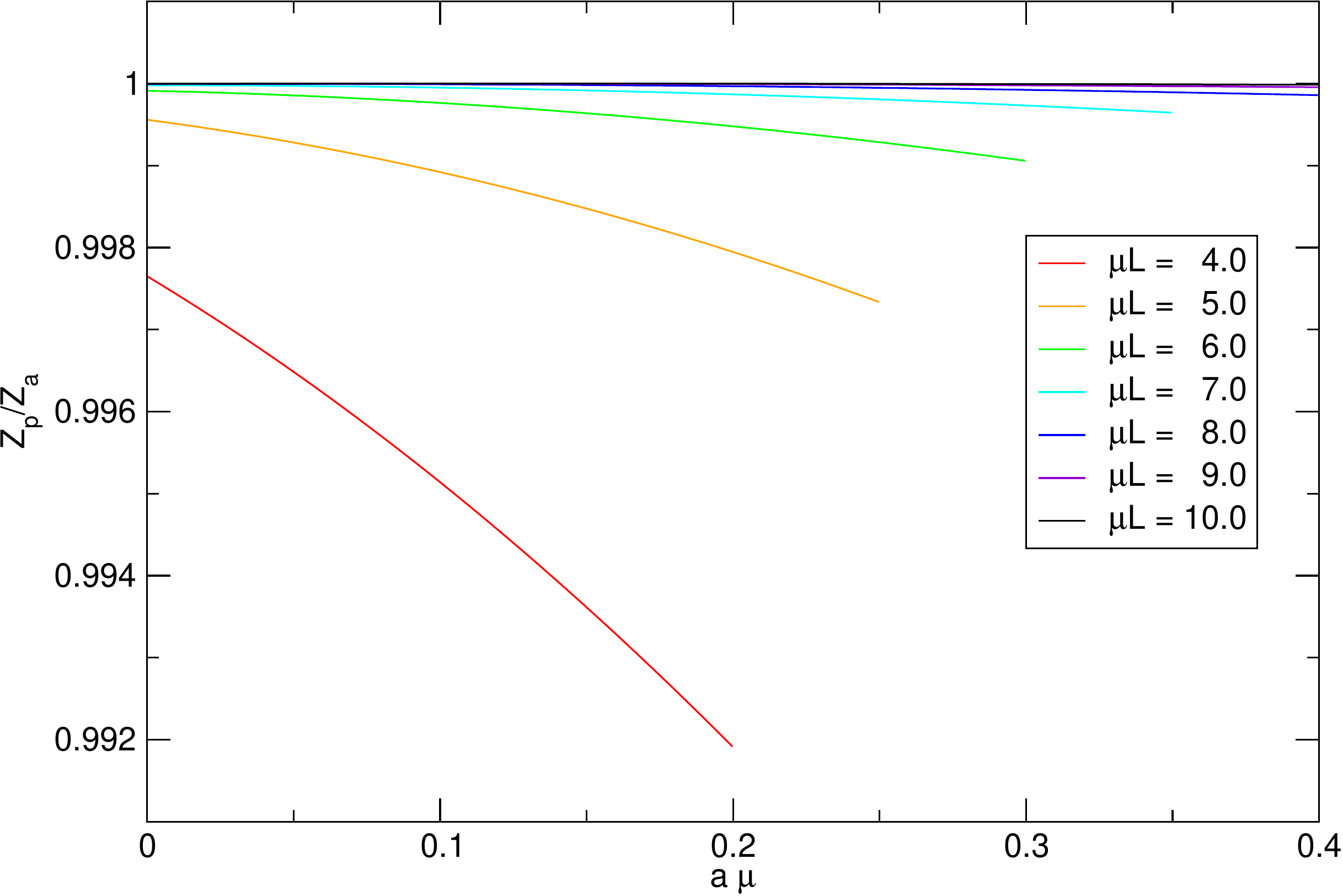}
 \caption{Unbroken supersymmetric quantum mechanics, standard
   discretisation. Continuum extrapolation of the ratio $Z_{p}/Z_{a}$
   for different values of $\mu L$ at fixed coupling $f_u = 1$.}
  \label{fig_Zp_Za_u_count}
\end{figure}

\begin{figure}
 \centering
 \includegraphics[width = 0.8\textwidth]{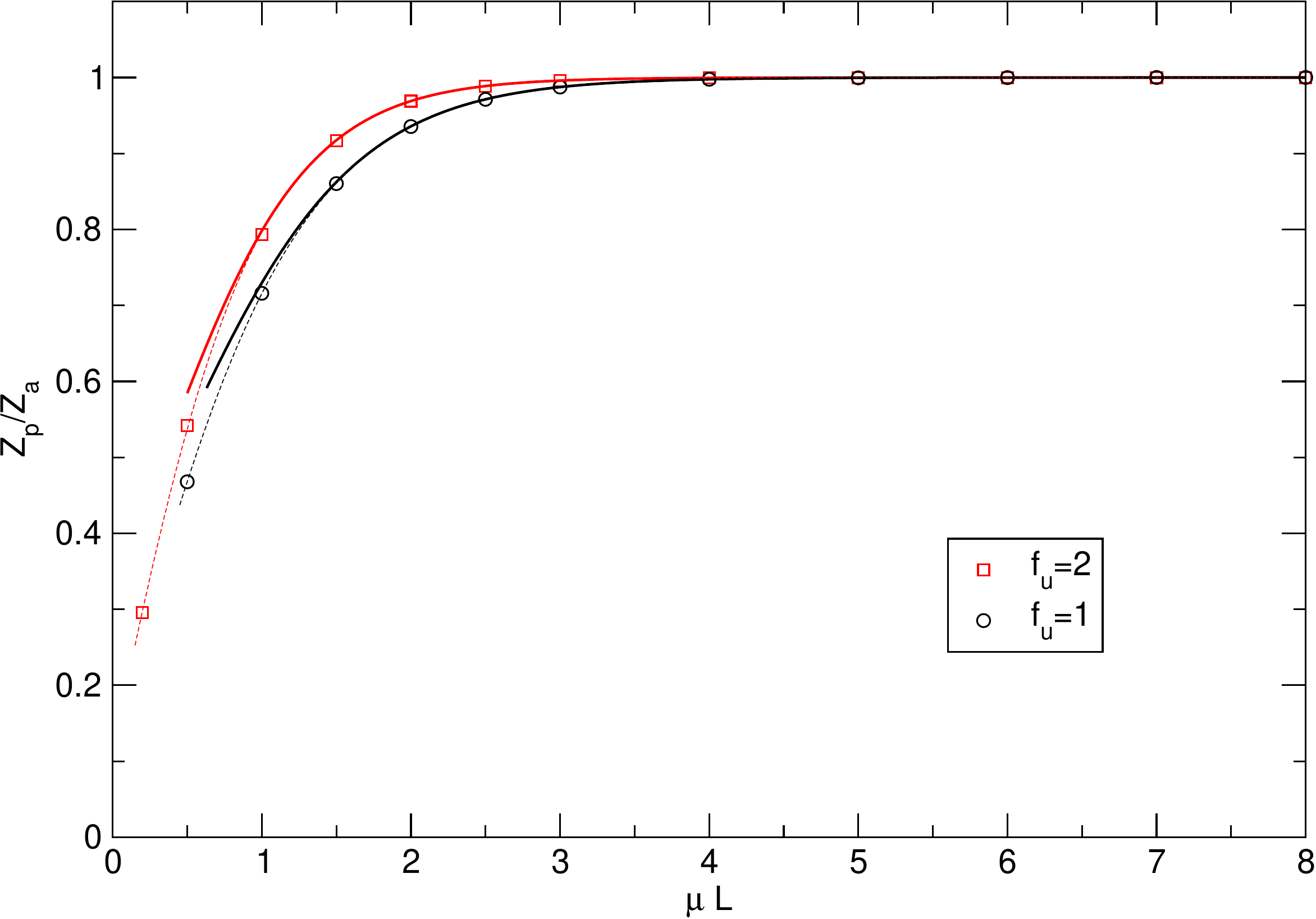}
 \caption{Unbroken supersymmetric quantum mechanics. The continuum
   values of the ratio $Z_{p}/Z_{a}$ versus $\mu L$ for different
   couplings $f_u = 1$ (black circles) and $f_u = 2$ (red
   squares). The full lines describe the asymptotic behaviour
   according to eq.(\ref{eq:asymptotic Zp/Za}) while the dashed lines
   include additional higher order contributions.}
  \label{fig_Zp_Za_phase_diag}
\end{figure}

First, we consider unbroken supersymmetry. In figure
\ref{fig_Zp_Za_u_count}, we plot the ratio $Z_{p}/Z_{a}$ versus $a
\mu$ for different values of fixed $\mu L$ using the standard
discretisation for fixed coupling $f_u = 1$.
At nonzero temperature we observe leading order lattice artefacts
which are linear in $a$. In the zero temperature limit they are
suppressed and the leading artefacts eventually become
$\O(a^2)$. Moreover, the artefacts become very small in this limit,
simply because at zero temperature only the bosonic groundstate
contributes and the nondegeneracies of the excited states at finite
$a$, cf.~figure \ref{fig:energy_levels_u_c}, become irrelevant. As the
temperature increases, the system gets more sensitive to the excited
states since their contributions to the partition function grow
larger, and consequently the nondegeneracies between the bosonic and
fermionic energy levels crystallise in the growing lattice
artefacts. In the continuum limit, we observe the expected deviation
of the ratio $Z_{p}/Z_{a}$ from one as discussed above. In figure
\ref{fig_Zp_Za_phase_diag}, we show the continuum value of the ratio
$Z_{p}/Z_{a}$ as a function of the inverse temperature $\mu L$ for two
different couplings $f_u = 1$ and $f_u = 2$. The full lines indicate
the asymptotic behaviour for $\mu L \rightarrow \infty$ according to
eq.(\ref{eq:asymptotic Zp/Za}), while the dashed lines include
additional higher order contributions.
It can be seen that the system reaches the asymptotic zero temperature
behaviour already at moderate values of $\mu L$. Moreover,
contributions from the fermionic vacuum to the partition function are
essentially negligible for $\mu L \gtrsim 4$.

\begin{figure}[t]
 \centering
 \includegraphics[width = 0.9\textwidth]{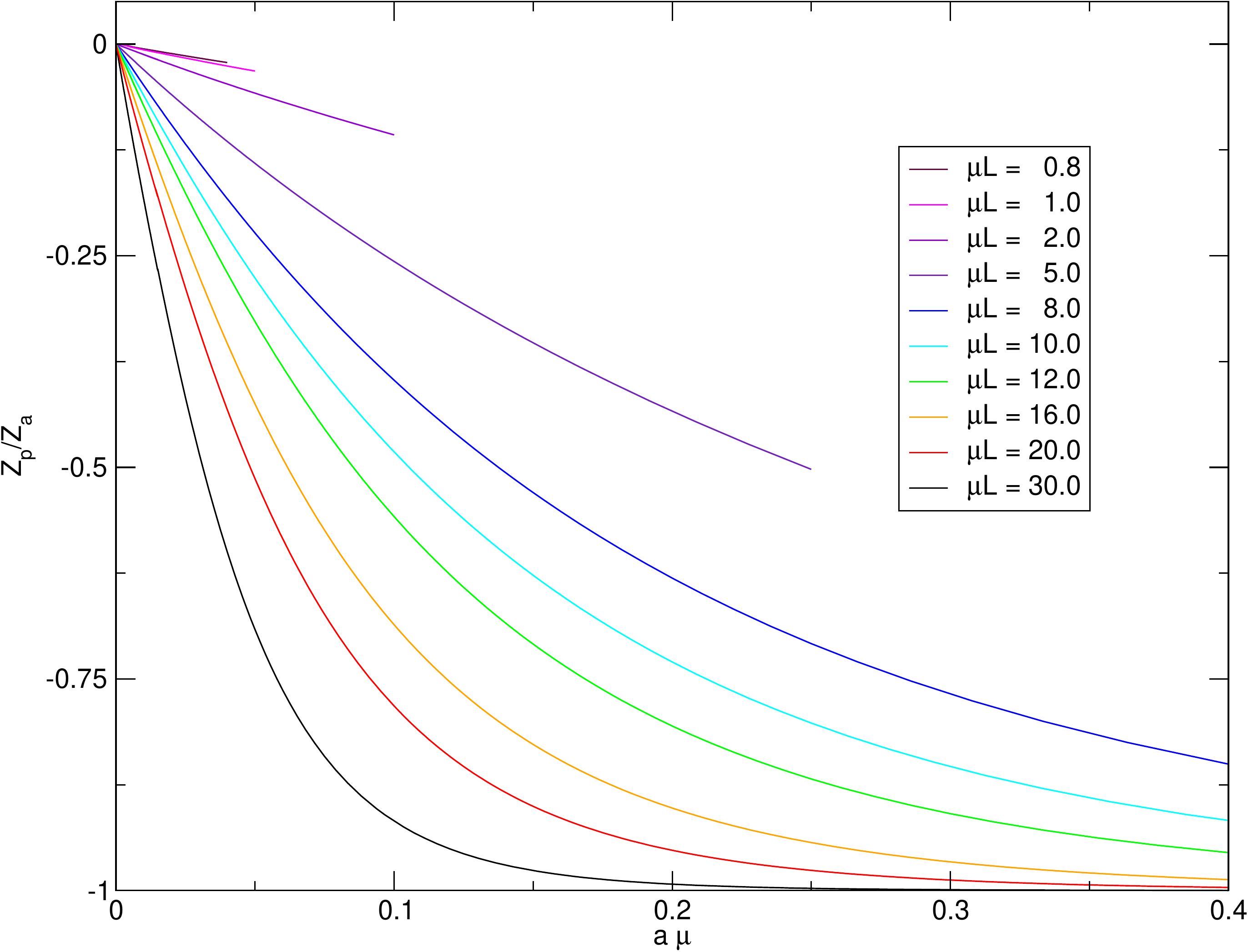}
 \caption{\label{fig:Zp_Za_b_count}Broken supersymmetric quantum
   mechanics, standard discretisation. Continuum extrapolation of the
   ratio $Z_{p}/Z_{a}$ for different values of $\mu L$ at fixed
   coupling $f_b = 1$.}
\end{figure}

For broken supersymmetry we plot the continuum limit of the ratio
$Z_{p}/Z_{a}$ versus $ a\mu$ for different values of $\mu L$ at fixed
coupling $f_b = 1$ using the standard discretisation in figure
\ref{fig:Zp_Za_b_count}.
First we note that the ratio goes to zero towards the continuum limit
indicating a vanishing Witten index in that limit independent of the
temperature. This is the expected continuum behaviour as argued above
in eq.(\ref{eq:Zp/Za broken continuum}) and relies on the fact that
the bosonic and fermionic energy levels become degenerate in
pairs. Since the lattice discretisation breaks this degeneracy
explicitly, cf.~figure \ref{fig:mass_extr_b}, the ratio is nonzero at
finite lattice spacing. One can think of the finite lattice spacing as
regulating the Goldstino zero mode and the energy difference between
the two vacua simply corresponds to the regulated Goldstino mass. As a
consequence the associated vanishing Witten index is regulated,
too. As explained in detail in the first paper of our series
\cite{Baumgartner:2014nka} a vanishing Witten index leads to a fermion
sign problem for Monte Carlo simulations. Since the finite lattice
spacing regulates the index one could argue that the sign problem is
avoided in this way, but of course it is not clear whether the lattice
artefacts and the statistical fluctuations can be kept under control.
In fact it turns out that the lattice artefacts for the ratio
$Z_p/Z_a$ can become extremely strong. While the leading artefacts are
evidently $\O(a)$, they grow exponentially large as the temperature is
lowered, i.e., at low temperature artefacts of all orders in $a$
become relevant such that the finite lattice spacing corrections in
the ratio are exponentially enhanced towards low temperatures.

The rather peculiar behaviour of the lattice corrections for small
temperatures can be explained as follows. Considering the illustration
of the supersymmetry broken spectrum at finite $a$ in figure
\ref{fig:mass_extr_b}, it is clear that the degeneracy between the
bosonic and fermionic vacuum is lifted. For small temperatures (large
values of $\mu L$) the tunnelling from the energetically lower to the
higher vacuum are exponentially suppressed with growing $\mu L$. On
the other hand, exactly these tunnellings are needed in order for the
higher vacuum to contribute to the partition function, eventually
leading to the vanishing Witten index. Only once the temperature is
large enough compared to the energy difference between the two vacua,
i.e.~the regulated Goldstino mass, the tunnelling becomes effective
enough to drive the Witten index to zero. Equivalently, at fixed
temperature the Goldstino mass, which to leading order is proportional
to $a \mu$, needs to become sufficiently small, and from figure
\ref{fig:Zp_Za_b_count} it becomes evident when this is the case.

The exponentially enhanced lattice artefacts have a rather dramatic
consequence for the Witten index concerning the order of the limits
$\mu L \rightarrow \infty$ and $a\rightarrow 0$. As is evident from
our discussion and the data in figure \ref{fig:Zp_Za_b_count},
extrapolating the index to $\mu L \rightarrow \infty$ always yields
$W=-1$ at {\it any} finite lattice spacing. Therefore the subsequent
continuum limit of the index at zero temperature comes out incorrectly
and the expectation in eq.(\ref{eq:Zp/Za broken continuum}) is hence
not confirmed. So in contrast to unbroken supersymmetry, here the
order of the limits is crucial and has to be taken into account for
the correct interpretation of the results.  Finally, from the plot we
infer that the fermionic vacuum has a lower energy than the bosonic
one, hence the Witten index tends to $-1$ for finite lattice spacing,
i.e., the picture at finite $a$ is exactly as depicted in Fig.~\ref{fig:mass_extr_b}.

\begin{figure}
 \centering
  \includegraphics[width = 0.8\textwidth]{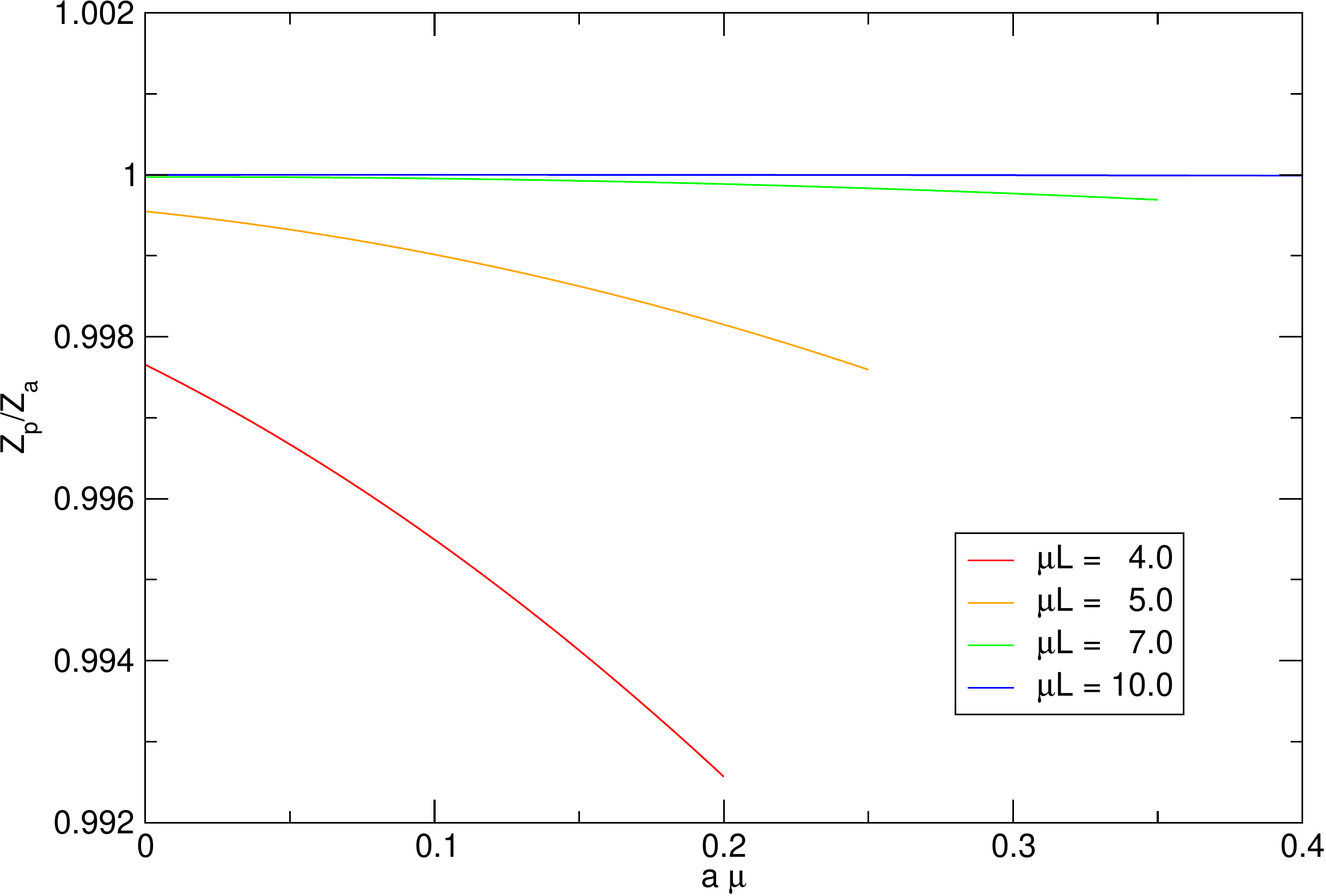}
  \caption{Unbroken supersymmetric quantum mechanics, $Q$-exact
    discretisation. Continuum extrapolation of the ratio $Z_{p}/Z_{a}$
    for different values of $\mu L$ at fixed coupling $f_u = 1$.}
  \label{fig_Zp_Za_u_q}
\end{figure}

Next, we consider the results for the Witten index using the $Q$-exact
discretisation.  In figure \ref{fig_Zp_Za_u_q} we plot the ratio
$Z_{p}/Z_{a}$ versus $a \mu$ for different values of $\mu L$ at fixed
coupling $f_u = 1$ for the case when supersymmetry is unbroken.
We observe lattice artefacts which are almost identical to the ones
found with the standard discretisation.  In addition, in the continuum
the ratios converge to the same values for any given inverse
temperature $\mu L$ and the temperature dependence in the continuum is
therefore given exactly as in figure \ref{fig_Zp_Za_phase_diag}. Of
course the agreement is a consequence of the universality of lattice
calculations in the continuum which is nicely confirmed by our
results. Turning to the case when supersymmetry is broken, the results
for the $Q$-exact discretisation are rather boring. Since the
degeneracy between the bosonic and fermionic energy levels is
maintained exactly at any value of the lattice spacing $a$, the
contributions from the bosonic and fermionic sector are always exactly
equal and cancel precisely, hence the Witten index is zero independent
of the temperature. Note however that the exact degeneracy of the
energy levels does not exclude lattice artefacts in the spectrum. In
fact, they are rather large as we will see in Section
\ref{subsec:mass_gaps_results}, but the Witten index is not sensitive
to it as long as the degeneracy between the bosonic and fermionic
levels is maintained at finite lattice spacing.

\subsection{Correlation functions}
\label{subsec:correlation functions results}
In this section, we present some exact results for two-point
correlation functions, merely as qualitative illustrations of how they
are affected by lattice artefacts. A more quantitative discussion will
follow in Section \ref{subsec:mass_gaps_results}, where we consider
the energy gaps, and in Section \ref{sec:Ward identitities} where we
investigate Ward identities relating fermionic and bosonic correlation
functions.

\begin{figure}
    \centering
        \subfigure[$\mu L = 2$]{%
            \label{fig_C_c_1_mL_2_L_60}
            \includegraphics[width=0.8\textwidth]{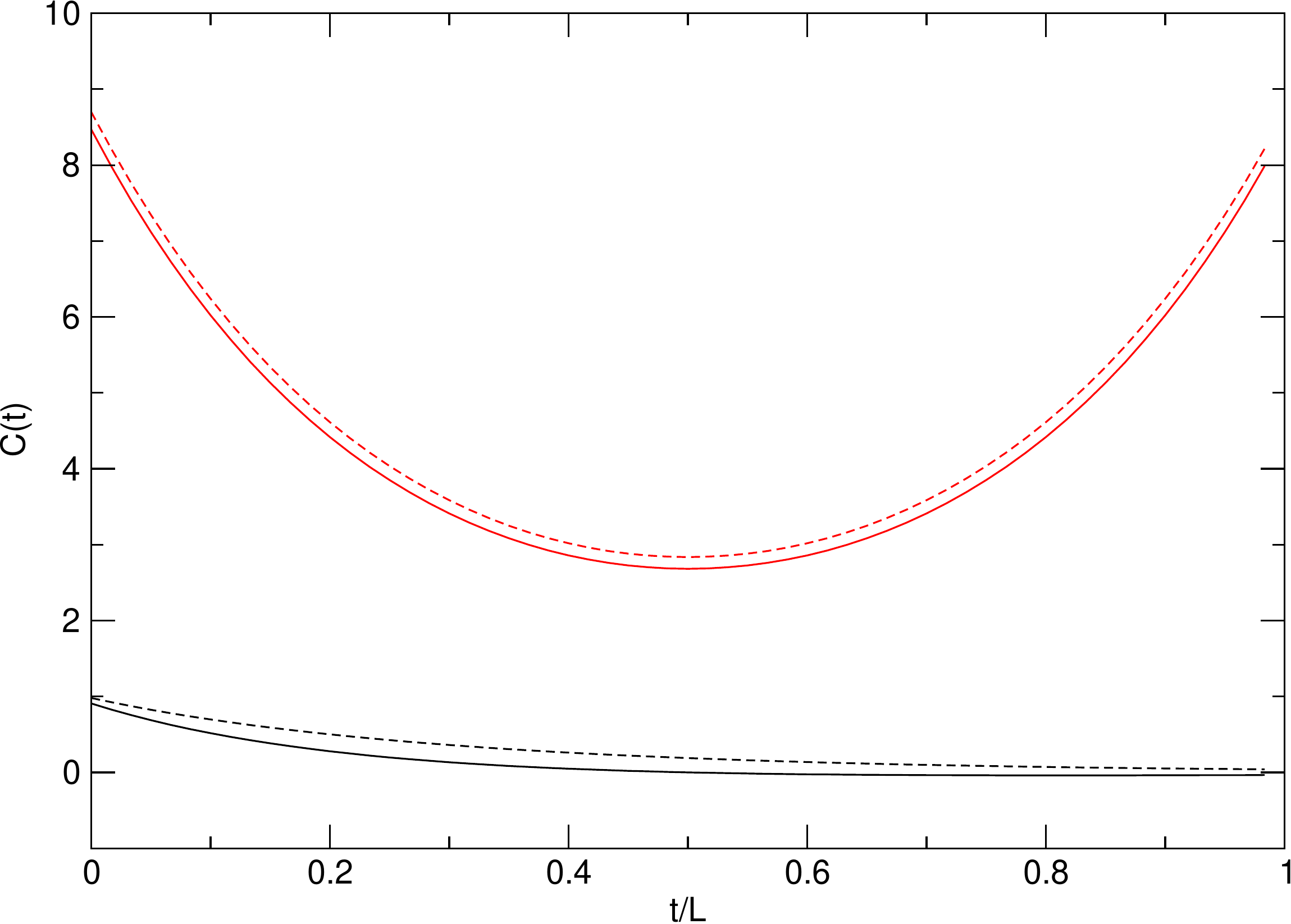}
        }\\
        \subfigure[$\mu L = 10$]{%
           \label{fig_C_c_1_mL_10_L_60}
           \includegraphics[width=0.8\textwidth]{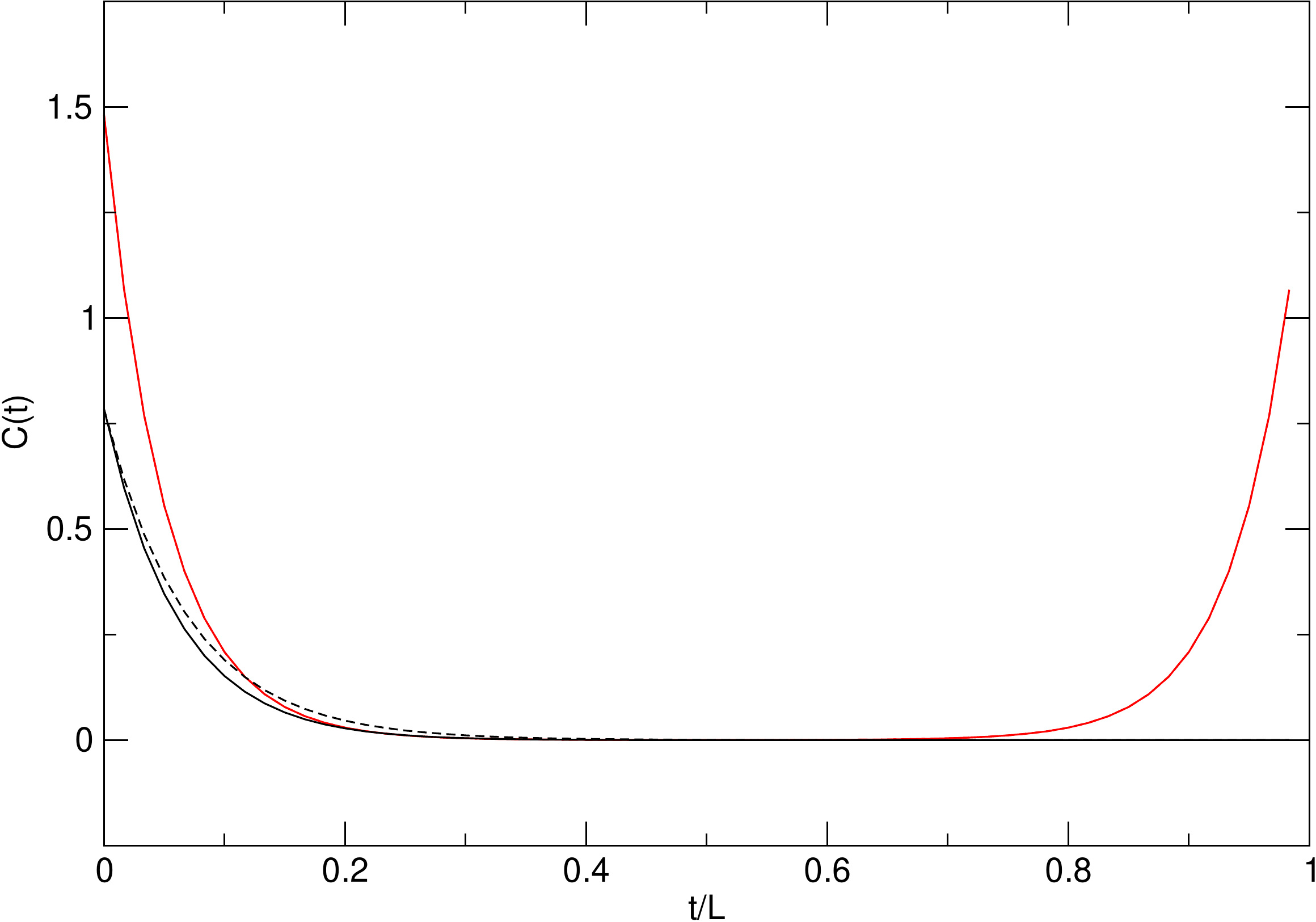}
        }
        \caption{Unbroken supersymmetric quantum mechanics, standard
          discretisation. The bosonic (red) and fermionic (black)
          correlation functions for periodic (dashed) and antiperiodic
          boundary conditions (solid) at fixed coupling $f_u =
          1$. Note that in plot (b) the bosonic correlation functions
          for periodic and antiperiodic b.c.~are indistinguishable.}
    \label{fig_correlation functions}
\end{figure}

\begin{figure}
 \centering
 \includegraphics[width = 0.8\textwidth]{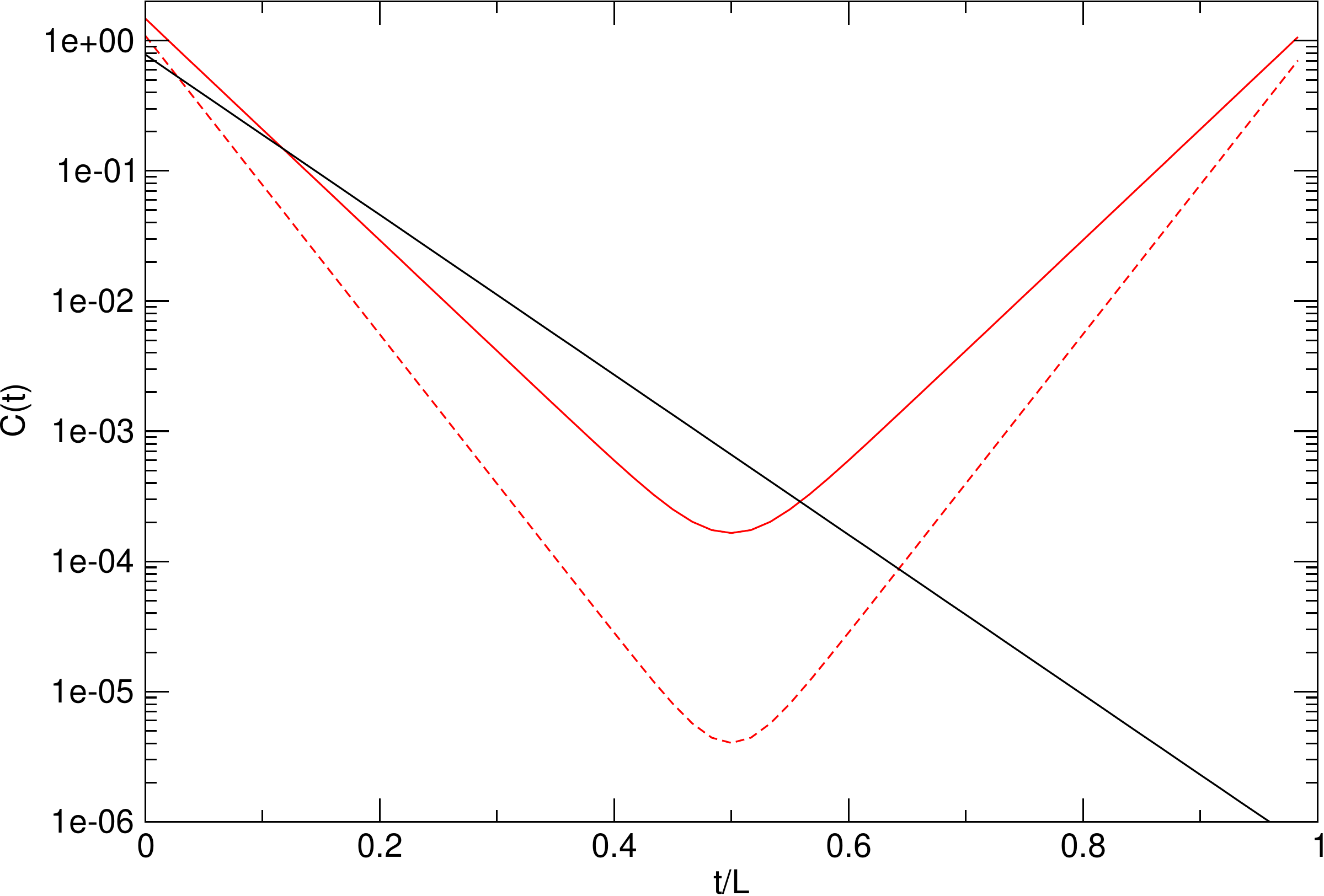}
 \caption{Unbroken supersymmetric quantum mechanics, standard
   discretisation. The bosonic (red) and fermionic (black) correlation
   functions in the bosonic sector $F = 0$ (solid) and the bosonic
   correlation function in the fermionic sector $F = 1$ (dashed) for
   $\mu L = 10$ and coupling $f_u = 1$.}
  \label{fig_C_sectors_c_1_mL_10_L_60}
\end{figure}

First, we show the bosonic and the fermionic correlation function for
unbroken supersymmetry using the standard discretisation. In figure
\ref{fig_C_c_1_mL_2_L_60} we display the bosonic and fermionic
two-point correlation functions $C^{b,f}(t)$ for periodic and
antiperiodic b.c., respectively, at fixed coupling $f_u = 1$ for $\mu
L = 2$ corresponding to a high temperature. In figure
\ref{fig_C_c_1_mL_10_L_60} the same correlation functions are
displayed for $\mu L = 10$ corresponding to a low temperature.
For $\mu L = 2$, we know from Section \ref{subsec:Zp_Za_results} that
finite temperature effects are not negligible. In figure
\ref{fig_C_c_1_mL_2_L_60} these effects are reflected by the fact that
the correlation functions for periodic and antiperiodic b.c.~are
clearly distinguishable, i.e., they are sensitive to the boundary
conditions. For $\mu L = 10$, we are in a regime where the system
behaves as being close to zero temperature where the system is
dominated by the bosonic vacuum. Thus, the bosonic correlation
functions receive contributions only from the bosonic sector and are
hence no longer distinguishable for periodic and antiperiodic b.c.~as
illustrated in figure \ref{fig_C_c_1_mL_10_L_60}. The fermionic
correlation functions on the other hand are different for periodic and
antiperiodic b.c.~even for this choice of parameters. This difference
originates from the specific implementation of the boundary conditions
via eq.(\ref{corr_ferm_t}) which takes into account how many times an
open fermion string of length $t$ can cross the boundary when the
translational invariance of the correlation function is incorporated.
To complete our discussion for unbroken supersymmetry, in figure
\ref{fig_C_sectors_c_1_mL_10_L_60} we display the correlation
functions for the same calculation as above, but individually for each
sector $F$ according to eq.(\ref{corr_sectors_bos}) and (\ref{corr_sectors_ferm}).
Note that we only plot the fermionic correlation function $C^f_0(t)$
in the bosonic sector $F = 0$, but not the antifermionic correlation
function $C_1^{\overline f}(t)$ in the fermionic sector $F=1$, cf.~our
discussion in \cite{Baumgartner:2014nka} and above in Section
\ref{subsec:mass_gaps} and \ref{subsec:correlation functions}. The
bosonic correlation functions are shown in both the bosonic and
fermionic sector. However, in this temperature regime $Z_0 \gg Z_1$
and therefore, the bosonic correlation function in the fermionic
sector $C_1^b(t)$ is heavily suppressed with respect to the one in the
bosonic sector $C_0^b(t)$ when contributing to the correlation
function $C^b_{p,a}(t)$ with fixed fermionic boundary conditions. It
is also interesting to note that the correlation functions consist of
a single exponential term only, i.e., the overlap of the operators
$\phi$ and $\psi$ with the state corresponding to the lowest mass gap
is maximal.

\begin{figure}
 \centering
 \includegraphics[width = 0.8\textwidth]{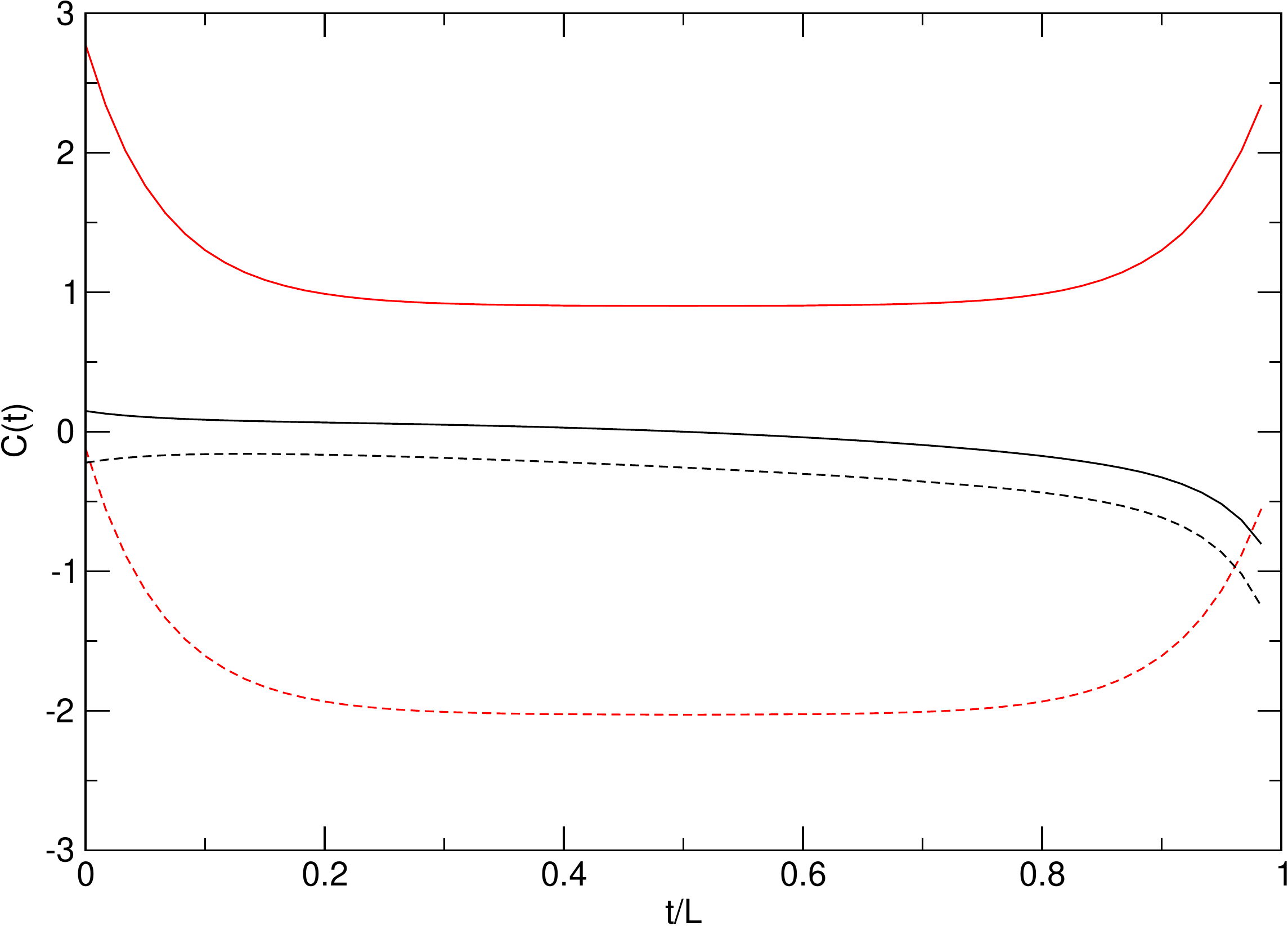}
 \caption{Broken supersymmetric quantum mechanics, standard
   discretisation. The bosonic (red) and fermionic (black) correlation
   functions for periodic (dashed) and antiperiodic b.c.~(solid) for
   $\mu L = 10$ and coupling $f_b = 1$.}
  \label{fig_C_c_1_mL_10_L_60_b}
\end{figure}

\begin{figure}
 \centering
 \includegraphics[width = 0.8\textwidth]{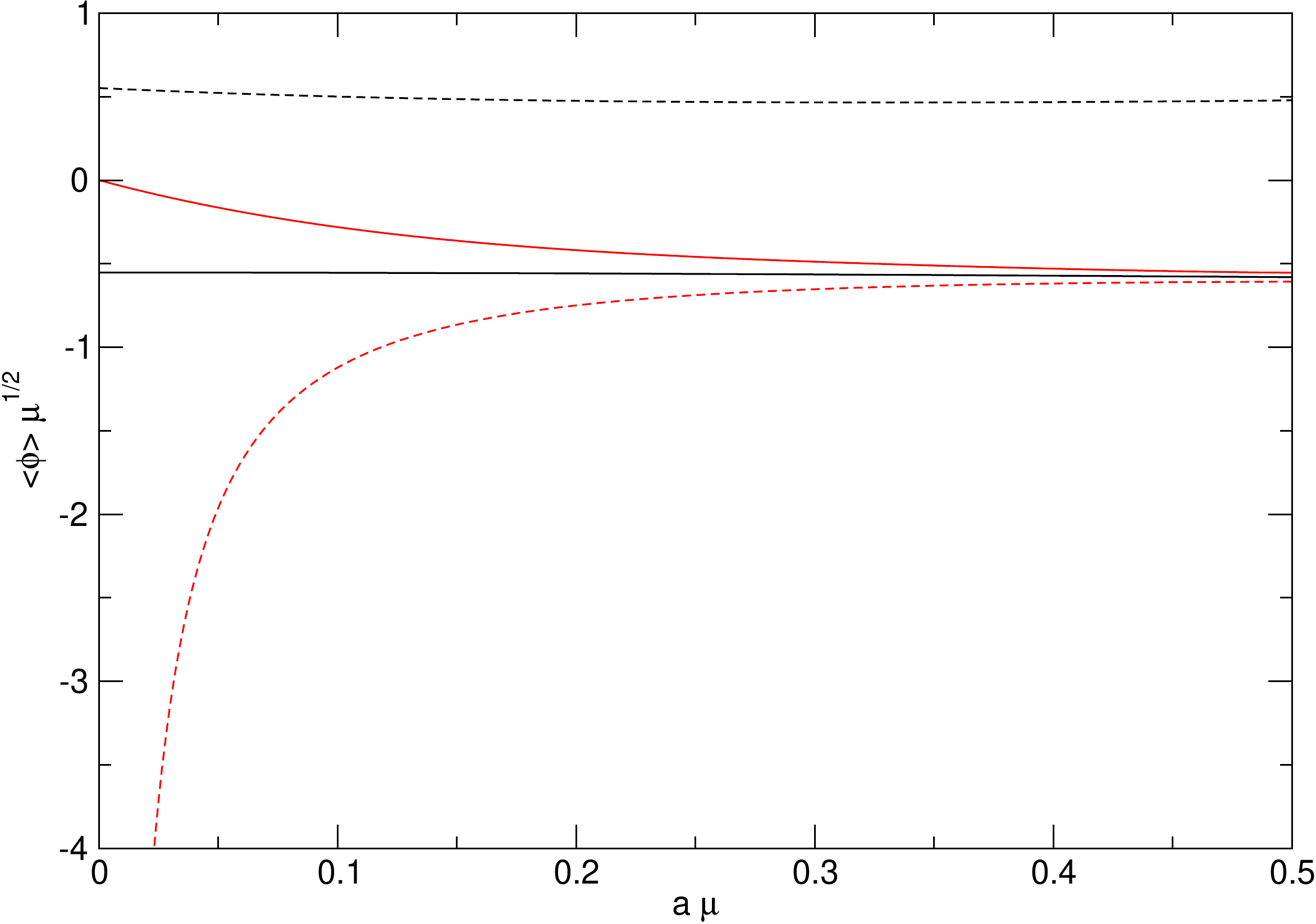}
 \caption{Broken supersymmetric quantum mechanics, standard
   discretisation. Continuum extrapolation of $\langle \phi \rangle$
   in the sectors $Z_0$ (black dashed line) and $Z_1$ (black solid
   line), for antiperiodic (solid red line) and periodic boundary
   conditions (dashed red line) at $\mu L = 10$ and $f_b = 1$.}
  \label{fig_phi_ill_b}
\end{figure}

We now turn to the analogous correlation functions for broken
supersymmetry. In figure \ref{fig_C_c_1_mL_10_L_60_b}, the bosonic and
the fermionic correlation functions are displayed for periodic and
antiperiodic b.c.~for $\mu L = 10$ at fixed coupling $f_b = 1$.
In contrast to unbroken supersymmetry, the bosonic correlation
functions do not approach zero for $t/L \sim 1/2$. To get an
understanding for this, we first need to consider figure
\ref{fig_phi_ill_b} where we show the continuum extrapolation for
$\langle \phi \rangle$ in the same physical situation, i.e.~at $\mu L
= 10$ and $f_b = 1$.
For each sector the expectation value $\langle \phi \rangle_F$
extrapolates to the same value but with opposite sign. This is
expected because of the additional $\mathbb{Z}_2$-symmetry for the
superpotential $P_b$. Furthermore, figure \ref{fig:Zp_Za_b_count}
suggests that both sectors $Z_0$ and $Z_1$ are weighted equally in the
continuum. Therefore, on the one hand, $\langle \phi \rangle_{a}
\rightarrow 0$. On the other hand, for periodic b.c., the numerator
takes a fixed value while the denominator goes to zero and the
expectation value $\langle \phi \rangle_{p}$ is thus ill-defined in
the continuum\footnote{Note, however, that for example $\langle \phi^2
  \rangle_{p}$ is well defined in the continuum, cf.~Section
  \ref{sec:Ward identitities}.}.

\begin{figure}
 \centering
  \subfigure[][Linear plot]{
 \includegraphics[width = 0.8\textwidth]{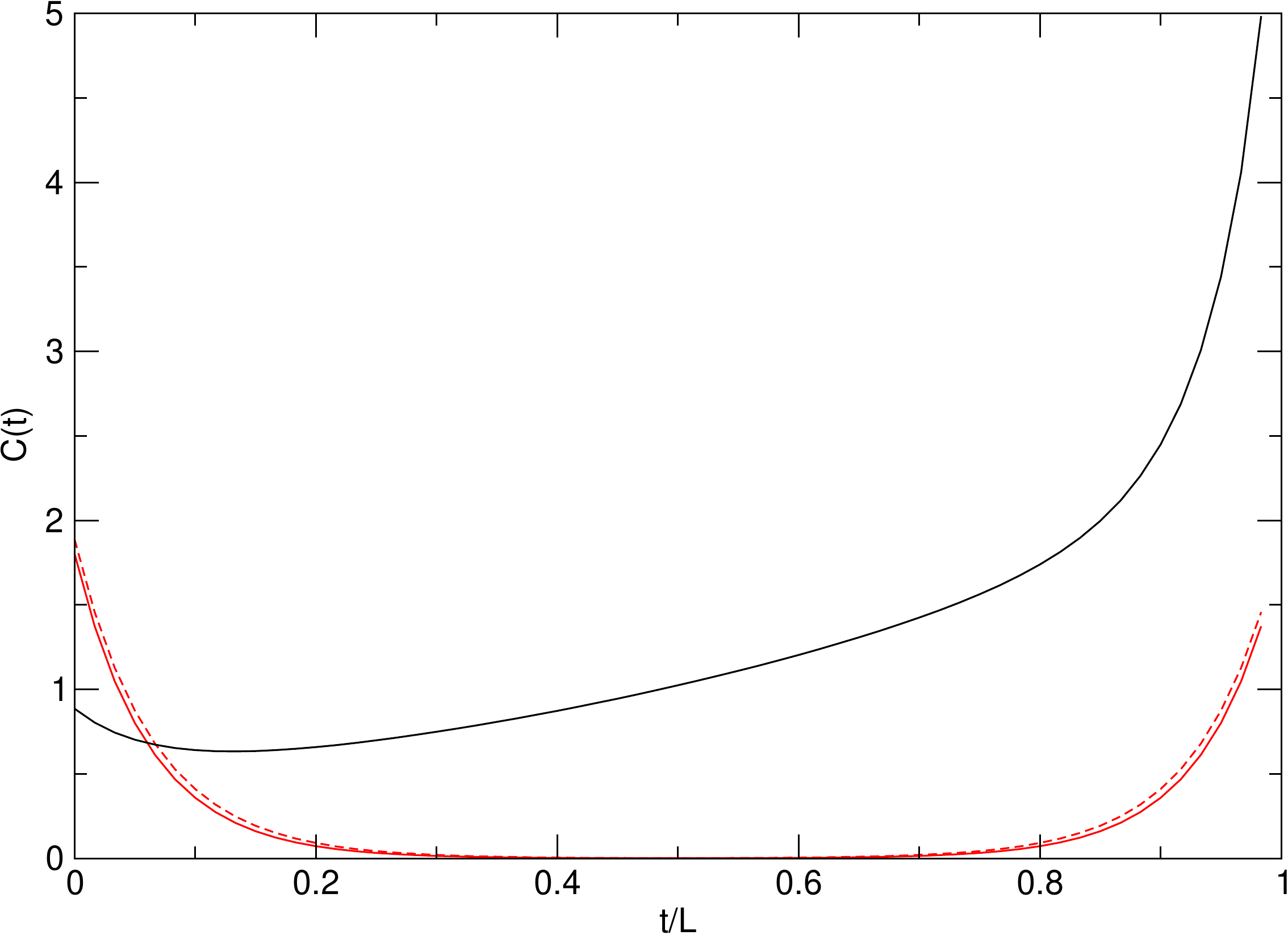}}\\
  \subfigure[][Logarithmic plot]{
\hspace{-0.75cm}
 \includegraphics[width = 0.85\textwidth]{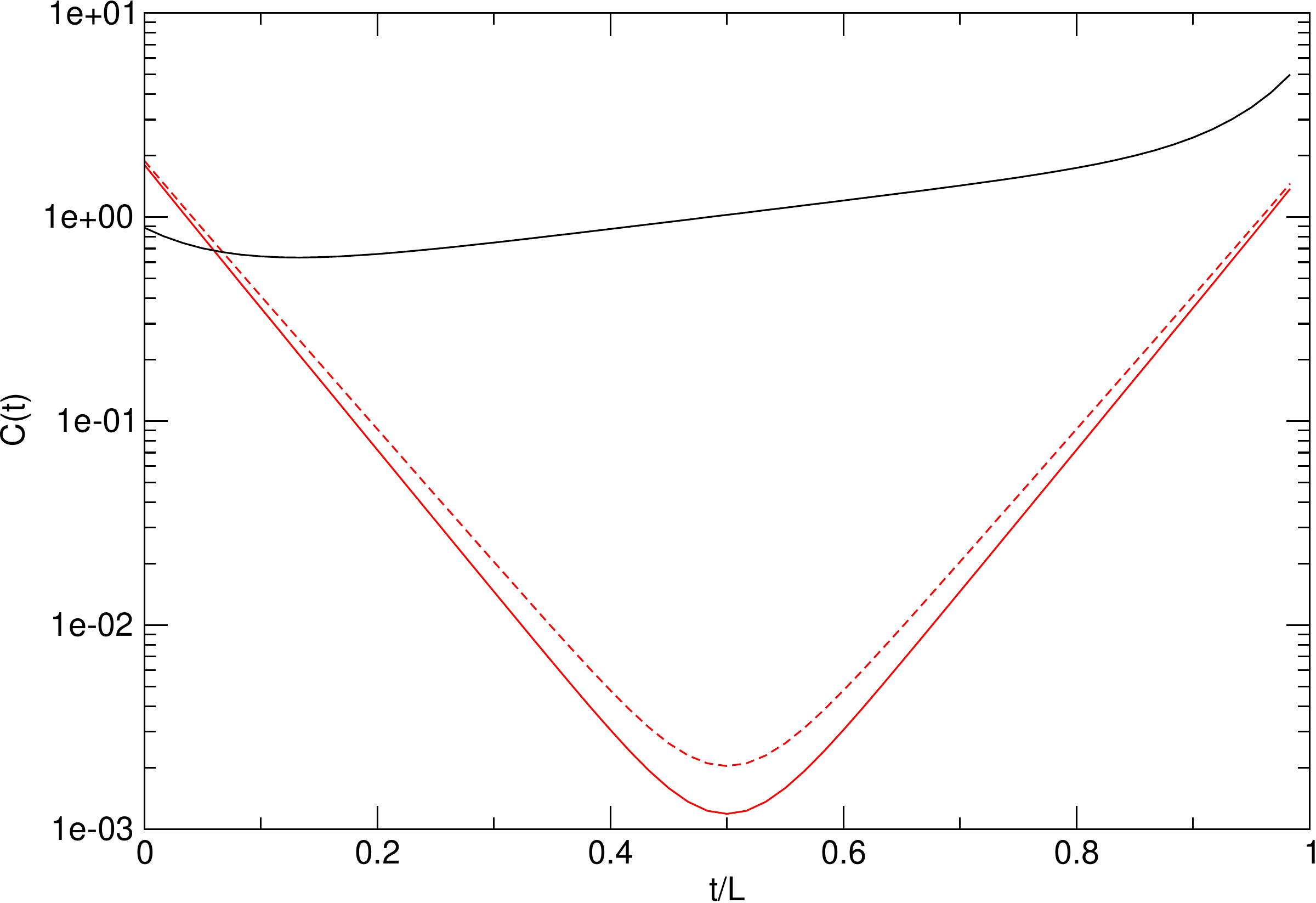}}
\caption{Broken supersymmetric quantum mechanics, standard
  discretisation. The bosonic (red) and fermionic (black) correlation
  functions in the sector $Z_0$ (solid line) and $Z_1$ (dashed line)
  at $\mu L = 10$ and $f_b = 1$.}
  \label{fig_C_sectors_c_1_mL_10_L_60_b}
\end{figure}

After these considerations concerning the expectation value of $\phi$,
we are now able to explain the rather strange fact that the bosonic
correlation function has a negative offset. Instead of removing a
possible constant offset in the connected bosonic two-point function,
the term $\langle \phi \rangle^2_{p}$ shifts the correlation function
to negative values. This problem of the shift into the negative
worsens closer to the continuum because the term $-\langle \phi
\rangle_{p}^2$ takes larger negative values for smaller lattice
spacings. The bosonic correlation function is therefore an ill-defined
observable in the continuum for periodic boundary conditions. It is
therefore necessary to look at the correlation function in each sector
individually and we do so in figure
\ref{fig_C_sectors_c_1_mL_10_L_60_b}.
We observe that the bosonic correlation functions are very similar in
each sector. Note, that the term $\langle \phi \rangle^2_F$ for the
correlation functions measured independently in the bosonic and
fermionic sector indeed removes the additional constant shift such
that the connected bosonic correlator is close to zero for $t/L \sim
1/2$. It is also worth discussing the rather oddly shaped fermionic
correlation function. The figure reveals that there are contributions
of four dominant exponentials instead of only one as in the unbroken
case. For $t/L \sim 0$ and for $t/L \sim 1$, there are two separate
exponentials with large slopes, one in forward direction and one in
backward direction. We can interpret these parts as coming from the
second mass gap of the fermion yielding an exponential decrease for
small $t$ and of the antifermion yielding an exponential increase for
large $t$. In addition, we have two rather flat exponential
contributions around $t/L \sim 1/2$. These can be interpreted as the
first mass gaps for the fermion and antifermion. As discussed before,
for broken supersymmetry the fermionic vacuum has a lower energy than
the bosonic one due to lattice artefacts, and therefore the lowest
mass gap for the fermion is in fact \emph{negative} and leads to the
increase around $t/L \sim 1/2$. The effective masses which are
extracted in this region are very small. In fact, these are the first
indications for the mass of the Goldstino which appears in the
spectrum for broken supersymmetry. We will elaborate further on this
when we discuss the exact results for the mass gaps in the following
section.

The correlation functions using the $Q$-exact discretisation do not
reveal anything qualitatively different, hence we directly proceed to
the discussion of the mass gaps where we can compare the lattice
artefacts for the standard and $Q$-exact discretisation on a more
quantitative level.

\subsection{Mass gaps}\label{subsec:mass_gaps_results}

The derivation of the mass gaps using the eigenvalues of the transfer
matrices in Section \ref{subsec:mass_gaps} suggests to calculate the
bosonic mass gaps in each sector separately. The fermionic mass gaps
are measured via ratios of eigenvalues of $T^1$ and $T^0$. This is a
reflection of the fermionic correlation function being defined in the
bosonic sector, but by reinterpreting the open fermion string in the
bosonic sector as describing the antifermion string in the fermionic
sector, one can also define the mass of an antifermion.  In addition
to our exact results at finite lattice spacing we also calculate the
spectra directly in the continuum using Numerov's algorithm as a
crosscheck and a benchmark for the lattice results. Since the spectrum
is a property of the transfer matrix independent of the system size,
the results do not depend on $\mu L$.

\begin{figure}
 \centering
  \includegraphics[width = 0.8\textwidth]{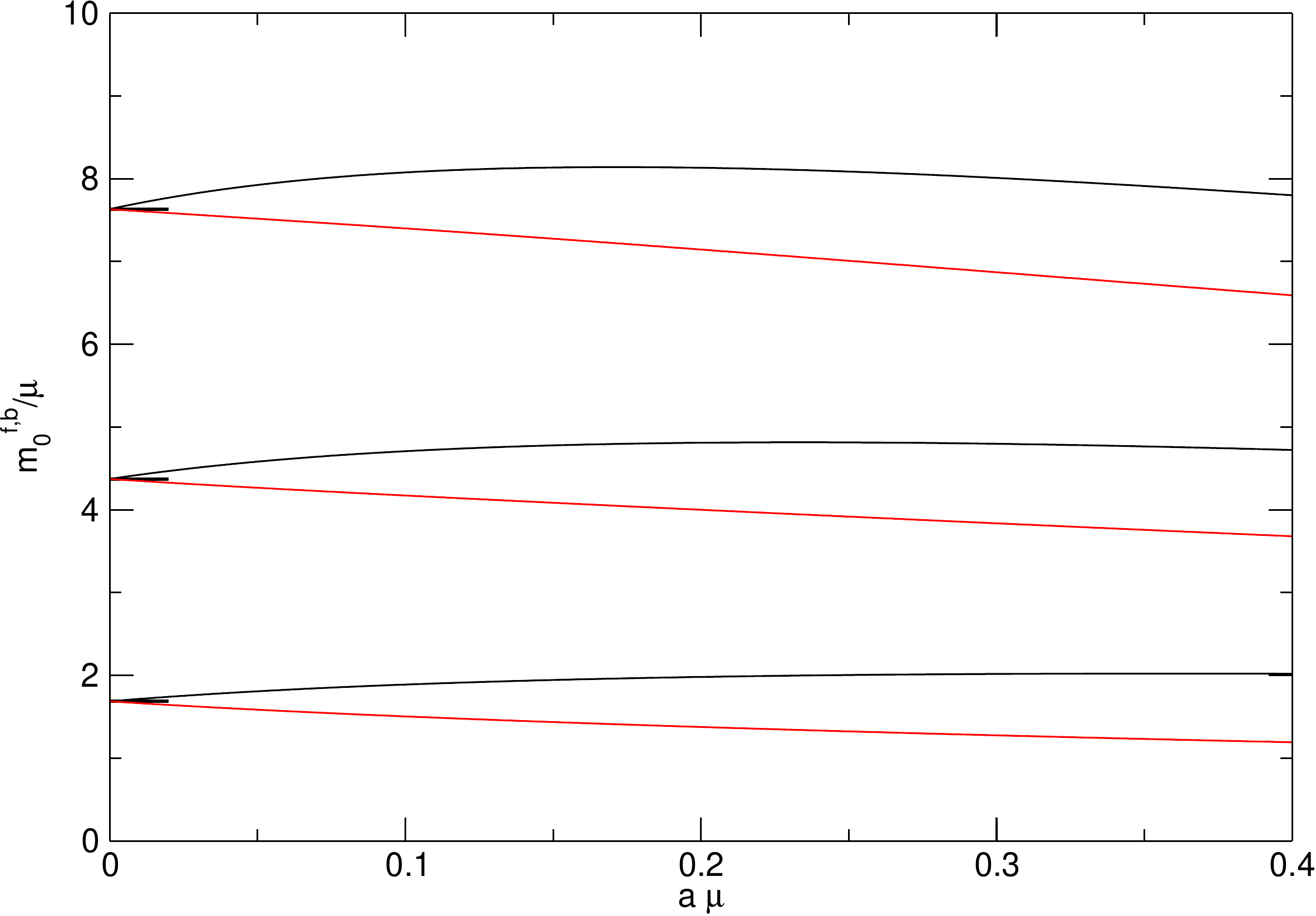}
  \caption{Unbroken supersymmetric quantum mechanics, standard
    discretisation. Continuum extrapolation of the bosonic (black) and
    fermionic (red) masses with respect to the bosonic vacuum at $f_u
    = 1$.}
  \label{fig_masses_u}
\end{figure}

We start as usual with unbroken supersymmetry using the standard
discretisation. In figure \ref{fig_masses_u} we plot the bosonic and
fermionic masses with respect to the bosonic vacuum at a coupling $f_u
= 1$.
First, we note that the mass gaps indeed extrapolate to the expected
continuum values indicated by the horizontal lines at the left side of
the plot. It turns out that the leading lattice artefacts are $\O(a)$
for both the fermion and boson masses but with opposite signs, and are
reasonably small even at rather coarse lattice spacings.  The mass
gaps relative to the fermionic vacuum can of course also be
calculated, but the information is redundant and we refer to
\cite{PhD_Baumgartner:2012} for the detailed results.

\begin{figure}
 \centering
  \includegraphics[width = 0.8\textwidth]{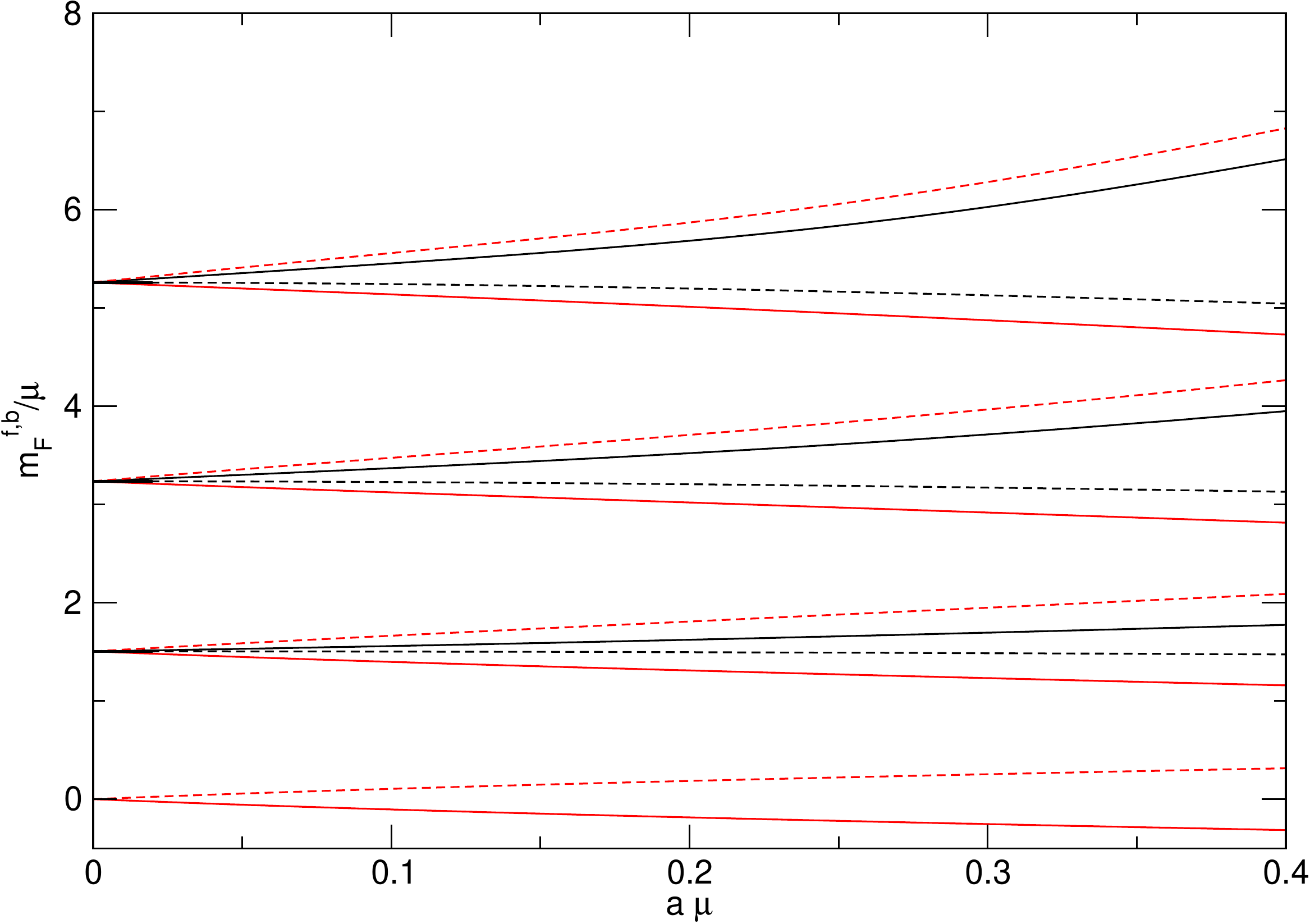}
  \caption{Broken supersymmetric quantum mechanics, standard
    discretisation. Continuum extrapolation of the bosonic (black) and
    fermionic (red) masses with respect to the bosonic (solid lines)
    and the fermionic (dashed lines) vacuum at $f_b = 1$.}
  \label{fig_masses_b}
\end{figure}

Next we discuss our exact results for broken supersymmetry using the
standard discretisation. In figure \ref{fig_masses_b}, we display the
results for the bosonic and the fermionic masses.
While the fermionic vacuum is preferred over the bosonic one at finite
lattice spacing, cf.~figure \ref{fig:Zp_Za_b_count}, in the continuum
they are on equal footing and contribute equally to the partition
functions and observables. Hence in figure \ref{fig_masses_b} we show
the results for all the energy gaps, bosonic ones in black and
fermionic ones in red, both with respect to the bosonic (solid lines)
and fermionic vacuum (dashed lines), despite the fact that the results
are partly redundant. In order to distinguish the lines we follow the
notation in figure \ref{fig:mass_extr_b} where the energy levels for
both sectors are depicted schematically for finite lattice spacing and
in the continuum. It is important to note that the bosonic and the
fermionic mass gaps extrapolate to the expected continuum values also
for broken supersymmetry, and that the supersymmetry in the spectrum,
i.e., the degeneracy between the bosonic and fermionic excitations, is
restored in the continuum limit. This is in contrast to supersymmetric
quantum field theories with spontaneously broken supersymmetry, where
the spectrum becomes nondegenerate, see e.g.~\cite{Steinhauer:2014yaa}
for a nonperturbative demonstration in the two-dimensional ${\cal
  N}=1$ Wess-Zumino model.

When supersymmetry is broken, one expects a fermionic zero-energy
excitation, the Goldstino mode \cite{Salam:1974}, which is responsible
for the tunnelling between the bosonic and the fermionic vacuum and
hence for the fact that $Z_p=0$. From figure \ref{fig_masses_b} it
becomes clear how the lattice acts as a regulator for the Goldstino
mode, namely by giving it a small mass of $\O(a)$, hence making it a
would-be Goldstino. As a consequence, the Witten index $W$ is
regulated.  This allows to give meaning to observables even in the
system with broken supersymmetry and periodic boundary conditions by
defining them at finite lattice spacing, where $Z_p$ is nonzero, and
then taking the continuum limit. If the observable couples to the
would-be Goldstino mode in the same way as $Z_p$ does, both vanish in
the continuum but their ratio is well defined.
Note that since the fermionic vacuum has a lower energy than the
bosonic one, the would-be Goldstino with positive mass is actually the
antifermionic excitation $m^{\overline f}_1$ in the fermion sector,
while the fermionic excitation $m^{f}_0$ in the bosonic sector has a
negative energy. A posteriori, this explains the rather odd shape of
the fermionic correlation function in the bosonic sector displayed in
figure \ref{fig_C_sectors_c_1_mL_10_L_60_b} where the slope of the
slowly increasing correlator corresponds to the small negative mass of
the Goldstino fermion.

Finally, we make the observation that the leading lattice artefacts of
the spectral mass gaps are all $\O(a)$. This is expected since we use
a discretisation of the derivative with $\O(a)$ discretisation errors,
both for the bosonic and fermionic degrees of freedom. However, it is
intriguing that the linear artefacts of the higher lying bosonic mass
gaps $m^b_{i,1}$ in the $F = 1$ sector become very small, and the
corrections are eventually dominated by artefacts of $\O(a^2)$, i.e.,
some interesting conspiracy of lattice artefacts appears to cancel the
$\O(a)$ artefacts.

\begin{figure}
 \centering
  \includegraphics[width = 0.8\textwidth]{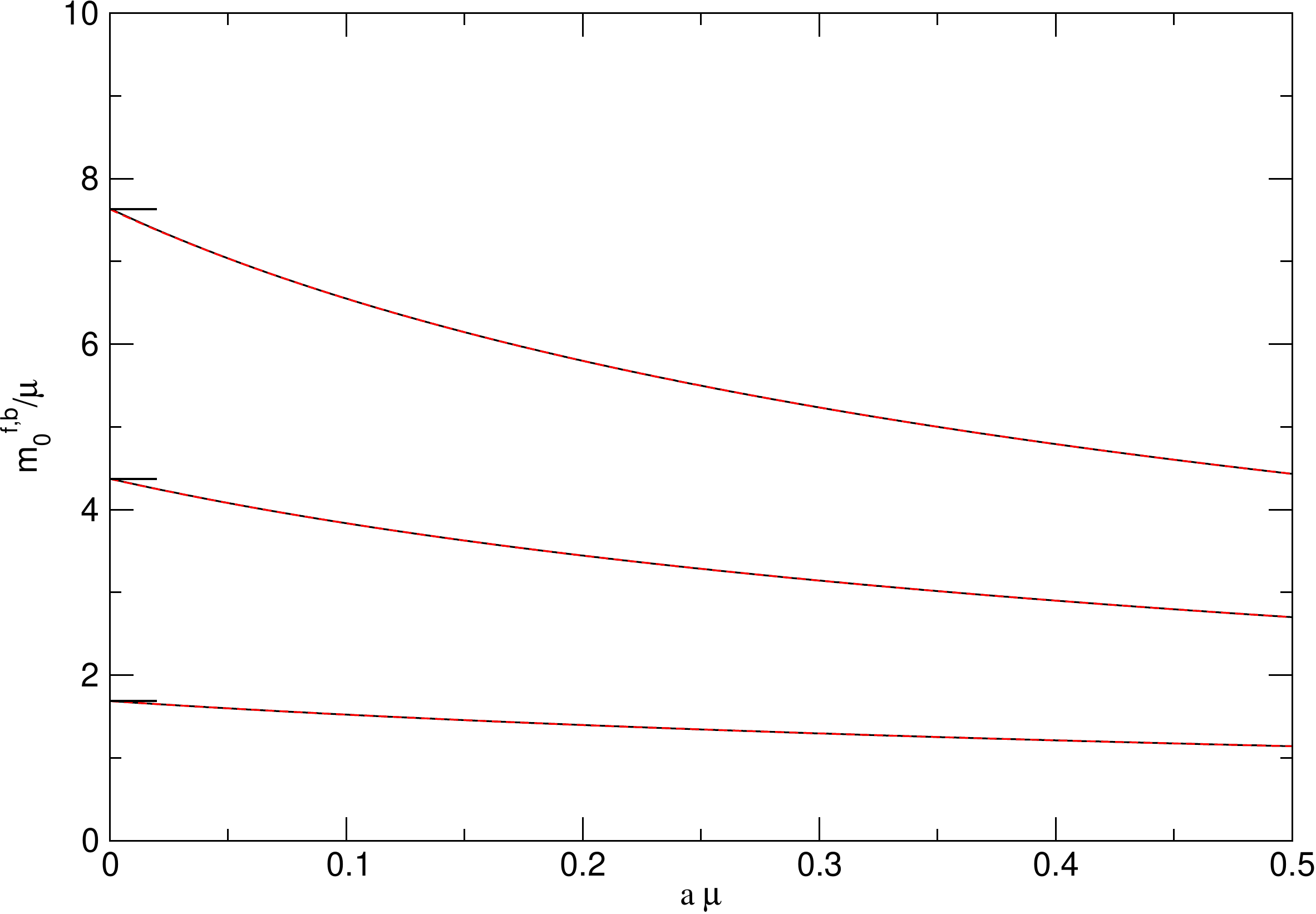}
  \caption{Unbroken supersymmetry, $Q$-exact discretisation. Continuum
    extrapolation of the bosonic masses measured with respect to the
    bosonic (black solid) and the fermionic (black dashed) vacuum and
    the fermionic masses measured with respect to the bosonic (red
    dashed) and the fermionic (red dotted) vacuum at $f_u = 1$.}
  \label{fig_Masses_mL_10_u_q}
\end{figure}

Next we consider the spectrum using the $Q$-exact discretisation. In
figure \ref{fig_Masses_mL_10_u_q} we plot the fermionic (red) and
bosonic mass gaps (black) with respect to the bosonic (full lines) and
the fermionic vacuum (dashed lines) for unbroken supersymmetry with
coupling $f_u = 1$. The characterisation of the lines is as in the
previous figures for the mass gaps.
From the figure it is clear that the degeneracy between the bosonic
and fermionic excitations is maintained for any finite value of the
lattice spacing. Apparently, keeping only half of the original
symmetries in eq.(\ref{susy_transf_cont}), as realised by the
$Q$-exact discretisation, is sufficient to guarantee the complete
degeneracy. However, the lattice artefacts are rather different from
the ones observed in the spectrum of the standard
discretisation. While the lattice artefacts in the lowest excitation
are quantitatively comparable to the ones in the lowest fermionic
excitation in the standard scheme, cf.~figure \ref{fig_masses_u}, they
turn out to be much larger for the higher excited states. As an
example we find lattice corrections of up to 45\% at a lattice spacing
of $a\mu=0.5$ for the third excited state. A possible explanation is
that in order to maintain the exact degeneracy between the bosonic and
fermionic energies, essentially aligning the lattice artefacts of the
bosonic and fermionic states, the eigenvalues have to rearrange in a
particular way and push the artefacts into the higher states.  So
while the $Q$-exact discretisation is an extremely useful scheme due
to its improved symmetry properties, one has to be aware that the
lattice artefacts may be dramatically enhanced for certain
observables.

The spectrum of the $Q$-exact action for broken supersymmetry turns
out to be very difficult to handle. On the one hand, using the
superpotential $P_b$ the transfer matrices $T^0$ and $T^1$ come out to
be exactly similar when using a reasonably large cutoff for the
bosonic occupation numbers and, hence, the energy levels are exactly
degenerate for any lattice spacing.  The similarity transformation
relating the two transfer matrices can be understood as the
supersymmetry transformation relating the bosonic and the fermionic
sector and is exactly maintained at finite lattice spacing. As a
consequence of the exact similarity we have an exactly massless
Goldstino mode and hence also $Z_{p}/Z_{a} = 0$, independently of both
the reasonably large occupation number cutoff and the lattice spacing
$a$.  On the other hand, however, results do not become independent of
the cutoff even for reasonably large bosonic occupation numbers. Even
on very small lattices and for coarse lattice spacings the occupation
numbers necessary to produce stable transfer matrix eigenvalues appear
to be extremely large. Hence, despite the fact that the properties of
the transfer matrices qualitatively yield the correct physics in terms
of the spectrum and the Witten index, we are not able to further
investigate the system with broken supersymmetry using the $Q$-exact
action.

\subsection{Ward identities}
\label{sec:Ward identitities}
In this section we present our exact results for the Ward identities
which we introduced in Section \ref{subsec:WI}. We discuss the
identities $W_0, W_1(t)$ and $W_2(t)$ in turn.

As usual we start with the discussion of the system with unbroken
supersymmetry using the standard discretisation. In that case, the
Ward identity $W_0$ in eq.(\ref{eq:WI unbroken}) is supposed to vanish
in the continuum limit. However, it turns out that the expectation
value is trivially zero at any value of the lattice spacing, simply
because in the bond formulation the $\mathbb{Z}_2$-symmetry $\phi
\rightarrow -\phi$ is exactly maintained for each bond
configuration. This can most easily be seen from the fact that the
site weights for this action are zero for an odd site occupation
number, $Q_0(2n + 1) = Q_1(2n + 1) = 0, \ n \in \mathbb{N}_0$, and
hence the expectation value of an odd power of $\phi$ trivially
vanishes.

\begin{figure}
 \centering
  \includegraphics[width = 0.8\textwidth]{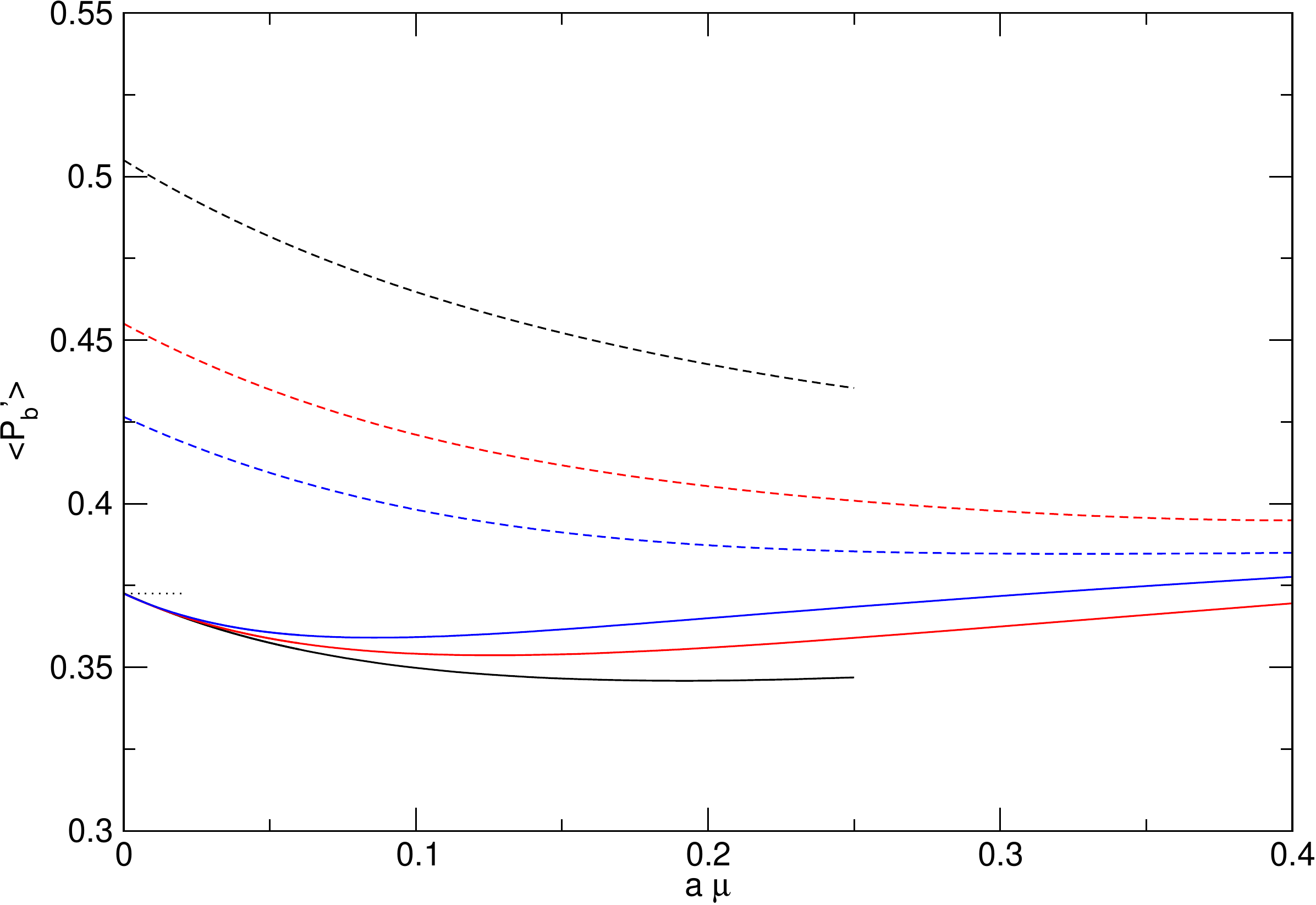}
  \caption{Broken supersymmetric quantum mechanics, standard
    discretisation. Continuum extrapolation of $\langle P_b^\prime
    \rangle/\sqrt{\mu}$ for $\mu L = 5$ (black), $\mu L = 8$ (red),
    and $\mu L = 12$ (blue) using periodic (dashed lines) and
    antiperiodic b.c.~(solid lines) at $f_b = 1$.}
  \label{fig_exp_P_prime_b}
\end{figure}

\begin{figure}
 \centering
 \includegraphics[width =  0.8\textwidth]{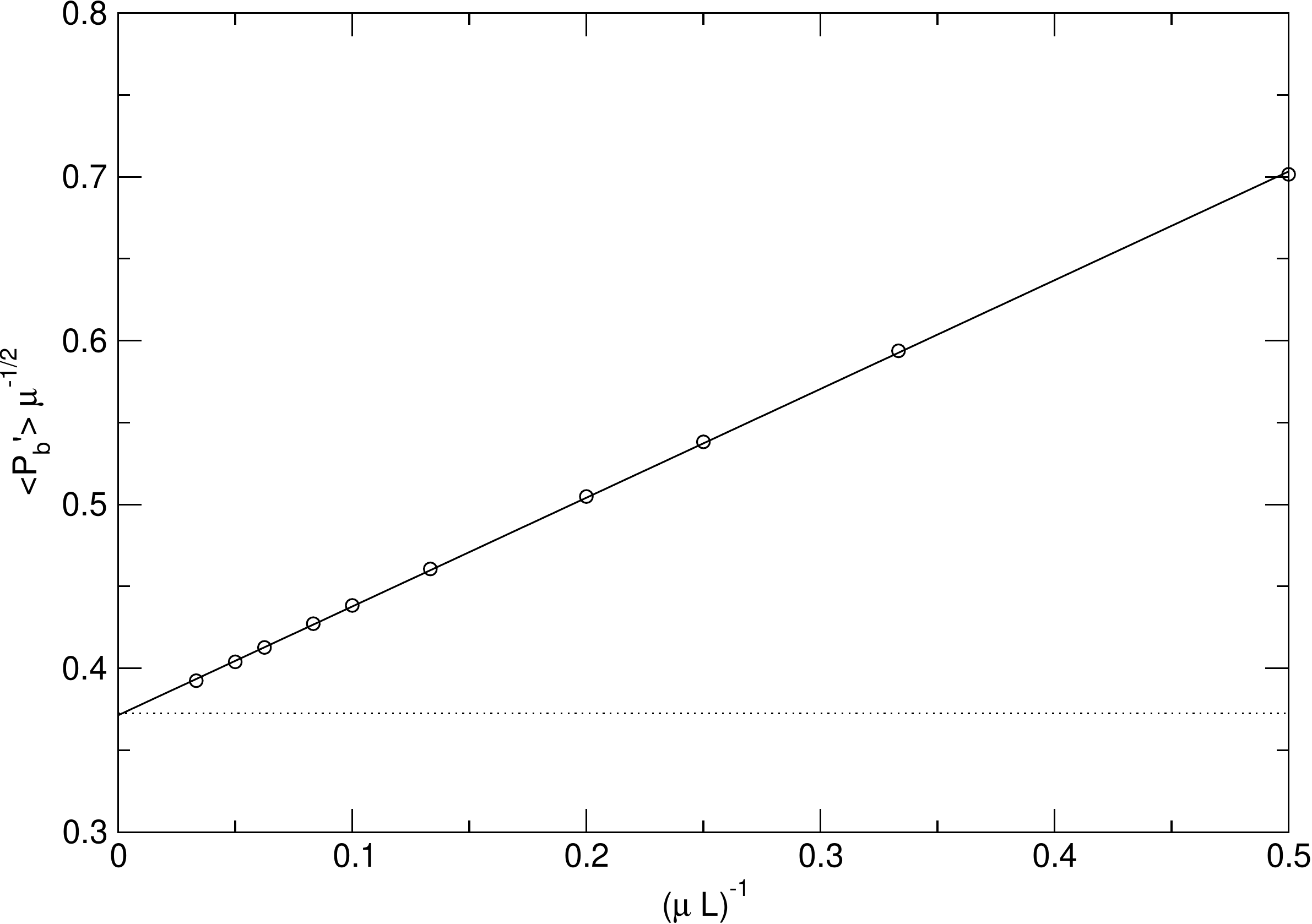}
 \caption{Broken supersymmetric quantum mechanics. The continuum
   values for $\langle P_b^\prime \rangle/\sqrt{\mu}$ as a function of
   $1/\mu L$ for periodic b.c.~at $f_b = 1$. The continuum value for
   antiperiodic b.c.~is indicated with the black dotted line. The
   solid black line is a fit linear in $1/\mu L$.}
  \label{fig_exp_P_prime_b_ex}
\end{figure}

For broken supersymmetry, we need to check whether or not the Ward
identity in eq.(\ref{eq:WI broken}) vanishes.  In figure
\ref{fig_exp_P_prime_b}, we plot the continuum extrapolation of
$\langle P^\prime_b \rangle$ for different values of $\mu L$ at fixed
coupling $f_b = 1$.
For antiperiodic b.c., the Ward identity extrapolates to the value
$\langle P^\prime \rangle /\sqrt{\mu} = 0.3725 \dots$ independently of
the chosen $\mu L$. This value is in agreement with a continuum
calculation in the operator formalism \cite{Wozar:2011gu} denoted by
the horizontal dotted line at the left side of the plot. The continuum
limit for periodic b.c., however, depends on the chosen $\mu L$ and
approaches the continuum value for antiperiodic b.c.~only at large
$\mu L$ where the effects from the boundary become smaller and
smaller.  Note that the continuum limit for this quantity is well
defined despite the fact that $Z_p=0$ in that limit.  In figure
\ref{fig_exp_P_prime_b_ex} we show the continuum values of $W_0$ for
periodic b.c.~as a function of $(\mu L)^{-1}$, i.e., the temperature
in units of $\mu$.
The figure reveals that for large $\mu L$, the values for periodic
b.c.~indeed approach the ones for antiperiodic b.c.~denoted by the
dotted line. Eventually, the values agree in the zero temperature
limit, or rather in the limit of infinite extent of the
system. Interestingly, the finite temperature corrections seem to be
described by the form $1/\mu L$ up to rather large values of $\mu
L$. A corresponding fit is shown in figure \ref{fig_exp_P_prime_b_ex}
as the solid line. In conclusion, the Ward identity $W_0$ serves us
indeed to verify that supersymmetry is broken in the continuum for the
superpotential $P_b$.

The results for the $Q$-exact action do not provide any new
interesting insights, because for unbroken supersymmetry $W_0$
vanishes trivially as for the standard discretisation. For broken
supersymmetry we are not able to achieve stable results using the
$Q$-exact action, as already discussed at the end of Section
\ref{subsec:mass_gaps_results}.

\begin{figure}

    \centering

        \subfigure[$\mu L = 4, Z_{a}$]{%
            \label{fig_11WI_I_4_u}
            \includegraphics[width=0.48\textwidth]{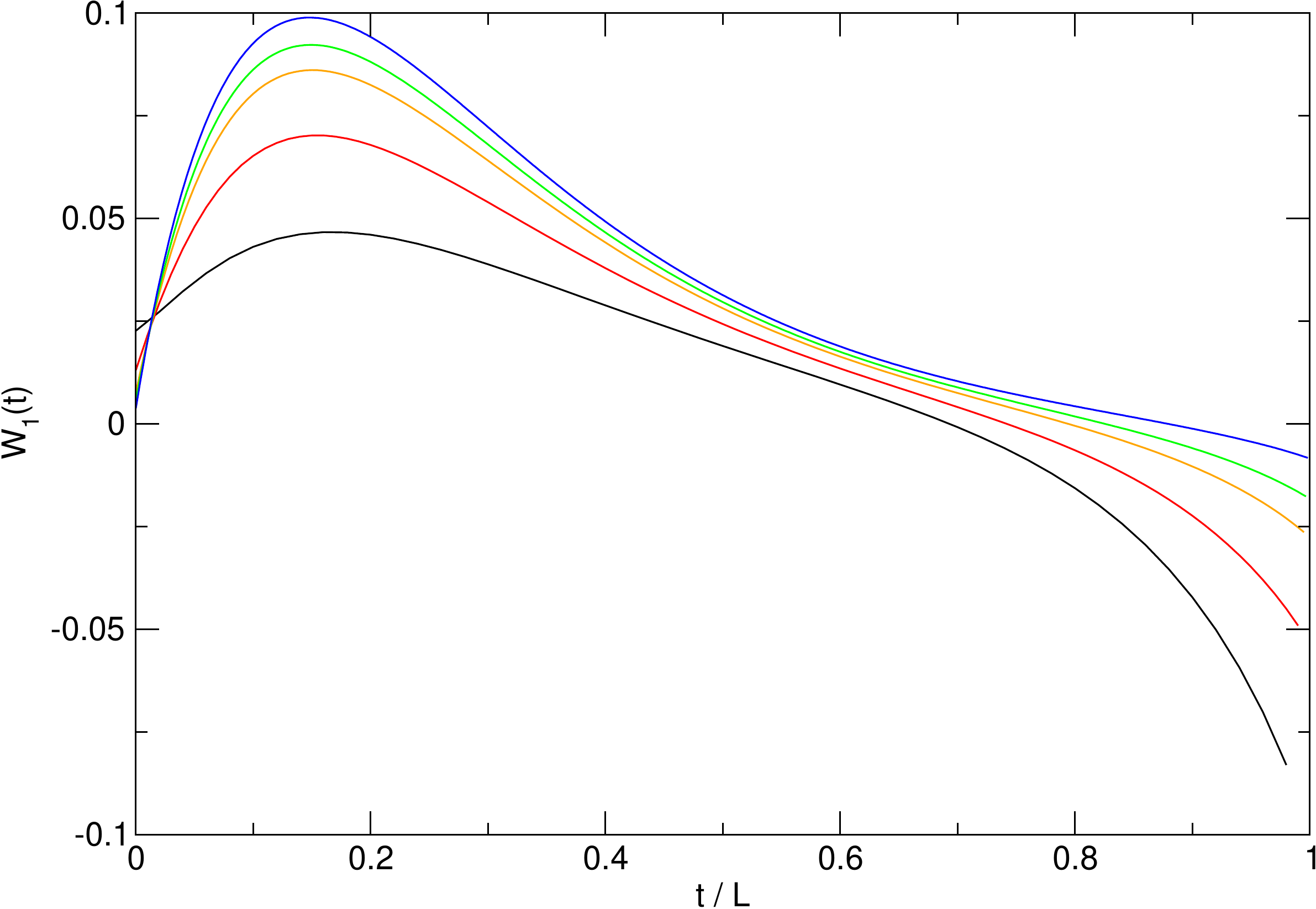}
        }%
        \subfigure[$\mu L = 4, Z_{p}$]{%
           \label{fig_12WI_I_4_u}
           \includegraphics[width=0.48\textwidth]{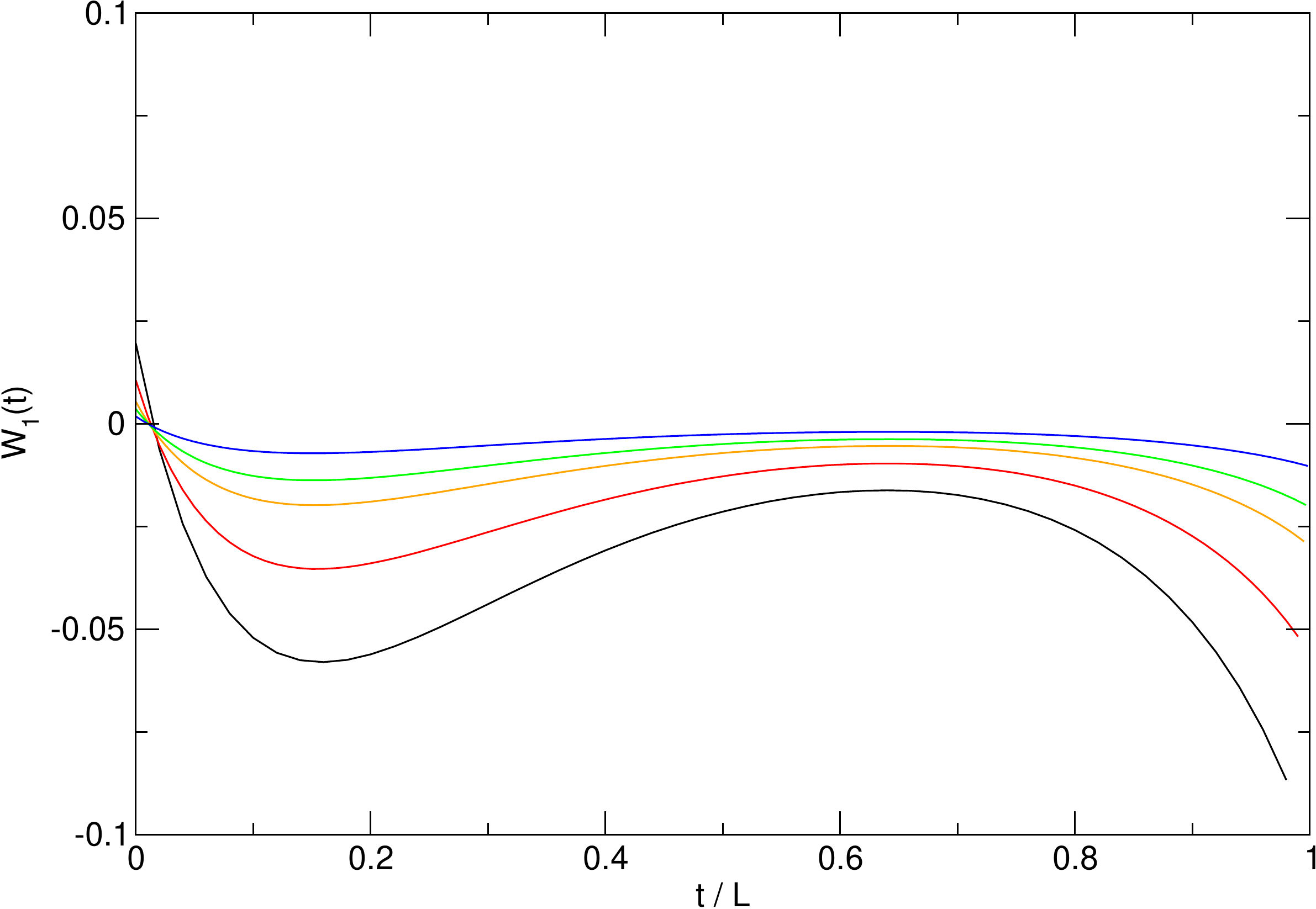}
        }\\%
        \subfigure[$\mu L = 10, Z_{a}$]{%
            \label{fig_121WI_I_u}
            \includegraphics[width=0.48\textwidth]{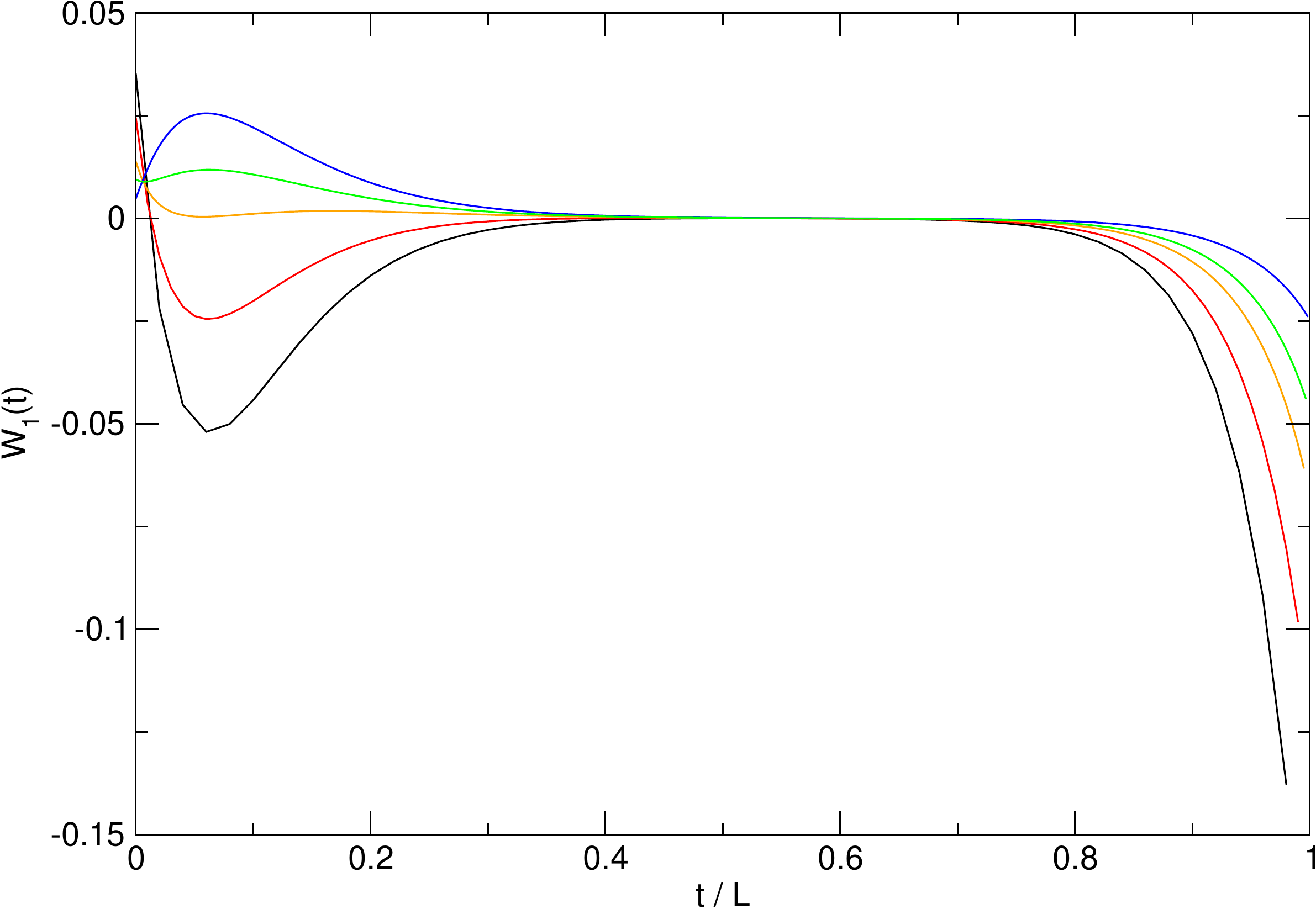}
        }%
        \subfigure[$\mu L = 10, Z_{p}$]{%
           \label{fig_22WI_I_u}
           \includegraphics[width=0.48\textwidth]{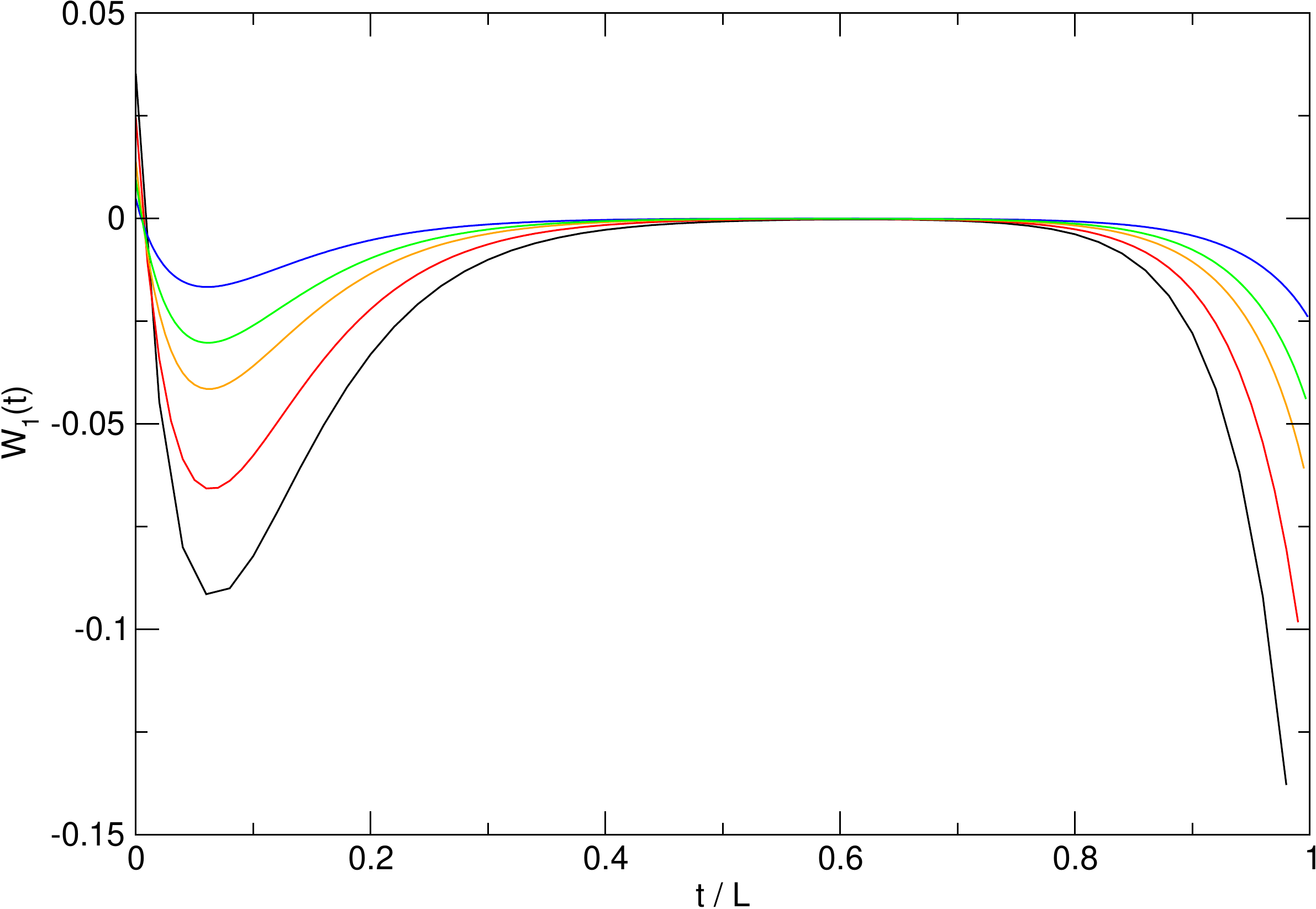}
        }
        \caption{Unbroken supersymmetric quantum mechanics, standard
          discretisation. The Ward identity $W_1$ for $L/a = 50$
          (black), $L/a = 100$ (red), $L/a = 200$ (orange), $L/a =
          300$ (green) and $L/a = 600$ (blue) for $\mu L=4$ and
          $\mu L=10$ at fixed coupling $f_u = 1$.}
    \label{fig_WI_I_u}
\end{figure}

\begin{figure}
 \centering
  \includegraphics[width = 0.8\textwidth]{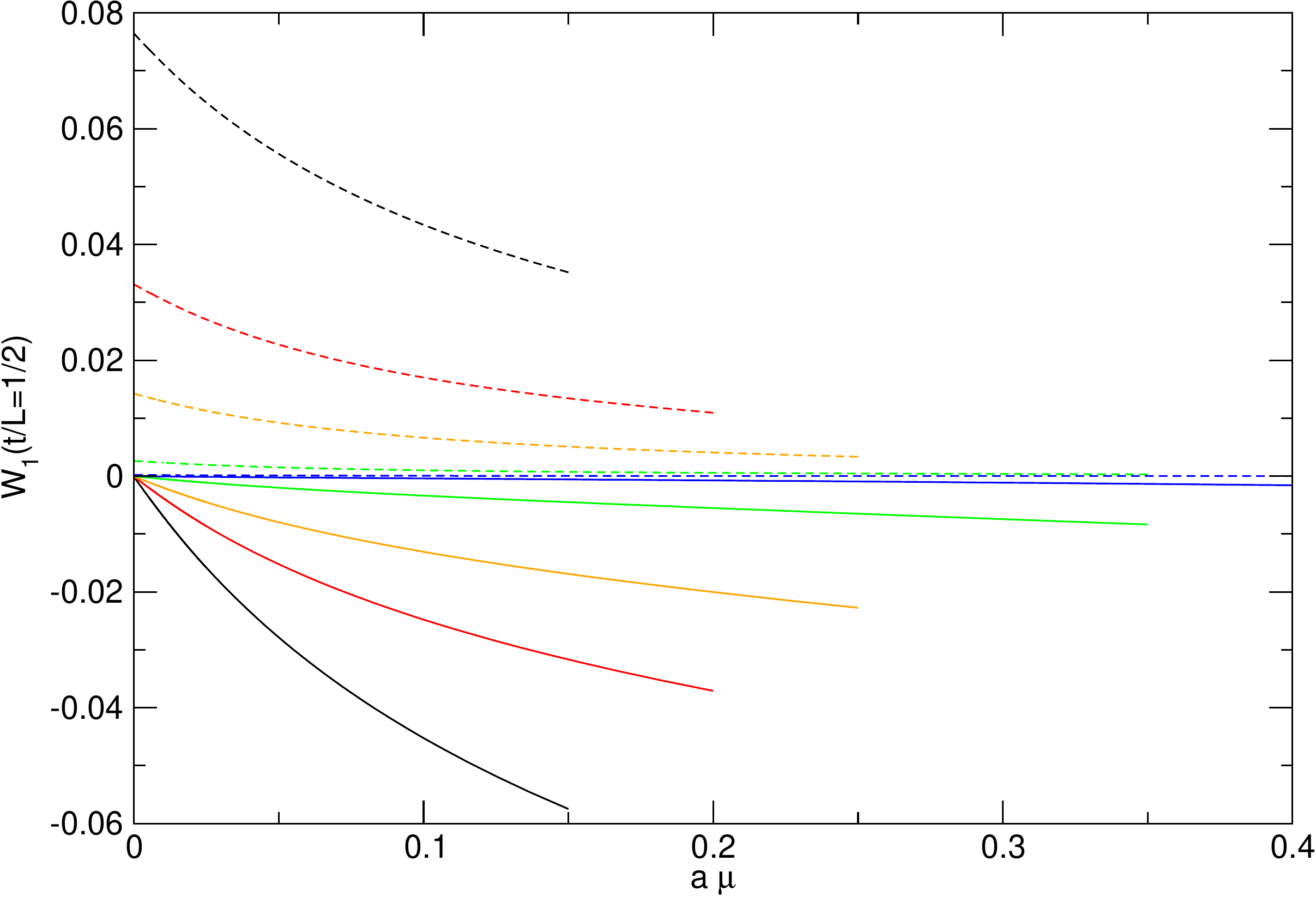}
  \caption{Unbroken supersymmetric quantum mechanics, standard
    discretisation. Continuum extrapolation of $W_1(t/L=1/2)$ for $\mu
    L = 3$ (black), $\mu L = 4$ (red), $\mu L = 5$ (orange), $\mu L =
    7$ (green) and $\mu L = 10$ (blue) for periodic (sold lines) and
    antiperiodic b.c.~(dashed lines) at fixed coupling $f_b = 1$.}
  \label{fig_WI_I_extra_u}
\end{figure}

We now turn to the Ward identity $W_1$ to verify supersymmetry
restoration and breaking for the corresponding superpotentials.  We
start again with the discussion of the results using the standard
discretisation in the system with unbroken supersymmetry.  In figure
\ref{fig_WI_I_u}, we show the Ward identity $W_1(t)$ for $\mu L = 4$
and $\mu L = 10$ for a range of lattice spacings $a/L$ at fixed
coupling $f_u = 1$ for both periodic and antiperiodic boundary
conditions.
The figure illustrates how the Ward identity $W_1(t)$ is violated for
finite lattice spacing. It can be seen that the violation for periodic
b.c.~becomes less severe as $a \rightarrow 0$, whereas for
antiperiodic b.c.~it does not.  In figure \ref{fig_WI_I_extra_u}, we
plot the continuum extrapolation of $W_1(t/L=1/2)$ at the coupling
$f_b = 1$ for different values of $\mu L$.
We find that $W_1(t/L=1/2)$ extrapolates to zero for periodic b.c.,
independent of the value of $\mu L$. Supersymmetry is therefore
restored in the continuum for periodic b.c., even at a finite extent
of the quantum mechanical system. For antiperiodic b.c.~on the other
hand, $W_1(t/L=1/2)$ does not extrapolate to zero for small $\mu L$,
i.e., high temperature. However, as the temperature decreases, the
violation weakens and for $\mu L \rightarrow \infty$ $W_1(t/L=1/2)$
extrapolates to zero, implying that supersymmetry is restored in the
zero temperature limit. On the level of the Ward identities $W_1$ in
the continuum, our results hence confirm all expected features of
unbroken supersymmetry at finite as well as at zero temperature.
Moreover, our results tell us how the system behaves at finite lattice
spacing. First we note that at any fixed lattice spacing, $W_1$
extrapolates to zero in the limit $\mu L \rightarrow \infty$
independently of the boundary conditions. This is a reflection of the
fact that the violation of the supersymmetry in the action from using
the standard discretisation is just a surface term which obviously
becomes irrelevant in the limit $\mu L \rightarrow \infty$. On the
other hand, we note that the decoupling of this artefact seems to
happen faster in a system with antiperiodic boundary conditions. In
other words, the convergence to $W_1=0$ is slower for periodic b.c.~as
can be seen by comparing the limit $\mu L\rightarrow \infty$ for
example at fixed $a\mu=0.15$.

\begin{figure}

    \centering

        \subfigure[$\mu L = 5, Z_{a}$]{%
            \label{fig_11WI_I_5_b}
            \includegraphics[width=0.48\textwidth]{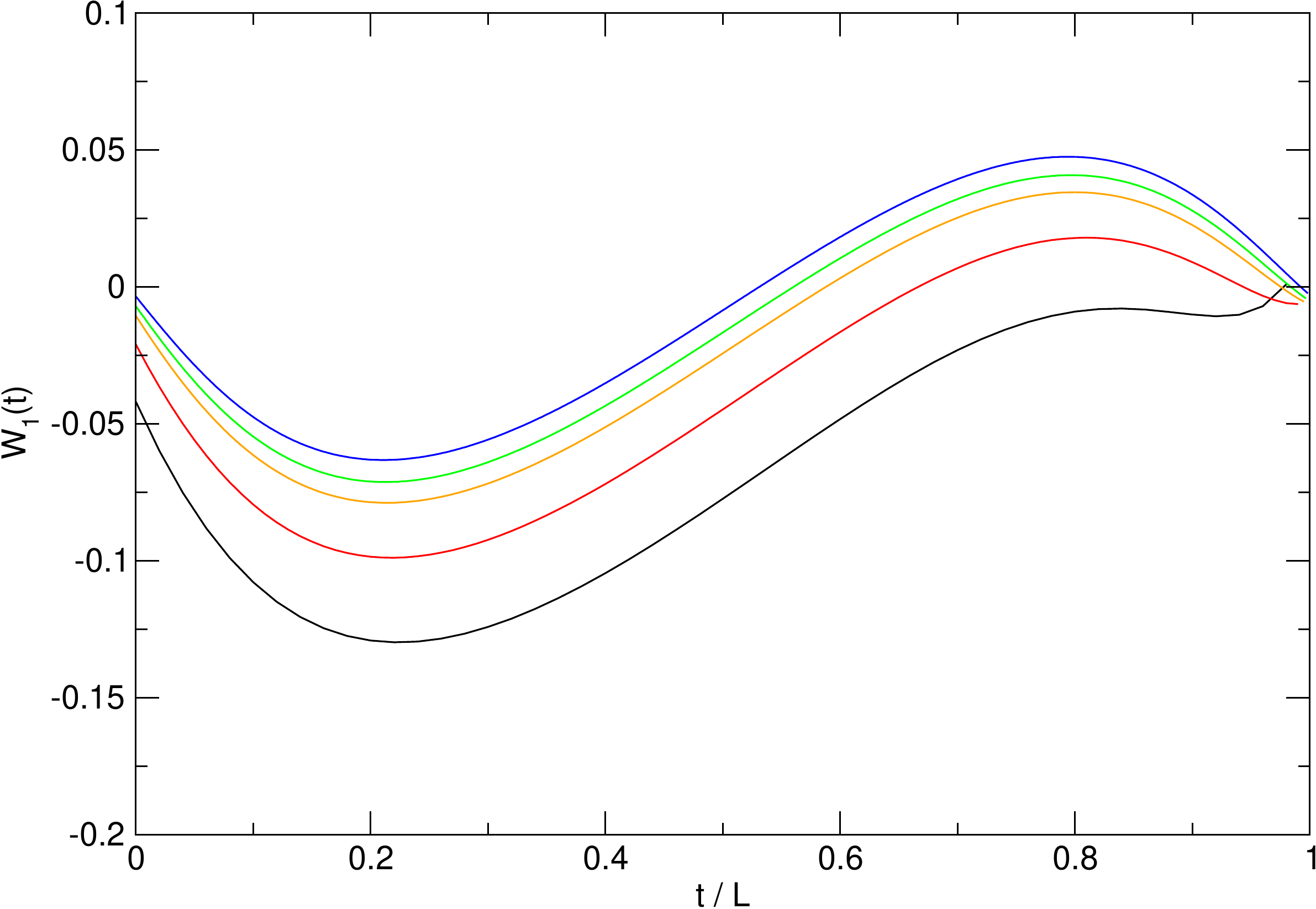}
        }%
        \subfigure[$\mu L = 5, Z_{p}$]{%
           \label{fig_12WI_I_5_b}
           \includegraphics[width=0.48\textwidth]{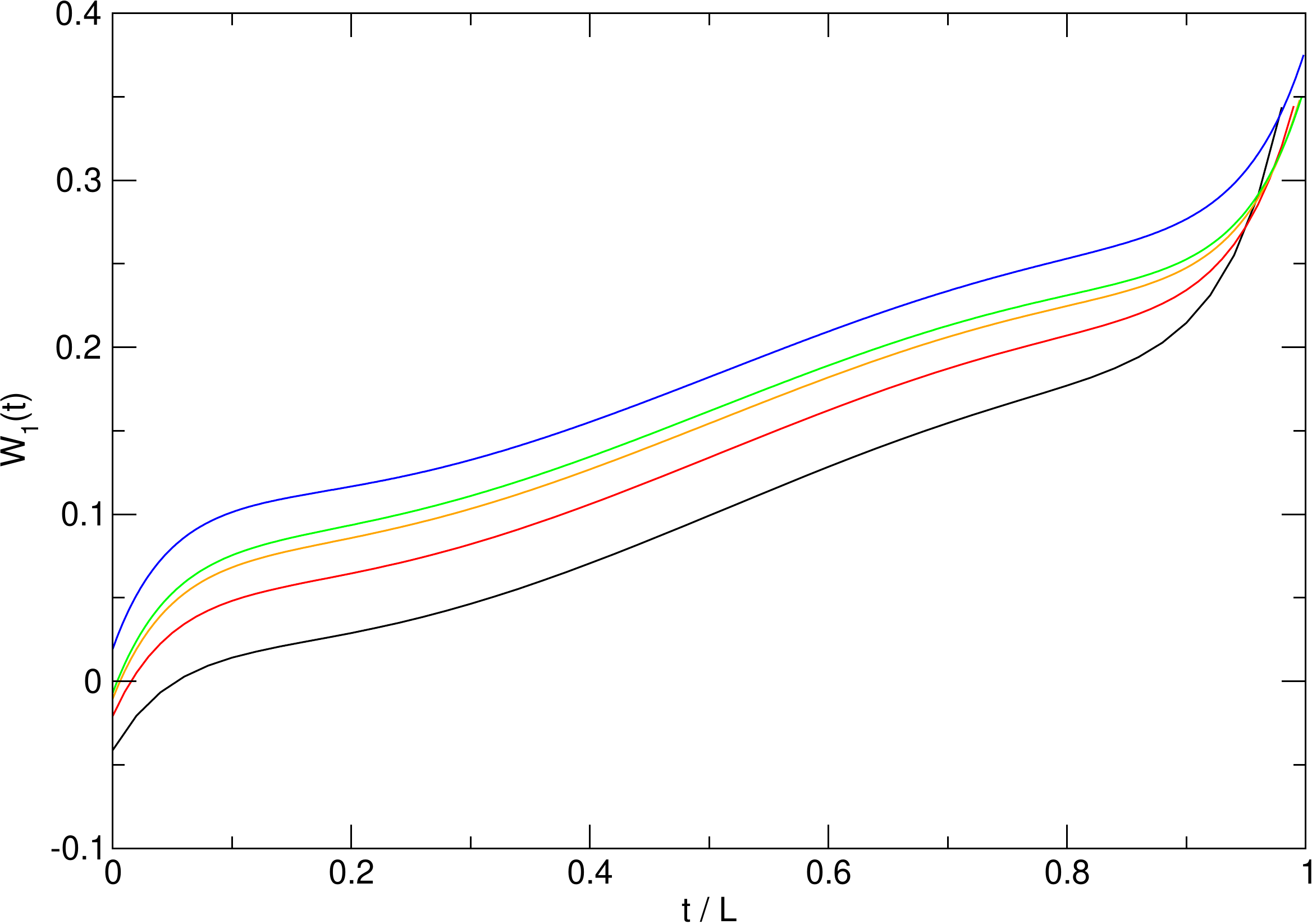}
        }\\%
        \subfigure[$\mu L = 10, Z_{a}$]{%
            \label{fig_121WI_I_b}
            \includegraphics[width=0.48\textwidth]{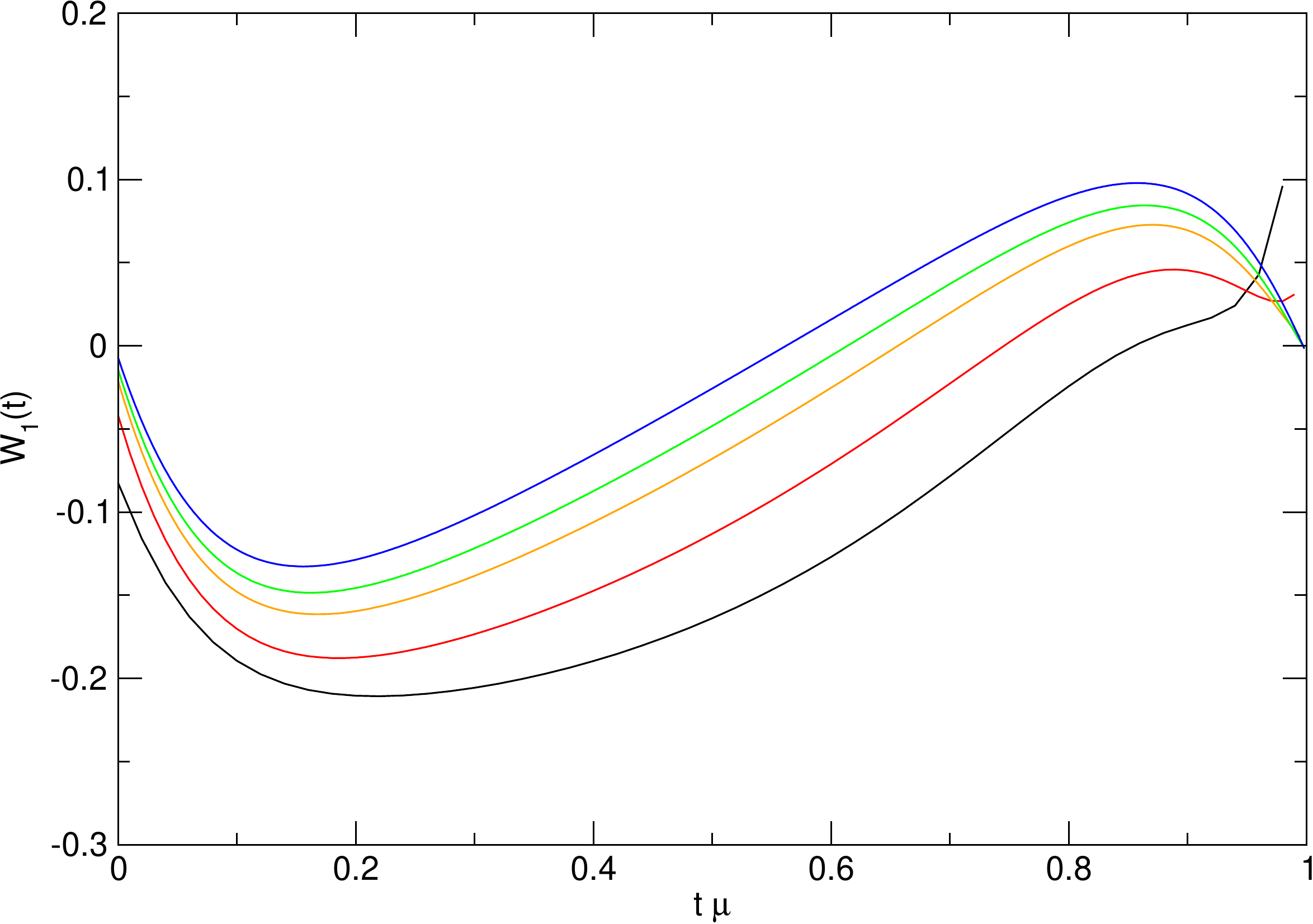}
        }%
        \subfigure[$\mu L = 10, Z_{p}$]{%
           \label{fig_22WI_I_b}
           \includegraphics[width=0.48\textwidth]{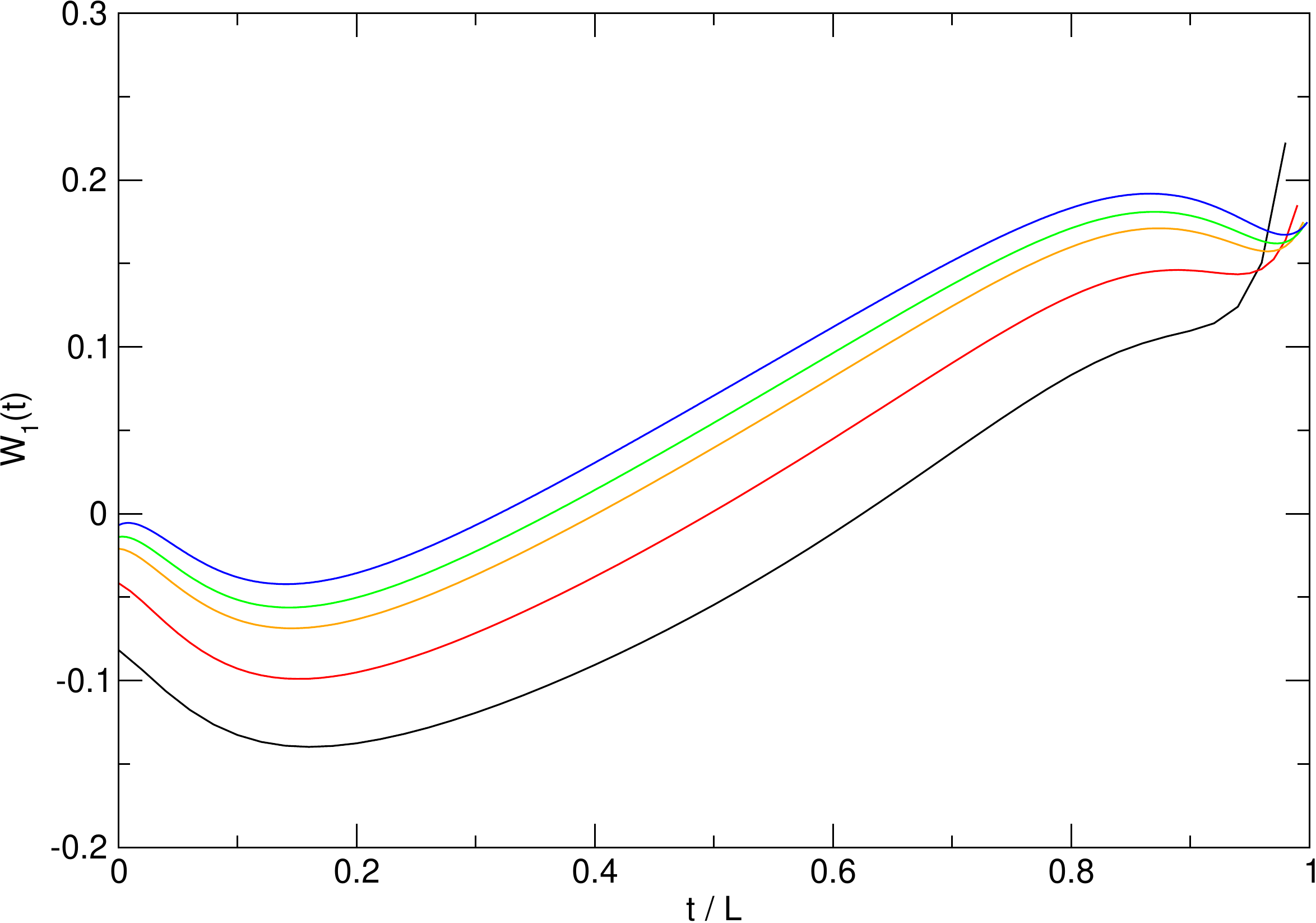}
        }
        \caption{Broken supersymmetric quantum mechanics, standard
          discretisation. The Ward identity $W_1$ for $L/a = 50$
          (black), $L/a = 100$ (red), $L/a = 200$ (orange), $L/a =
          300$ (green) and $L/a = 600$ (blue) for $\mu L=5$ and $\mu
          L=10$ at fixed $f_b = 1$.}
    \label{fig_WI_I_b}
\end{figure}

\begin{figure}
 \centering
  \includegraphics[width = 0.8\textwidth]{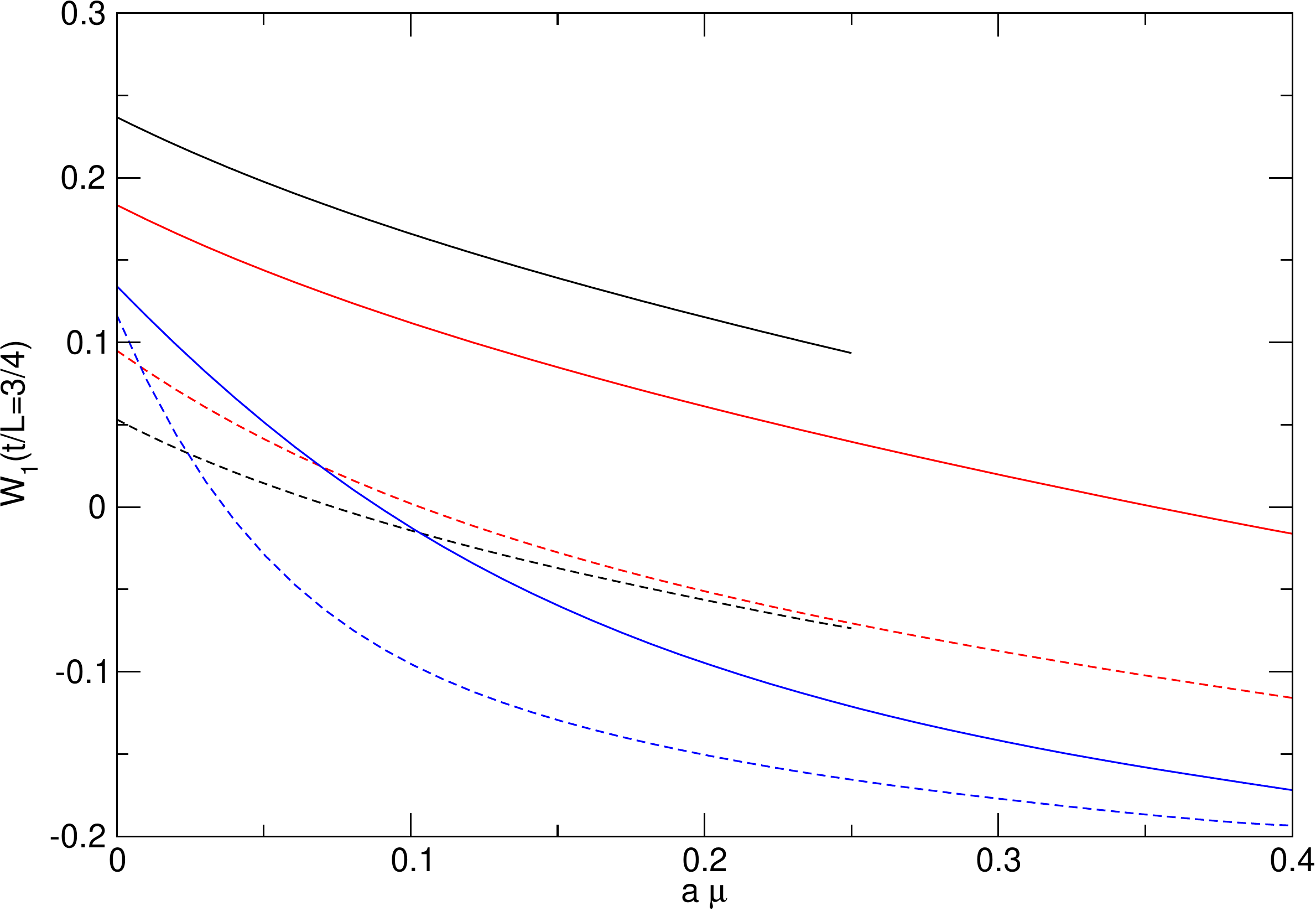}
  \caption{Broken supersymmetric quantum mechanics, standard
    discretisation. Continuum extrapolation of $W_1(t/L=3/4)$ for $\mu L = 5$
    (black), $\mu L = 10$ (red) and $\mu L = 20$ (blue) for periodic
    (solid lines) and antiperiodic b.c.~(dashed lines) at fixed
    coupling $f_b = 1$.}
  \label{fig_WI_I_b_c}
\end{figure}

Next, we consider the Ward identity $W_1$ for broken supersymmetry
using the standard discretisation.  In figure \ref{fig_WI_I_b}, we
show $W_1$ for $\mu L = 5$ and $\mu L = 10$ for a range of lattice
spacings $a/L$ at fixed coupling $f_b = 1$ both for periodic and
antiperiodic boundary conditions.
This figure illustrates how the Ward identity $W_1$ is violated for
broken supersymmetry at finite lattice spacing. However, unlike in the
previous case of unbroken supersymmetry, the violation of the Ward
identity $W_1$ remains finite even when the lattice spacing or the
temperature goes to zero. To illustrate this further, we trace the
Ward identity $W_1(t/L=3/4)$ into the continuum for different $\mu L$
in figure \ref{fig_WI_I_b_c}.
Clearly, the violation persists in the continuum, independently of the
boundary conditions and the size or temperature of the system. In this case too, all features of broken supersymmetry are numerically confirmed on the level of the Ward identity $W_1$.

\begin{figure}
 \centering
  \includegraphics[width = 0.8\textwidth]{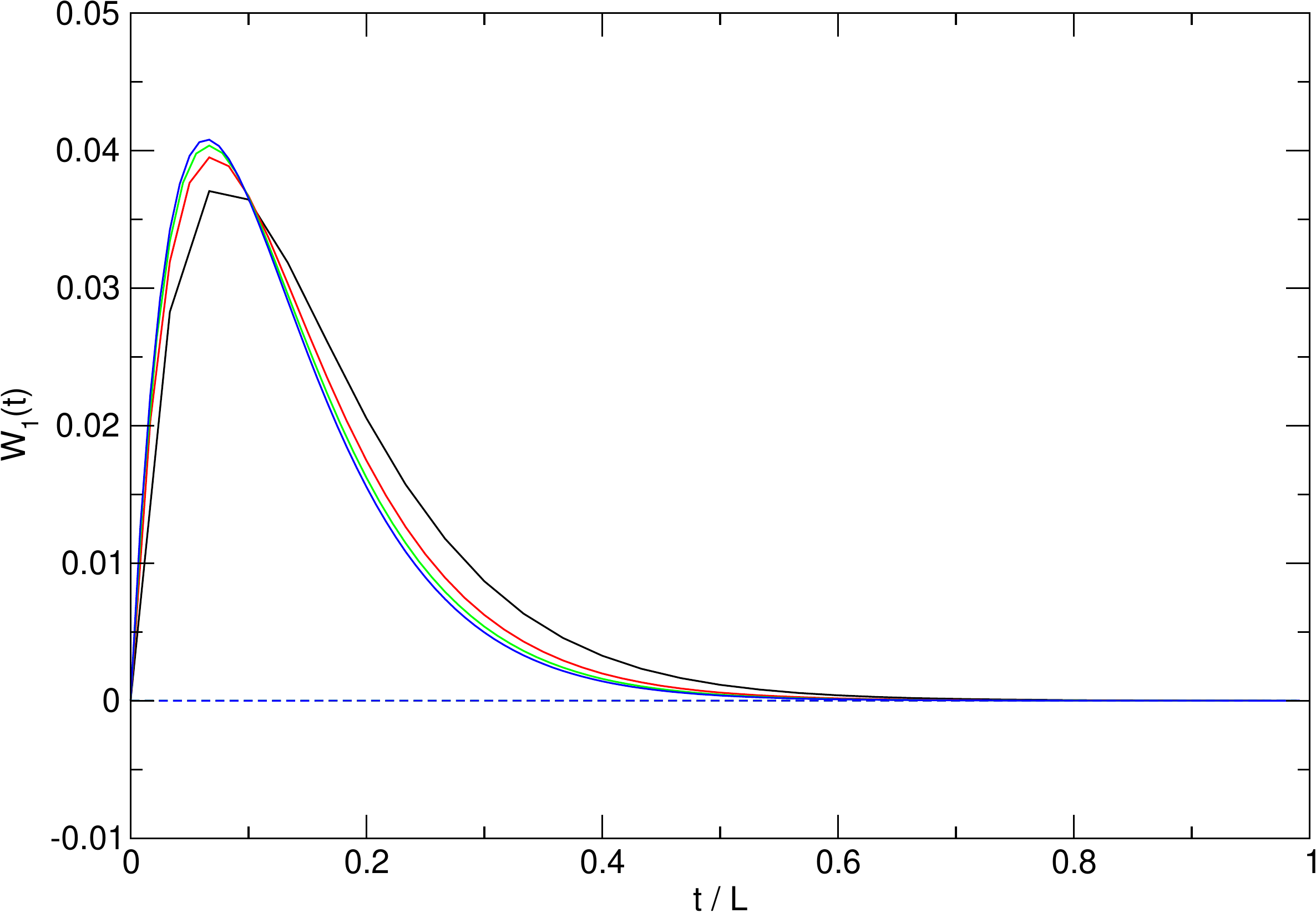}
  \caption{Unbroken supersymmetric quantum mechanics, $Q$-exact
    discretisation. The Ward identity $W_1(t)$ for $L/a = 30$ (black),
    $L/a = 60$ (red), $L/a = 90$ (green) and $L/a = 120$ (blue) for
    antiperiodic (solid lines) and periodic b.c.~(dashed lines) for
    $\mu L = 10$ at fixed coupling $f_u = 1$.}
  \label{fig_WI_I_mL_10_u_q}
\end{figure}

\begin{figure}
 \centering
  \includegraphics[width = 0.8\textwidth]{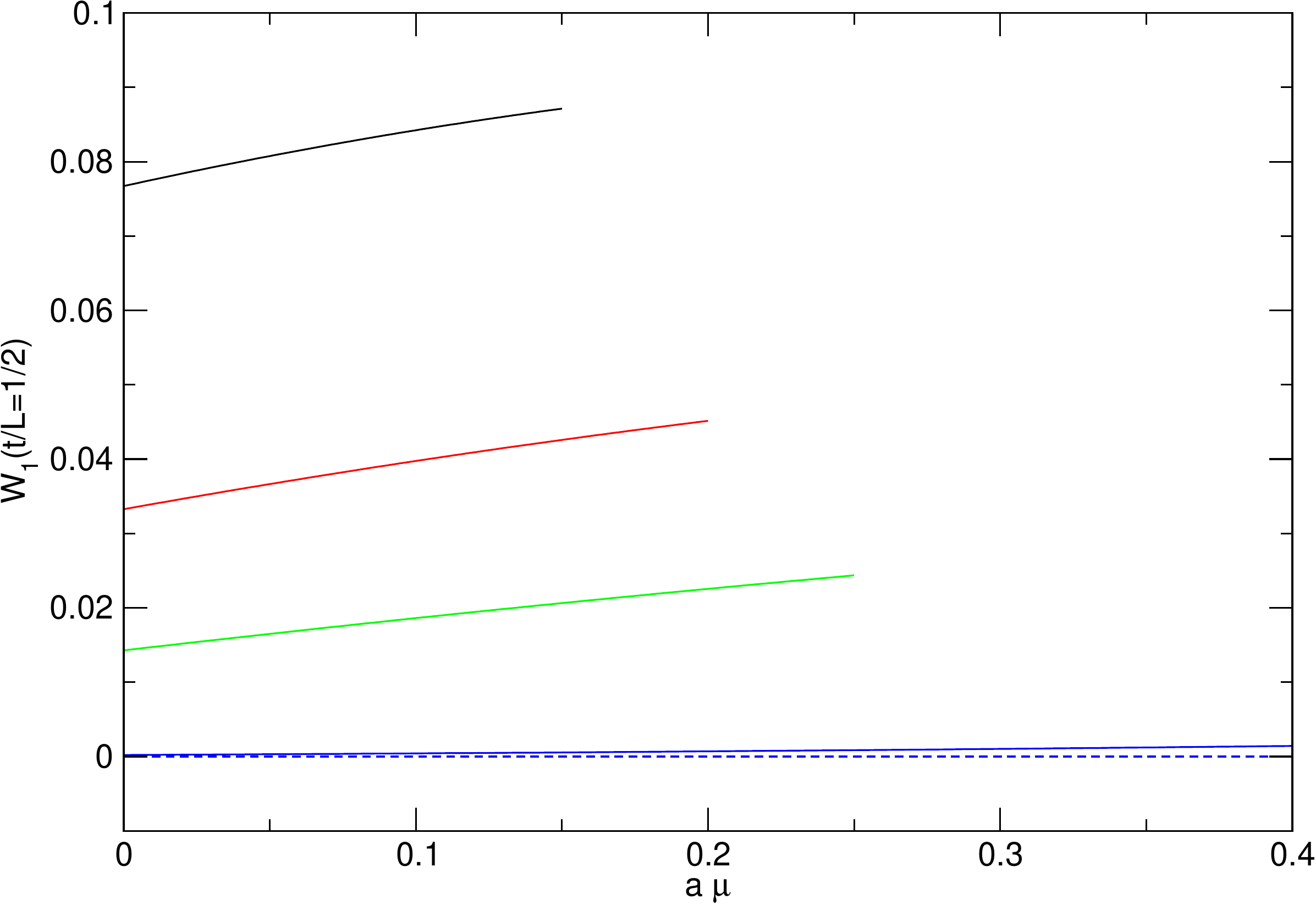}
  \caption{Unbroken supersymmetric quantum mechanics, $Q$-exact
    discretisation. Continuum extrapolation of $W_1(t/L=1/2)$ for $\mu
    L = 3$ (black), $\mu L = 4$ (red), $\mu L = 5$ (green) and $\mu L
    = 10$ (blue) for antiperiodic (solid lines) and periodic
    b.c.~(dashed lines) at fixed coupling $f_b = 1$.}
  \label{fig_WI_I_u_q}
\end{figure}

Next we consider the Ward identity $W_1$ for unbroken supersymmetry
using the $Q$-exact discretisation. In this case $W_1$ is of special
interest since for this action $\delta_1 S_L^{Q} = 0$ at finite
lattice spacing, and we should hence be able to confirm that $W_1(t) =
0$ exactly $\forall t$ at finite lattice spacing for unbroken
supersymmetry and periodic boundary conditions.
In figure \ref{fig_WI_I_mL_10_u_q} we show the Ward identity $W_1$ at
different values of the lattice spacing $a/L$ at fixed coupling $f_u =
1$ and fixed extent $\mu L=10$ for both periodic and antiperiodic boundary conditions.
The plot shows that the Ward identity $W_1$ represented by the dashed
line is indeed zero $\forall t$ using periodic boundary conditions at
any finite lattice spacing. Note, that $W_1(t)$ is composed of the
bosonic and fermionic correlators as given in eq.(\ref{eq:WI
  unbroken}) and is in fact nontrivially zero. For antiperiodic
b.c.~on the other hand, the violation of the Ward identity at finite
temperature is evident. To observe the behaviour of the Ward identity
$W_1$ in the zero temperature limit, in Fig.~\ref{fig_WI_I_u_q} we again trace $W_1(t/L=1/2)$
into the continuum for different $\mu L$. 
Of course, for periodic b.c.~$W_1(t/L=1/2)$ is zero for any finite $a
\mu$ and any value of $\mu L$ (dashed lines). However, for
antiperiodic b.c., the extrapolation of $W_1(t/L=1/2)$ shows a
dependence on $\mu L$, but in the limit $\mu L \rightarrow \infty$
this violation also vanishes, as expected.

\begin{figure}
 \centering
  \includegraphics[width = 0.8\textwidth]{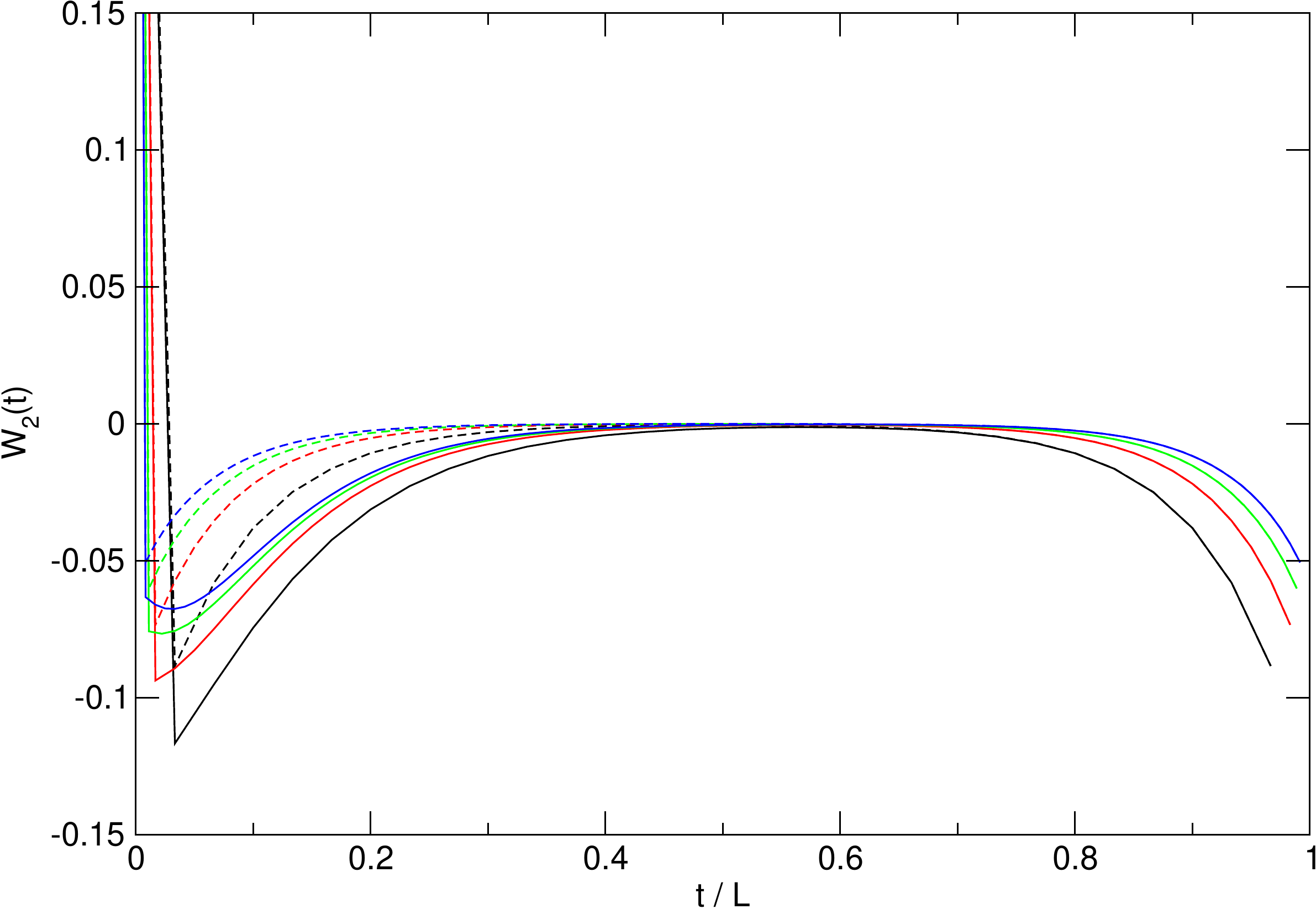}
  \caption{Unbroken supersymmetric quantum mechanics, $Q$-exact
    discretisation. The Ward identity $W_2$ for $L/a = 30$ (black),
    $L/a = 60$ (red), $L/a = 90$ (green) and $L/a = 120$ (blue) for
    antiperiodic (solid lines) and periodic b.c.~(dashed lines) for
    $\mu L = 10$ at fixed coupling $f_u = 1$.}
  \label{fig_WI_II_mL_10_u_q}
\end{figure}

\begin{figure}
 \centering
  \includegraphics[width = 0.8\textwidth]{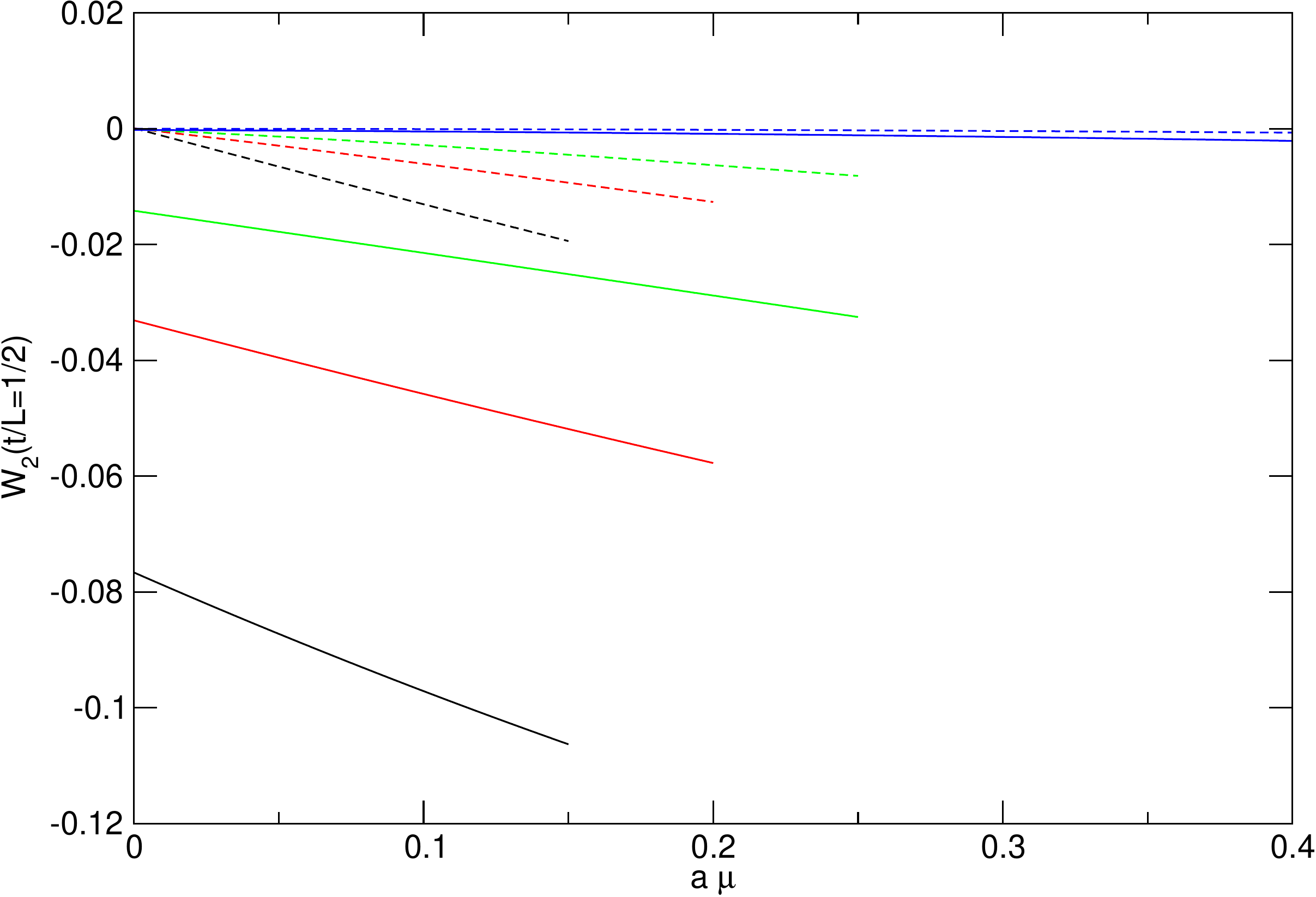}
  \caption{Unbroken supersymmetric quantum mechanics, $Q$-exact
    discretisation. Continuum extrapolation of $W_2(t/L=1/2)$ for $\mu
    L = 3$ (black), $\mu L = 4$ (red), $\mu L = 5$ (green) and $\mu L
    = 10$ (blue) for antiperiodic (solid lines) and periodic
    b.c.~(dashed lines) at fixed coupling $f_u = 1$.}
  \label{fig_WI_II_u_q}
\end{figure}

We now perform the same analysis for the Ward identity $W_2$ given in
eq.(\ref{eq_WI_II}).  This Ward identity is not expected to vanish for
finite lattice spacing, since the action $S_L^{Q}$ is not invariant
under the supersymmetry transformation $\delta_2$. In figure
\ref{fig_WI_II_mL_10_u_q} we show $W_2(t)$ for different lattice
spacings $a/L$ for $\mu L = 10$ at fixed coupling $f_u = 1$.
As expected, this Ward identity is violated for both periodic and
antiperiodic b.c.~at finite lattice spacing and for finite
temperature. To observe the continuum behaviour, we trace
$W_2(t/L=1/2)$ in this case, too. The continuum extrapolation for
different $\mu L$ is shown in figure \ref{fig_WI_II_u_q}.
For periodic b.c., the violation of the Ward identity $W_2(t/L=1/2)$
vanishes in the continuum independently of the chosen $\mu L$. The
restoration of supersymmetry in the continuum is thus also confirmed
via the Ward identity $W_2(t/L=1/2)$. For antiperiodic b.c.~however,
the violation does not vanish for small $\mu L$. Again, this is just a
reflection of the fact that the finite temperature breaks the
supersymmetry, and it is only restored in the zero temperature limit.
Hence, on the level of the Ward identities $W_1$ and $W_2$, all the
features of unbroken supersymmetry formulated with the $Q$-exact
action are numerically confirmed. Analogously to the standard
discretisation $W_2(t/L=1/2)$ extrapolates to zero at any value of the
lattice spacing, independent of the employed boundary conditions. This
confirms that the violation of the supersymmetry $\delta_2$ in the
$Q$-exact formulation is just a boundary term which decouples from the
system in the limit $\mu L \rightarrow \infty$.
\subsection{The ground state energy $E_0$}
\label{subsec:ground state energy results}
\begin{figure}
 \centering
  \includegraphics[width = 0.8\textwidth]{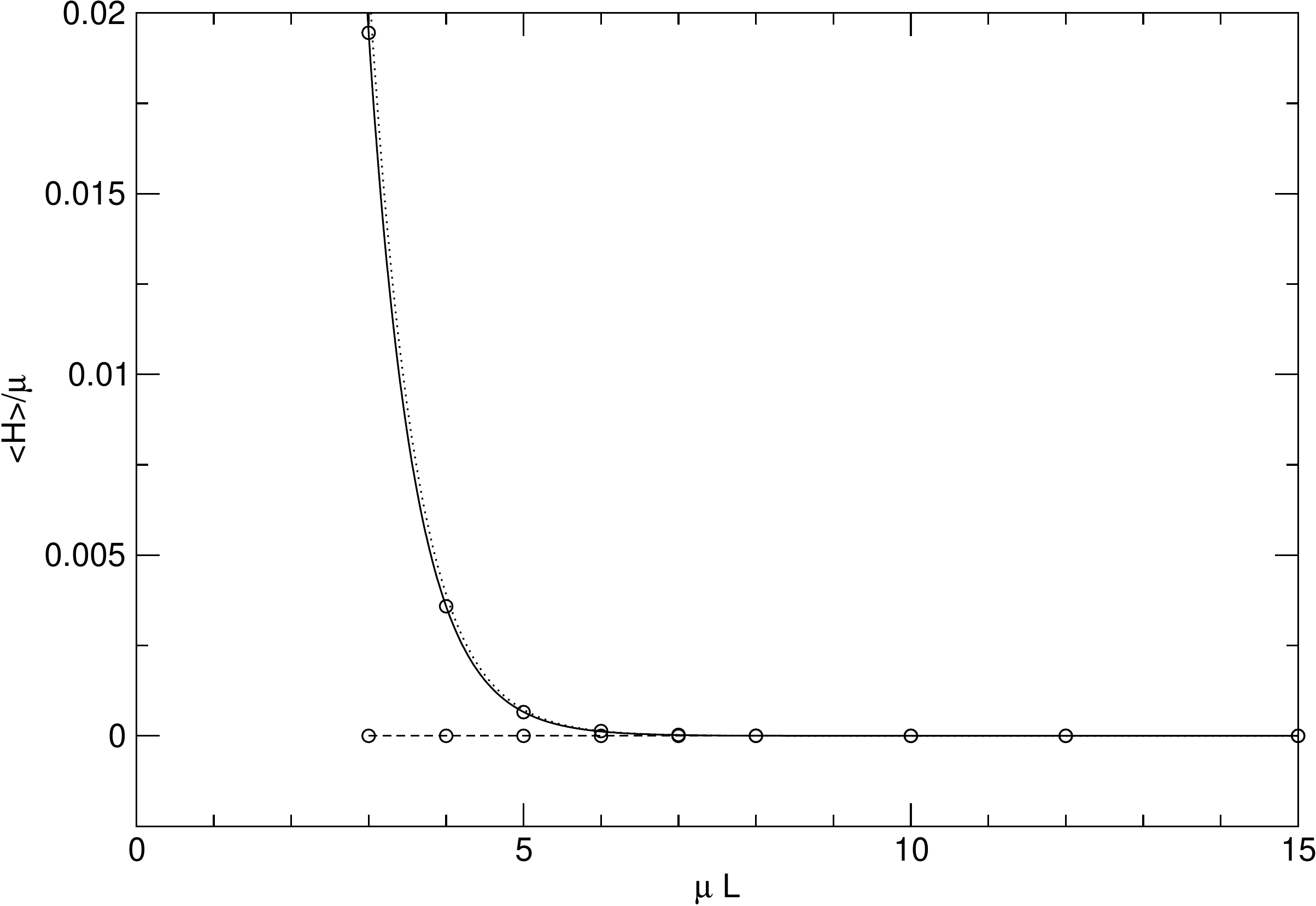}
  \caption{Unbroken supersymmetric quantum mechanics, $Q$-exact
    action. Continuum values of $\langle H \rangle / \mu = E_0/\mu$
    for periodic (dashed line) and antiperiodic b.c.~(solid line) for
    a range of system sizes $\mu L$ at fixed coupling $f_u = 1$. The
    dotted line describes the leading asymptotic behaviour for large
    $\mu L$ while the full line is a phenomenological fit.}
  \label{fig_H}
\end{figure}

In this section we follow \cite{Kanamori:2007yx} and measure the
ground state energy $E_0$ for the $Q$-exact action via the expectation
value of an appropriate Hamilton operator $H$.  In a field theory it
is a priori not clear how to measure an absolute energy and there are
in fact several possible candidate Hamilton operators which differ
from each other by constant shifts. However, the authors of
\cite{Kanamori:2007yx} argue via the off-shell formulation of the
theory, that constructing the Hamilton operator from the $Q$-exact
action leads to the correct measurement of the ground state energy.
In the lattice formulation, it reads
\begin{equation}
  H = -\frac{1}{2}(\Delta^- \phi)^2 + \frac{1}{2}\left( P^\prime
  \right)^2 - \frac{1}{2}\psibar(\Delta^- - P^{\prime \prime}) \psi \, .
\end{equation}
Using the superpotential $P_u$ in eq.(\ref{eq:superpotential P_u}),
the expectation value of this Hamilton operator is explicitly given by
\begin{eqnarray}\label{H_op}\nonumber
  \langle H \rangle & = & \frac{1}{2}(\mu^2 - 2)\langle \phi^2 \rangle + \mu g \langle \phi^4 \rangle + \frac{1}{2}g^2\langle \phi^6\rangle + \langle \phi_1 \phi_0 \rangle \\
  && + \frac{1}{2}(\mu - 1) \langle \psibar \psi \rangle + \frac{1}{2} \langle \psibar_1 \psi_0 \rangle + \frac{3}{2}g^2\langle \psibar \psi \phi^2 \rangle.
\end{eqnarray}
In figure \ref{fig_H} we show the continuum values for $\langle H
\rangle /\mu$ for different $\mu L$ for both periodic and antiperiodic
b.c.~at a coupling $f_u = 1$.
For periodic b.c.~the operator $H$ yields zero independently of the
lattice spacing $a/L$ and $\mu L$. Here, too, this zero is nontrivial,
since it emerges from an exact cancellation of the various expectation
values in eq.(\ref{H_op}).  For antiperiodic b.c.~the continuum values
show an exponentially decreasing behaviour with $\mu L$ and the
expectation value $\langle H \rangle/\mu$ goes to zero only in the
limit $\mu L \rightarrow \infty$. The exponential behaviour can easily
be inferred from expanding the expectation value in terms of the
energy states. Taking only the lowest mass gap into account one
obtains
\begin{equation}
\label{eq:lowest order ansatz for H}
\langle H \rangle = \frac{2 \, m_1^b \, e^{-m_1^b L}}{1+2 \, e^{-m_1^b
    L}} \, .
\end{equation}
The dotted line in figure \ref{fig_H} corresponds to this expression
with $m_1^b/\mu = 1.6865$, in agreement with our results in section
\ref{subsec:mass_gaps_results}. The full line is a phenomenological
fit using $m_1^b$ in eq.(\ref{eq:lowest order ansatz for H}) as an
effective fit parameter which also takes into account additional
contributions from higher excitations. In conclusion, our exact
results confirm the arguments presented in \cite{Kanamori:2007yx}.

\section{Conclusions and outlook}
\label{sec:conclusion}
In this paper we have presented exact results for ${\cal N}=2$
supersymmetric quantum mechanics discretised on the
lattice. Expressing the bosonic and fermionic degrees of freedom in
terms of bosonic and fermionic bonds, respectively, allows to
completely characterise the system by means of transfer matrices
defined separately in the bosonic and fermionic sector. From the
properties of the transfer matrices one can derive exact results for
all observables at finite lattice spacing and we present such results
for a variety of interesting observables using two different
discretisation schemes. The first is the standard discretisation which
involves a Wilson term for the fermions and a counterterm which
guarantees the restoration of supersymmetry in the continuum. The
second discretisation is a $Q$-exact one which maintains one of the
two supersymmetries exactly at finite lattice spacing
\cite{Catterall:2000rv}.  The exact calculations allow to study in
detail how the continuum limit $a\mu\rightarrow 0$ as well as the
thermodynamic limit $\mu L \rightarrow \infty$ are approached and how
the two limits interfere with each other. The latter of the two limits
can be interpreted as the zero temperature limit in a system with
antiperiodic b.c.~for the fermion. Since the supersymmetry of the
system can be broken both by the finite lattice spacing and the finite
temperature the interplay of the two limits is of particular interest
in order to gain a complete understanding of the various lattice
discretisation schemes.

For the ratio of partition functions $Z_p/Z_a$, which is proportional
to the Witten index, we find for example the interesting result that
in a system with broken supersymmetry, where the Witten index is
supposed to vanish, it extrapolates to $-1$ in the zero temperature
limit at any finite lattice spacing. On the other hand, it
extrapolates to 0 in the continuum limit for any finite temperature or
extent of the system. In fact it turns out that the lattice spacing
corrections are exponentially enhanced towards the low temperature
limit, so in this case the order of the limits is crucial to describe
the correct physics in the continuum.  It is also interesting to study
the influence of the finite lattice spacing on the fermionic and
bosonic two-point correlation functions. In particular, for broken
supersymmetry one expects the emergence of a massless Goldstino mode
and within our approach we can study in detail how the mode expresses
itself in the fermionic correlation function. Moreover, we study the
bosonic and fermionic spectrum of the theory which allows to better
quantify the lattice corrections. We demonstrate how the degeneracy
between the bosonic and fermionic excitations is restored in the
continuum both for broken and unbroken supersymmetry when the standard
discretisation scheme is used. For broken supersymmetry we see how the
finite lattice spacing regulates the Goldstino mode and hence also the
vanishing Witten index. Although the coupling strengths we study are
well in the nonperturbative regime, the leading lattice corrections in
the spectrum turn out to be reasonably small and follow the usual
expectations of being $\O(a)$ to leading order. For the $Q$-exact
discretisation scheme we find exact degeneracy between the fermionic
and bosonic excitations at any finite lattice spacing. It seems that
maintaining only one of the supersymmetries on the lattice is
sufficient to guarantee the exact degeneracy. In this case, too, the
lattice artefacts are $\O(a)$ to leading order, but appear to be
enhanced with respect to the standard discretisation, in particular
for the higher lying excitations.

We are also able to study in detail the behaviour of various Ward
identities towards the continuum and thermodynamic limits for both
discretisation schemes. Our exact results show that the finite lattice
spacing and finite temperature effects can sometimes be rather large,
but nevertheless both supersymmetries are completely restored in the
appropriate limits without any surprises. The Ward identities $W_1$
and $W_2$ play a particularly important role for the $Q$-exact
discretisation. Since in that case half of the supersymmetries is
exactly maintained, some of the Ward identities are expected to be
fulfilled at finite lattice spacing for periodic boundary
conditions. We prove numerically that this is indeed the
case. Finally, for the $Q$-exact discretisation we also demonstrate
the correctness of the conjecture in
\cite{Kanamori:2007yx,Kanamori:2010gw} which provides a scheme to
calculate the ground state energy.

In conclusion, we now have a rather complete qualitative and
quantitative understanding of the interplay between infrared and
ultraviolet effects in supersymmetric quantum mechanics regulated on a
lattice of finite extent and finite lattice spacing. Moreover, our
exact results provide a benchmark for any attempt to deal with
supersymmetric field theories using a new discretisation scheme, or in
fact even for any new regularisation scheme such as, e.g., the one
described in \cite{Hanada:2007ti}. In addition, new simulation
algorithms specific to supersymmetric theories can be tested against
our exact results. For example, there exist particular algorithms
which are tailored to efficiently simulate bond occupation numbers, be
they bosonic \cite{Prokof'ev:2001} or fermionic
\cite{Wenger:2008tq}. In fact, in the third paper of our series
\cite{Baumgartner:2015zna} we present the practical application of the
so-called open fermion string algorithm to supersymmetric quantum
mechanics in the bond formulation and prove its feasibility to deal
numerically with the sign problem associated with broken
supersymmetry. Our exact results here provide the necessary background
to assess the validity and success of the numerical simulations using
the fermion loop approach. Similarly, alternative approaches which
attempt or claim to solve fermion sign problems, such as the ones in
\cite{Chandrasekharan:2009wc,Chandrasekharan:2012fk,Chandrasekharan:2013rpa},
can be tested in supersymmetric quantum mechanics and gauged against
the exact results presented here.

An important question is of course whether the bond formulation and
transfer matrix approach outlined here can be extended and applied to
more complicated systems. This is indeed possible as we demonstrated
in \cite{Steinhauer:2014oda} where the fermion loop approach is
applied to supersymmetric Yang-Mills quantum mechanics. In that
system, transfer matrices describing the fermionic degrees of freedom
can also be constructed explicitly in each sector with fixed fermion
number and it is shown how they are related to the standard canonical
approach. It is interesting to note that, despite the model involving
a gauge degree of freedom, the transfer matrix approach can be
extended to handle also this situation. Concerning the extension of
the approach to higher dimensions the perspectives are not so
bright. Up to a few exceptions, it is in general not possible or
practical to construct transfer matrices for systems in higher
dimensions. In contrast, the fermion loop formulation can be used on
its own, e.g.~\cite{Maillart:2011hz,Steinhauer:2013tba}, and in some
cases even provides the basis for the solution of the fermion sign
problem such as in the ${\cal N}=1$ Wess-Zumino model
\cite{Baumgartner:2011cm,Steinhauer:2014yaa}.

\begin{appendix}
\section{Technical aspects}
\label{app:technical aspects}
\subsection{Cutoff procedure for the bond occupation numbers}
\label{app:technical aspects cutoff}
In this appendix we briefly describe and illustrate our procedure to
choose an appropriate cutoff for the bond occupation numbers. The
introduction of the cutoff is necessary in order to construct transfer
matrices of finite size, such that they can be handled numerically. In
the bond formulation the weights involving large bond occupation
numbers are suppressed by factors of $1/n^b_i!$, so their
contributions become irrelevant as the occupation numbers grow. The
truncation of the hopping expansion hence provides a natural and
systematic scheme to limit the size of the transfer matrices.

\begin{figure}
 \centering
 \includegraphics[width = 0.8\textwidth]{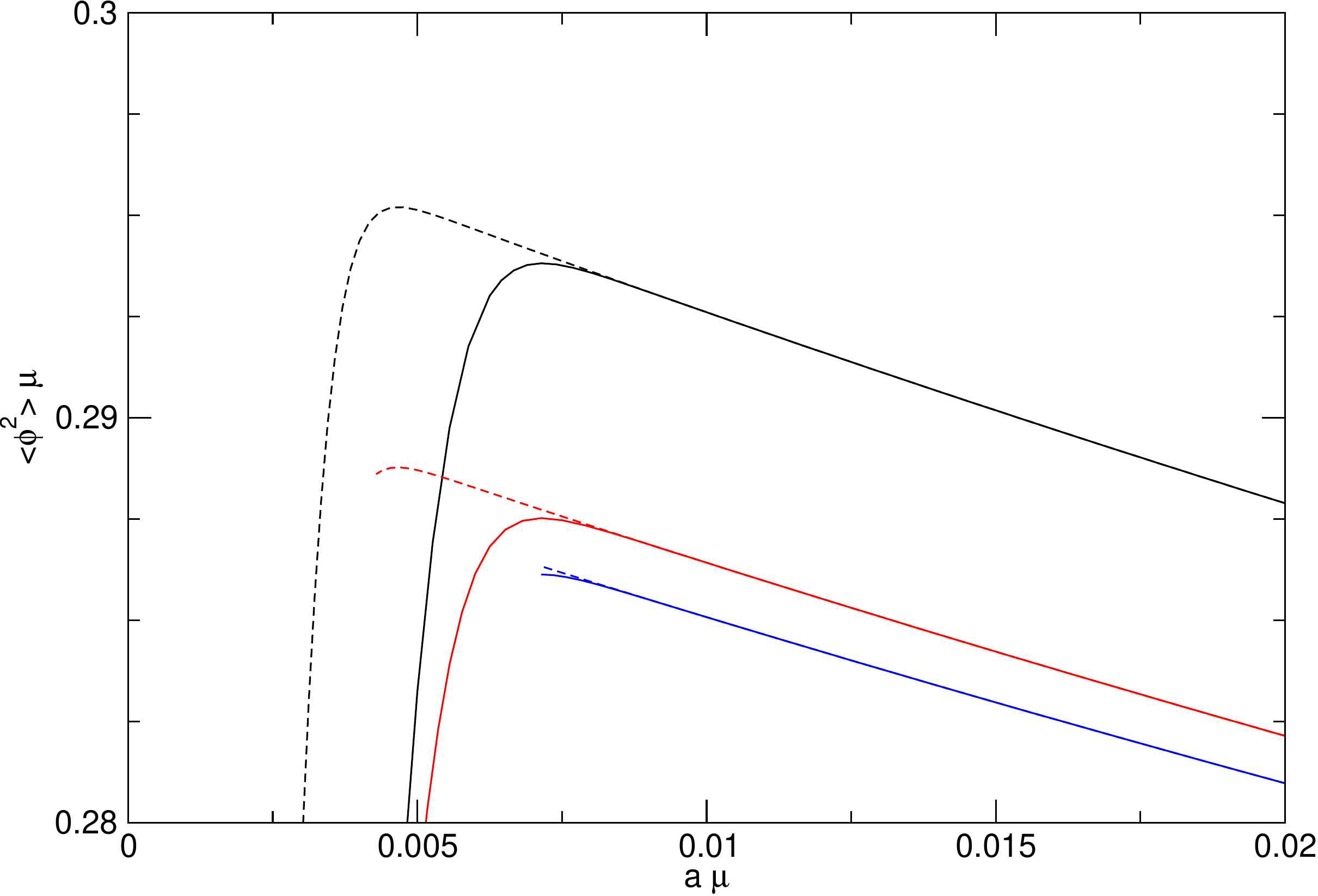}
 \caption{$\langle \phi^2 \rangle_{a} \cdot \mu$ as a function of $a
   \mu$ at fixed coupling $f_u = 1$ for different $\mu L = 2$ (black),
   $\mu L = 3$ (red), and $\mu L = 5$ (blue) and $N^{cut}_{1
     \rightarrow 1} = 800$ (dashed line) and $N^{cut}_{1 \rightarrow
     1} = 500$ (solid line). Effects from the finite cutoff on the
   bosonic bond occupation numbers near the continuum are clearly
   visible.}
  \label{fig_ex_breakdown}
\end{figure}

For the standard discretisation we only have one type of bosonic bond
$b_{1\rightarrow 1}$ and hence the size of the transfer matrix grows
linearly with the cutoff $N^{cut}_{1\rightarrow 1}$ on the occupation
number $n_{1\rightarrow 1}^b$. Calculating an observable at different
lattice spacings with varying cutoff $N^{cut}_{1\rightarrow 1}$,
effects from the finite bosonic cutoff manifest themselves as a sudden
bend in an otherwise linear curve close to the continuum. In figure
\ref{fig_ex_breakdown} we show an example of this effect by means of
the expectation value $\langle \phi^2 \rangle_{a} \cdot \mu$ for
antiperiodic b.c.~and unbroken supersymmetry as a function of the
lattice spacing $a \mu$ for different values of $\mu L$ at fixed
coupling $f_u = 1$. The effect of the finite cutoff for the bond
occupation numbers is illustrated by comparing the observable for two
different cutoffs, $N_{1 \rightarrow 1}^{cut} = 800$ and $N_{1
  \rightarrow 1}^{cut} = 500$ close to the continuum.
The curves for the expectation value are indistinguishable for $a \mu
\gtrsim 0.075$, but closer to the continuum, the curve for the smaller
cutoff suddenly diverges from the curve for the larger cutoff, and the
values obtained using the lower cutoff are no longer reliable. For the
larger cutoff a similar effect appears at a smaller lattice spacing,
but is again clearly visible. So for any given cutoff, the results are
reliable only down to a specific lattice spacing, which however is
easy to determine since the cutoff effects are so dramatic. It turns
out that for the observables considered in this paper, a cutoff $N_{1
  \rightarrow 1}^{cut} = 800$ is sufficient to safely reach a lattice
spacing $a \mu \sim 0.005$, well in the regime where the dominating
lattice artefacts are of order $\O(a)$ and the corrections of $O(a^2)$
are very small. It is then safe to extrapolate the data to the
continuum by fitting a quadratic function
\begin{equation}
 f(x) = c_0 + c_1 x + c_2 x^2
\end{equation}
to the data with $a\mu \gtrsim 0.005$ while making sure that the data
is not affected by a change of the cutoff around $N_{1 \rightarrow
  1}^{cut} = 800$. For almost all observables, these fits can be
performed without any difficulties. In cases where the lattice
artefacts turn out to be particularly large, higher corrections can be
taken into account without any problems and we indicate in the
discussion when we do so.

For the $Q$-exact discretisation, we have two types of bosonic bonds
$b_{1\rightarrow 1}$ and $b_{1\rightarrow \nu}$ and we need to
introduce two cutoffs $N_{1\rightarrow 1}^{cut}$ and $N_{1 \rightarrow
  \nu}^{cut}$ on the corresponding occupation numbers, hence the size
of the transfer matrices grows quadratically in the cutoff, i.e., as
$N_{1\rightarrow 1}^{cut} \cdot N_{1\rightarrow \nu}^{cut}$.
Nevertheless, it turns out that also in this case the onset of cutoff
effects in the observables is clearly indicated by a sudden bend away
from the leading (linear) behaviour expected towards the continuum $a
\mu \rightarrow 0$. Typically we choose the cutoffs $N_{1 \rightarrow
  1}^{cut} = 64$ and $N_{1 \rightarrow \nu}^{cut} = 16$ for our
calculations, yielding transfer matrices of size $1105 \times 1105$.
For the extrapolations, we proceed analogously to the case with the
standard discretisation and we find that extrapolating the exact
results with quadratic fits allows for reliable continuum results for
the $Q$-exact discretisation, too.

\subsection{Construction of the transfer matrix elements}
\label{app:technical aspects construction}
In this appendix we briefly comment on the construction of the
transfer matrices. As we emphasised at the end of Section
\ref{subsec:transfer matrices and partition functions}, the evaluation
of the site weights $Q_F(N)$ tends to be numerically unstable for
large values of $N$. In the third paper of our series
\cite{Baumgartner:2015zna} we will discuss an algorithm which allows to
reliably calculate the ratios
\begin{equation}
  R_F'(N)\equiv \frac{Q_F(N+1)}{Q_F(N)}, \quad R_F(N) \equiv
  \frac{Q_F(N+2)}{Q_F(N)}, \quad R_m(N) \equiv \frac{Q_0(N)}{Q_1(N)}
\end{equation}
for large $N={\cal O}(1000)$. Here we discuss how the transfer matrix
elements in eq.(\ref{eq:transfer matrix generic}) can be constructed
using these ratios while avoiding arithmetic over- or underflow.

First we note that for the calculation of the mass gaps or expectation
values, the overall normalisation of the transfer matrices is not
relevant, as long as all matrices are normalised consistently. We can
therefore rescale all matrix elements by a constant factor,
e.g., $Q_1(0)$ which is of order ${\cal O}(1)$. The contribution of the site
weight to the matrix elements then becomes $Q_F(N)/Q_1(0)$ and this can
be evaluated as
\begin{equation}
\frac{Q_F(N)}{Q_1(0)} = R_m(0)^{1-F} \cdot \prod_{n=0}^{N-1} R'_F(n) \, .
\end{equation}
In case the ratios $R'_F$ are ill-defined, e.g.~when employing the superpotential
$P_u$ for which only
$N=0\mod 2$ is allowed, they are replaced by $R_F(n)$ and the product
runs up to $N-2$. Even when the
ratios $R'_F$ or $R_F$ are ${\cal O}(1)$ the product can lead to
arithmetic overflow when $N$ is ${\cal O}(1000)$. In that case it is
advisable to evaluate the product logarithmically, i.e., as 
\begin{equation}
\ln \prod_{n} R'_F(n) = \sum_{n} \ln R'_F(n) 
\label{eq:logsum weights}
\end{equation}
and analogously for $R_F$.

In addition, the factorials in eq.(\ref{eq:transfer matrix
  generic}) may lead to arithmetic underflow when $n_i^b$ or $m_i^b$ is ${\cal
  O}(300)$. Also in this case it is advisable to calculate the
factorials logarithmically,
\begin{equation}
\ln \sqrt{\frac{1}{n_i^b!}}= -\frac{1}{2} \sum_{n=0}^{n_i^b} \ln n \, ,
\label{eq:logsum factorials}
\end{equation}
and possibly combine the result with the one from eq.(\ref{eq:logsum weights})
before exponentiating. Finally, it is important to note that the
logarithmic sums should be chosen only when the over- or underflow de facto
occurs, as otherwise the loss of precision in eq.(\ref{eq:logsum
  weights}) and (\ref{eq:logsum factorials}) might be sufficient to
yield inaccurate results.

\end{appendix}
%
\bibliography{susyQM_ExactResults}
\bibliographystyle{JHEP}
\end{document}